\documentclass[12pt,usenames,dvipsnames]{article}

\usepackage{latexsym}
\usepackage{amssymb,amsfonts,amsmath}
\usepackage{graphicx} 
\usepackage{indentfirst}
\usepackage{bbm}
\usepackage{tikz-qtree}
\usepackage{amssymb}
\usepackage[normalem]{ulem}
\usepackage{footmisc}
\usepackage{verbatim}
\usepackage{amsmath, amsthm, amssymb}
\usepackage{mathrsfs}
\usepackage{hyperref}
\usepackage{amsfonts}
\usepackage{dsfont}
\usepackage{cite}
\usepackage{xcolor}
\usepackage{enumitem}  
\usepackage[utf8]{inputenc}
\usepackage{xcolor}
\usepackage{hyperref} 

\hypersetup{backref=true,       
    pagebackref=true,               
    hyperindex=true,                
    colorlinks=true,                
    breaklinks=true,                
    urlcolor= black,                
    linkcolor= blue,                
    bookmarks=true,                 
    bookmarksopen=false,
    filecolor=black,
    citecolor=blue,
    linkbordercolor=blue
}

\AtBeginDocument{\hypersetup{pdfborder={0 0 1}}}

\parskip=\medskipamount

\newcommand {\cD}{{\cal D}}

\newcommand {\cN}{{\cal N}}

\newcommand {\cS}{{\cal S}}

\def\a{\alpha}

\def\b{\beta}
\def\c{\chi}
\def\d{\delta}

\def\f{\phi}
\def\g{\gamma}
\def\G{\Gamma}

\def\l{\lambda}

\def\r{\rho}
\def\s{\sigma}
\def\t{\tau}

\def\x{\xi}
\def\z{\zeta}
\def\D{\Delta}
\def\F{\Phi}

\def\L{\Lambda}
\def\O{\Omega}
\def\P{\Pi}
\def\Q{\Theta}

\def\U{\Upsilon}

\def\rd{{\rm d}}
\def\ri{{\rm i}}
\def\rT{{\rm T}}
\def\rL{{\rm L}}

\def\rLAL{{\rm LAL}}

\newcommand{\ad}{{\dot{\alpha}}}                           
\newcommand{\bd}{{\dot{\beta}}}                            
\newcommand{\ve}{\varepsilon}                            
\newcommand{\DB}{\bar{D}}

\newcommand{\pa}{\partial}                           

\newcommand{\be}{\begin{equation}}
\newcommand{\ee}{\end{equation}}
\newcommand{\bea}{\begin{eqnarray}}
\newcommand{\eea}{\end{eqnarray}}
\newcommand{\non}{\nonumber}
\newcommand{\bm}[1]{\mbox{\boldmath$#1$}}

\def\double #1{#1{\hbox{\kern-2pt $#1$}}}

\newcommand{\gd}{{\dot\g}}

\newif\ifdtup

\newcommand{\bsubeq}{\begin{subequations}}
\newcommand{\esubeq}{\end{subequations}}
\newcommand{\ba}{\begin{align}}
\newcommand{\eal}{\end{align}}

\numberwithin{equation}{section}

\newcommand{\sSU}{\mathsf{SU}}
\newcommand{\sSL}{\mathsf{SL}}

\newcommand{\mc}{\mathcal}
\newcommand{\mf}{\mathfrak}
\newcommand{\ms}{\mathscr}
\newcommand{\mb}{\mathbb}

\makeatletter
\DeclareFontFamily{OMX}{MnSymbolE}{}
\DeclareSymbolFont{MnLargeSymbols}{OMX}{MnSymbolE}{m}{n}
\SetSymbolFont{MnLargeSymbols}{bold}{OMX}{MnSymbolE}{b}{n}
\DeclareFontShape{OMX}{MnSymbolE}{m}{n}{
    <-6>  MnSymbolE5
   <6-7>  MnSymbolE6
   <7-8>  MnSymbolE7
   <8-9>  MnSymbolE8
   <9-10> MnSymbolE9
  <10-12> MnSymbolE10
  <12->   MnSymbolE12
}{}
\DeclareFontShape{OMX}{MnSymbolE}{b}{n}{
    <-6>  MnSymbolE-Bold5
   <6-7>  MnSymbolE-Bold6
   <7-8>  MnSymbolE-Bold7
   <8-9>  MnSymbolE-Bold8
   <9-10> MnSymbolE-Bold9
  <10-12> MnSymbolE-Bold10
  <12->   MnSymbolE-Bold12
}{}

\let\llangle\@undefined
\let\rrangle\@undefined
\DeclareMathDelimiter{\llangle}{\mathopen}%
                     {MnLargeSymbols}{'164}{MnLargeSymbols}{'164}
\DeclareMathDelimiter{\rrangle}{\mathclose}%
                     {MnLargeSymbols}{'171}{MnLargeSymbols}{'171}
\makeatother

\begin{document}


\begin{titlepage}
\begin{flushright}
July, 2024\\
\end{flushright}
\vspace{5mm}

\begin{center}
{\Large \bf Spin-$(s,j)$ projectors and gauge-invariant spin-$s$ actions\\[7pt] in maximally symmetric backgrounds  }

\end{center}

\begin{center}

{\bf 
Daniel Hutchings$^1$ and Michael Ponds
}
\vspace{5mm}

\footnotesize{
{\it ${}^1$Department of Physics M013, The University of Western Australia\\
35 Stirling Highway, Crawley W.A. 6009, Australia}}  
\vspace{2mm}
~\\
Email: \texttt{daniel.hutchings@uwa.edu.au, michaelponds95@gmail.com}\\
\vspace{2mm}

\end{center}

\begin{abstract}
\baselineskip=14pt

Given a maximally symmetric $d$-dimensional background with isometry algebra $\mf{g}$, a symmetric and traceless rank-$s$ field $\phi_{a(s)}$ satisfying the massive Klein-Gordon equation furnishes a collection of massive $\mf{g}$-representations with spins $j\in \{0,1,\cdots,s\}$. In this paper we construct the spin-$(s,j)$ projectors, which are operators that isolate the part of $\phi_{a(s)}$ that furnishes the representation from this collection carrying spin $j$. In the case of an (anti-)de Sitter ((A)dS$_d$) background, we find that the poles of the projectors encode information about (partially-)massless representations, in agreement with observations made earlier in $d=3,4$. We then use these projectors to facilitate a systematic derivation of two-derivative actions with a propagating massless spin-$s$ mode. In addition to reproducing the massless spin-$s$ Fronsdal action, this analysis generates new actions possessing higher-depth gauge symmetry.
In (A)dS$_d$ we also derive the action for a partially-massless spin-$s$ depth-$t$ field with $1\leq t \leq s$.  
The latter utilises the minimum number of auxiliary fields, and corresponds to the action originally proposed by Zinoviev after gauging away all St\"{u}ckelberg fields. 
Some higher-derivative actions are also presented, and in $d=3$ are used to construct (i) generalised higher-spin Cotton tensors in (A)dS$_3$; and (ii) topologically-massive actions with higher-depth gauge symmetry. 
Finally, in four-dimensional $\mc{N}=1$ Minkowski superspace, we provide closed-form expressions for the analogous superprojectors. 

\end{abstract}

\vfill

\vfill
\end{titlepage}

\newpage
\renewcommand{\thefootnote}{\arabic{footnote}}
\setcounter{footnote}{0}

{
  \hypersetup{linkcolor=black}
  \tableofcontents
}
\vspace{1cm}
\bigskip\hrule

\allowdisplaybreaks


\section{Introduction}

The unitary irreducible representations (UIRs) of the Poincar\'{e} group in four dimensions were classified in 1939 by Wigner \cite{Wigner}. 
Here the empirical properties mass and spin (or helicity in the massless case), carried by all elementary particles, were shown to originate as the parameters which label these UIRs. 
In higher dimensions, $d>4$, the UIRs of the connected Poincar\'{e} group $\mathsf{ISO}_0(d-1,1)$ have also been studied in the literature, see e.g. \cite{BB06} and references therein. 
These classifications form the foundation of all relativistic field theories and are the starting point when building actions for fields with any mass and spin.

The spin(s) of a massive representation is identified with the label(s) classifying the  representation of the little group $\mathsf{SO}(d-1)$ which induces it, and is therefore dimensionally dependent. 
However, common to all dimensions are the massive UIRs $D(m,s)$ labelled by the mass $m\in \mathbb{R}_{>0}$ and a single integer spin parameter $s$ with $s\geq 0$.
In this paper we focus our attention on representations of this type, and their massless cousins with $m=0$. 
In $d$-dimensional Minkowski space M$^d$, the representation $D(m,s)$  may be furnished by a totally symmetric rank-$s$ tensor field $\phi^{\perp}_{a_1\dots a_s} = \phi^{\perp}_{(a_1\dots a_s)} \equiv\phi^{\perp}_{a(s)}$ that is on the mass-shell,
\begin{align}
0 &= (\Box - m^2) \phi^{\perp}_{a(s)}~, \label{KG1}
\end{align}
and is also subject to the transverse and traceless (TT) constraints
\begin{subequations} \label{constraints}
	\begin{align}
	0 &= \pa^{b}\phi^{\perp}_{ba(s-1)}~,\label{C1}\\ 
	0 &= \phi^{\perp}_{a(s-2)b}{}^{b}~. \label{C2}
	\end{align}
\end{subequations}
The second and third constraints are absent for $s=0$ whilst the third is absent for $s=1$.
These equations were first discovered in $d=4$ by Dirac \cite{Dirac1936}, and Fierz \& Pauli \cite{Fierz1939,FierzPauli1939}.

Let us temporarily restrict our attention to symmetric traceless fields $\phi_{a(s)}$ that obey \eqref{KG1}. 
Such a field furnishes the following reducible representation of the Poincar\'{e} algebra $\mf{iso}(d-1,1)$
\begin{align}
D(m, s) ~\oplus~ D(m , s-1) ~\oplus~ \cdots ~\oplus~ D(m , 0) ~. \label{decomp1}
\end{align}
The constraint \eqref{C1} eliminates the lower spin modes and selects the UIR $D(m,s)$ with maximal spin from \eqref{decomp1}.\footnote{A traceful field $\F_{a(s)}$ realises the reducible representation \eqref{bigdecomp} of $\mf{iso}(d-1,1)$. To isolate the UIR $D(m,s)$ it is necessary to impose both the traceless \eqref{C2} and transverse \eqref{C1} constraints.  The former reduces the multiplicities of the UIRs in \eqref{bigdecomp} to one, while the latter removes the remaining lower spin modes.} A natural question to ask is: what constraints can be placed on $\phi_{a(s)}$ in order to isolate the part of $\phi_{a(s)}$ which furnishes the UIR $D(m, j)$ with $0\leq j\leq s$? 
To answer this question we utilise a set of tools we refer to as spin-$(s,j)$ projection operators.

The spin-$(s,j)$ projector $\Pi_{(s,j)}^{\perp}$, with fixed $s$ and $0\leq j \leq s$, is a non-local differential operator which, as we will see, enjoys many useful properties. Most important to the current discussion is that its image $\Pi_{(s,j)}^{\perp}\phi_{a(s)}$ on a symmetric and traceless field $\phi_{a(s)}$ obeying the KG equation \eqref{KG1} furnishes the UIR $D(m ,j)$. The most well-known member of this family of operators is the one with $j=s$, in which case the field $\Pi_{(s,s)}^{\perp}\phi_{a(s)}$ satisfies the constraints \eqref{constraints}. The operator $\Pi_{(s,s)}^{\perp}$ was first investigated in $d=4$ by Behrends and Fronsdal \cite{Fproj, BFproj}, and in arbitrary $d$ by Segal \cite{Se} (see also \cite{IP} for half-integer spins).
It has been used, for example, in the construction of (i) massive spin-$s$ propagators \cite{Singh1}; (ii) conformal spin-$s$ actions \cite{FT, Se}; and (iii) massive spin-$s$ actions with $s\leq 4$ \cite{Fproj, Chang}.  

The operators $\Pi_{(s,j)}^{\perp}$ with $0\leq j \leq s-1$ are much less well-known, and were constructed in $d=4$ by Aurilia and Umezawa \cite{AuriliaUmezawa1967, AuriliaUmezawa1969}. They have been used, for example, to build massive spin-$s$ wave equations \cite{AuriliaUmezawa1969}, and to quantise the massive spin-$2$ field \cite{Barnes, Rivers, vN73}. 
The spin-$(s,j)$ projectors with $j\leq s-1$ have not yet been derived in M$^d$ for $d\neq 4$, and in section \ref{secProjMd} we fill this gap for $d \geq 3$.\footnote{In $d=3$ a massive spin-$s$ field can have helicity $\pm s$, and the representation $D(m, s)$ is reducible (with respect to the \textit{connected} Poincar\'{e} group), $D(m, s) = D(m, s,+)\oplus D(m, s,-)$. In appendix \ref{AppHel} we derive spin-$(s,j)$ projectors which select a field with definite helicity. For $j=s$ the helicity projectors were derived in \cite{BKLFP}. } We use the method employed in \cite{AuriliaUmezawa1969} (see also \eqref{linalg}) in which $\Pi_{(s,j)}^{\perp}$ are formulated entirely in terms of Casimir operators of $\mf{iso}(d-1,1)$, which is in contrast to all other approaches (which use only $\pa_a$ and $\eta_{ab}$). 
We will exploit the fact that in this form the projectors may act on tensors $\phi_{a(s')}$ with $s'\neq s $, which allows us to deduce many useful properties that were not noticed in \cite{AuriliaUmezawa1967, AuriliaUmezawa1969}. 
In appendix \ref{AppExp} we provide explicit expressions for $\Pi_{(s,j)}^{\perp}$ when restricted to the space of rank-$s$ tensors $\phi_{a(s)}$.
The method we apply also allows us to obtain all the spin-$(s,j)$ projectors in $d$-dimensional (anti-)de Sitter space ((A)dS$_d$) in one fell swoop, which had previously been completed only for $j=s$ in $d=4$ \cite{KPproj} and $d=3$ \cite{HKP}. This analysis is performed in section \ref{secProjAdSd}. 
As was first observed in \cite{KPproj} and subsequently in \cite{BHKP, HKP, Hutchings2023}, the spin-$(s,j)$ projectors in (A)dS$_d$ derived here conform to the pattern of encapsulating information about massless and partially massless fields.
 
As mentioned above, it is relatively common knowledge that the spin-$(s,s)$ projectors may be used to construct the action for the massive spin-$s$ field. Far less common is the knowledge that the spin-$(s,j)$ projectors may be used to construct massless spin-$s$ actions. 
This is not surprising since this technique has been used on only two occasions \cite{SG2, GKP}, both within the context of constructing actions for off-shell supermultiplets.
More specifically, an algorithm to construct the supersymmetric analogue of the spin-$(s,j)$ projectors was described by Gates and Siegel \cite{SG} (see also \cite{GGRS}) in four-dimensional $\mc{N}$-extended Minkowski superspace M$^{4|4\mc{N}}$.
The authors of \cite{SG2} and \cite{GKP} made use of the corresponding $\mc{N}=1$ superspin-$(s,j)$ projectors to completely classify the possible linearised actions for the off-shell massless gravitino ($s=1$) and supergravity ($s=\frac{3}{2}$) multiplets in M$^{4|4}$, respectively.

 In a spirit similar to \cite{SG2, GKP}, in section \ref{secActMd} we provide a systematic derivation of the possible two-derivative (in the dynamical sector) actions in M$^d$ involving the field $\phi_{a(s)}$ that propagates a massless spin-$s$ mode, and which require a single auxiliary field for gauge invariance. We show how to reproduce the massless spin-$s$ Fronsdal action \cite{Fronsdal1, Fronsdal5} using the spin-$(s,j)$ projection operators, and derive a new family of two-derivative actions that are invariant under higher-depth gauge transformations of $\phi_{a(s)}$. 
 We use the projectors to analyse the on-shell content of these models and find that they propagate some lower-spin modes (the exact spectrum depends on the depth) in addition to a massless spin-$s$ mode. 
 In section \ref{secActAdSd} analogous results are obtained in (A)dS$_d$. In section \ref{secPM} we use our method to derive the two-derivative action for a partially massless spin-$s$ depth-$t$ field with  $1 \leq t \leq s$ using the minimum number of auxiliary fields. These actions coincide with those derived by Zinoviev \cite{Zinoviev} after gauging away all of the St\"{u}ckelberg fields present in \cite{Zinoviev} and expressing his doubly traceless fields in terms of traceless fields.
 
 All of the above actions require a varying spectrum of compensating gauge fields in order to ensure gauge invariance. However, by increasing the number of derivatives in the action, in section \ref{secNoCompMd} we use the projectors to propose two families of higher-derivative, manifestly gauge invariant actions which do not involve compensators. In section \ref{secGenCHSMd} we modify one family of actions such that they become conformally-invariant and correspond to the action for a conformal spin-$s$ depth-$t$ field, originally proposed  in  \cite{FT, Se} for $t=1$ and in \cite{Vasiliev2009} for $2 \leq t \leq s$. This provides a novel representation of higher-depth conformal spin-$s$ actions in terms of projection operators. The other family of higher-derivative actions without compensators is new. Finally, in $d=3$ we use our results to propose:\footnote{These are higher-depth generalisations of the results obtained in \cite{KP21} and \cite{Bergshoeff11, KP18, Dalmazi21, Boulanger2014vya} respectively. } (i) generalised higher-spin Cotton tensors in (A)dS$_3$ and; (ii) novel topologically massive models with higher-depth gauge symmetry.

Finally, in section \ref{secSUSY} we revisit the problem of building superprojectors for the four-dimensional  $\mc{N}=1$ Poincar\'{e} superalgebra $\mf{iso}(3,1|4)$.
This area of research is almost 50 years old, and amongst the most notable works are the seminal papers \cite{Sokatchev, SG}.
Given a real superfield $\bm{\phi}^{\perp}_{\a(s)\ad(s)}$ with arbitrary $s\geq 0$ that is constrained to be transverse, $0=\pa^{\b\bd}\bm{\phi}^{\perp}_{\b\a(s-1)\bd\ad(s-1)}$, ref. \cite{Sokatchev} derived the four superprojectors which select the four UIRs of $\mf{iso}(3,1|4)$ that such a superfield encodes (see eq. \eqref{SUSYFPrep}).
Six years later ref. \cite{SG} invented an algorithm which can be used to construct the superprojectors which select any UIR encoded within an \textit{unconstrained} superfield of arbitrary Lorentz type $\bm{\phi}_{\a(m)\ad(n)}$.\footnote{The superprojectors which act on a scalar superfield were first derived by Salam and Strathdee in \cite{SalamStrathdee1974}.}
Explicit examples of this algorithm were provided in \cite{SG} for $(m,n)=\{(0,0), (1,0), (1,1)\}$.
The novelty of our work is that we provide closed-form expressions for the superprojectors in the case $m=n=s$ with $s$ arbitrary (the case $m\neq n$ may be easily accommodated within our framework). 
The method we employ to construct the superprojectors is similar in spirit to \cite{Sokatchev}, where Casimir operators played a central role. 
In contrast, the authors of \cite{SG} purposely avoided the use of Casimir operators,\footnote{ Their rationale for this was the ``complicated form [of the superprojectors] which requires lengthy algebra for simplification". This is true if one wants to move from a representation in terms of Casimirs to one in terms of only (spinor) derivatives, which is sometimes necessary for practical calculations. However, using the novel properties we establish, we find that for most purposes it is suitable to remain in the former representation. } and centred their approach around the covariant expansion of a superfield into chiral superfields.  Our method allows us to establish novel properties of the superprojectors which we hope to be useful in constructing off-shell supersymmetric gauge invariant actions in the future.

\noindent\textbf{Notation and conventions}

For a group of $m$ totally symmetry indices we use the notation $\Phi_{a_1\dots a_m} = \Phi_{(a_1\dots a_m)} \equiv\Phi_{a(m)}$. Indices that are denoted by the same symbol (we consider the indices $a_1$ and $a_2$ to be distinct) are assumed to be symmetrised over,
\begin{align}
U_{a(m)} V_{a(n)} = U_{(a_1\cdots a_m} V_{a_{m+1}\cdots a_{m+n})} =\frac{1}{(m+n)!}\big(U_{a_1 \cdots a_m} V_{a_{m+1}\cdots a_{m+n}}+\cdots\big)~. \label{conventionVec}
\end{align}
We consider only totally symmetric bosonic fields and not fields of a mixed symmetry or fermionic type. 
Capital Greek symbols (usually $\Phi$) are reserved for symmetric fields that are reducible with respect to (w.r.t.) the Lorentz group (i.e. traceful), whilst lower-case Greek letters (usually $\phi$) are reserved for symmetric fields that are irreducible (i.e. traceless). If a field is transverse then it is endowed with the superscript `$\perp$', e.g. $\phi^{\perp}_{a(s)}$. The linear space of rank-$s$ symmetric traceless fields $\phi_{a(s)}$ will be denoted $\mathscr{V}_{(s)}$. Our conventions concerning the geometry of (A)dS$_d$ are described in section \ref{secProjAdSd}.


\section{Spin-$(s,j)$ projection operators}\label{SpinProjectionOperators}

The spin-$(s,s)$ projector $\Pi_{(s,s)}^{\perp}$ is traditionally defined through its action on a rank-$s$ field, or,  in other words, it is taken to have index structure $(\Pi_{(s,s)}^{\perp})_{a(s),}{}^{b(s)}$. For example, the most commonly used version of the spin-$(1,1)$ projector in M$^d$ is $\Pi_{a,}{}^{b}=(\delta_a{}^{b}-\frac{\pa_a\pa^b}{\Box})$. As the spin increases so too does the structural complexity of the corresponding projector. Though this form may have the advantage that its structure (in terms of $\pa_a$ and $\eta_{ab}$) is very explicit, it is disadvantageous for a number of reasons (other than its cumbersome appearance). 

A more useful form for the projectors can be obtained by using the quadratic and quartic Casimir operators (i.e. the mass and spin operators) of the relevant isometry algebra. This version of the projectors stems from the following basic fact in linear algebra (see e.g. \cite{BK}). If a linear operator $\mb{F}$ on a vector space $\ms{L}$ takes distinct eigenvalues $f_1,f_2,\dots,f_n$ such that $\ms{L}=\ms{L}_1\oplus\ms{L}_{2}\oplus\cdots\oplus\ms{L}_n$ with $\mb{F}|_{\ms{L}_i}=f_i\mathds{1}$, 
then the operators $\mc{P}_i$ with\footnote{In this paper we do not assume a summation convention for the indices labelling projectors. Additionally, 
whenever the upper bound of a product symbol is less than the lower bound, we define it to mean unity.
}
\begin{align}
\mc{P}_i = \prod_{1 \leq k \leq n}^{k\neq i}\frac{\big(\mb{F}-f_{k}\big)}{\big(f_i-f_k\big)}~,\qquad \mc{P}_i\mc{P}_j = \delta_{ij}\mc{P}_i~, \qquad \mc{P}_1+\cdots +\mc{P}_n =\mathds{1}~, \label{linalg}
\end{align}
form a complete set of orthogonal projection operators onto the eigenspaces $\ms{L}_i$.

This is the approach that Aurilia and Umezawa \cite{AuriliaUmezawa1967, AuriliaUmezawa1969} adopted when deriving the spin-$(s,j)$ projectors in M$^4$ and is the method we will employ in both M$^d$ and (A)dS$_d$.
The advantages of formulating $\Pi_{(s,j)}^{\perp}$ in this way are three-fold: (i) it is an index free operator and may be defined to act on rank-$s'$ fields with $s'\neq s$; (ii) it is expressed purely in terms of Casimir operators and hence commutes with the (covariant) derivative and the Lorentz generators; and (iii) it leads to compact expressions for all projectors on maximally symmetric backgrounds. These observations will lead us to obtain a number of useful properties for $\Pi_{(s,j)}^{\perp}$ not noticed in \cite{AuriliaUmezawa1967, AuriliaUmezawa1969}.

\subsection{Spin-$(s,j)$ projection operators in M$^d$}
In this section we work in $d$-dimensional Minkowski space M$^d$ which has $\mf{iso}(d-1,1)$ as its isometry algebra.  
In the field representation the generators of $\mf{iso}(d-1,1)$ are $P_a = -\ri\pa_{a}$ and $J_{ab}=-\ri M_{ab}$, where we have dropped the orbital part of $J_{ab}$ and kept only the spin contribution $M_{ab} = -M_{ba}$. They satisfy the commutation relations
\begin{subequations}\label{PoinAl}
	\begin{align}
	\big[\pa_a,\pa_b\big] &=0~,\\
	\big[M_{ab}, \pa_{c}\big] &= \eta_{ca}\pa_{b}-\eta_{cb}\pa_{a}~,\\
	\big[M_{ab},M_{cd}\big] &= \eta_{ac}M_{bd}-\eta_{bc}M_{ad}-\eta_{ad}M_{bc}+\eta_{bd}M_{ac}~,
	\end{align}
\end{subequations}
where $\eta_{ab}$ has signature $(-+\cdots+)$. The operator $M_{bc}$ acts on $\Phi_{a(s)}$ according to the rule
\begin{align}
M_{bc}\Phi_{a(s)} = s \big ( \eta_{a b}\Phi_{c a(s-1)} - \eta_{a c}\Phi_{b a(s-1)} \big )~. \label{LorAct}
\end{align}
We remind the reader that the implied convention \eqref{conventionVec} is imperative for the validity of  \eqref{LorAct}.

In addition to the quadratic and quartic Casimir operators of $\mf{iso}(d-1,1)$, 
\bsubeq
\begin{align}
\Box &:= \pa^b\pa_b~, &\big[\Box,\pa_{a}\big]=0~, \\
\mb{W}^2&:=-\frac{1}{2}\Box \mb{M}^2+\pa^a\pa^bM_{a}{}^{c}M_{bc}~, &\big[\mb{W}^2,\pa_{a}\big]=0 ~,\label{spinOp}
\end{align}
the quadratic Casimir of the $d$-dimensional Lorentz algebra $\mf{so}(d-1,1)$,\footnote{The operator $\mb{M}^2$ is, of course, not a Casimir operator of $\mf{iso}(d-1,1)$ since $[\mb{M}^2,\pa_a]=2(d-1)\pa_a-4\pa^bM_{ab}$.}
\begin{align}
\mb{M}^2 &:= M^{ab}M_{ab}~, \qquad\qquad\qquad\qquad\qquad\qquad~~\big[\mb{M}^2, M_{ab}\big] = 0~,
\end{align}
\esubeq
will also play a prominent role in our analysis. 
We note that our definition of $\mathbb{W}^2$ differs to others (e.g. \cite{SKAP}) by an overall factor of $(d-3)!$. Specifically, the operators $\mb{W}^2$ and $\mathbb{W}^{a_1\cdots a_{d-3}}\mathbb{W}_{a_1\cdots a_{d-3}}$ with $\mathbb{W}_{a_1\cdots a_{d-3}}=-\frac{1}{2}\ve_{a_1\cdots a_{d-3}fgh}M^{fg}\pa^h$ are related via $ \mathbb{W}^{a_1\cdots a_{d-3}}\mathbb{W}_{a_1\cdots a_{d-3}}=(d-3)!\mathbb{W}^2$.

Before moving on we introduce the massless spin-$s$ UIR of $\mf{iso}(d-1,1)$ denoted by $D(0,s)$. It may be furnished by a symmetric rank-$s$ field $\phi^{\perp}_{a(s)}$ satisfying the massless KG equation 
\begin{align}
0 &= \Box \phi^{\perp}_{a(s)}~, \label{masslessKG}
\end{align}
and the TT constraints \eqref{constraints}. Equation \eqref{masslessKG} and constraints \eqref{constraints} admit the gauge symmetry
\begin{align}
\delta \phi_{a(s)}^{\perp} = \pa_a\xi_{a(s-1)}^{\perp}~,
\end{align}
provided $\xi_{a(s-1)}^{\perp}$ is TT and also satisfies the massless KG equation. The representation $\phi_{a(s)}^{\perp}$ furnishes is only irreducible upon fixing the gauge freedom (i.e. factoring out the gauge modes). Nevertheless, with this understood, we will say that $\phi_{a(s)}^{\perp}$ furnishes the UIR $D(0,s)$.

\subsubsection{Traceless projectors} \label{secTrace}
Given a totally symmetric field $\Phi_{a(s)}$ it is not difficult to show using \eqref{LorAct} that
\begin{align}
\mb{M}^2\Phi_{a(s)} = \gamma_{(s,d)}\Phi_{a(s)} +2s(s-1)\eta_{a(2)}\Phi_{a(s-2)b}{}^{b}~,\qquad \gamma_{(s,d)}:=-2s(s+d-2)~. \label{M2red}
\end{align} 
It follows that traceless symmetric fields are eigentensors of $\mb{M}^2$
\begin{align}
0=\big(\mb{M}^2-\gamma_{(s,d)}\big)\phi_{a(s)}~,\qquad\qquad \phi_{a(s-2)b}{}^{b}=0~. \label{M2irred}
\end{align}
In other words, the field $\phi_{a(s)}$ furnishes an irreducible representation (irrep) of $\mf{so}(d-1,1)$, which we denote by $L(s)$. For the remainder of this paper we omit the label $d$ from the eigenvalue $\gamma_{(s,d)}$ and instead use $\gamma_s$, as the dimension (if it is fixed) is usually clear from context. 

The traceful field $\Phi_{a(s)}$ furnishes the reducible representation 
\begin{align}
L(s)\oplus L(s-2)\oplus \cdots \oplus 
\begin{cases} 
L(1) & ~~s \text{~odd}\\
L(0) & ~~s \text{~even}
\end{cases}
~~=~~ \bigoplus_{j=0}^{\lfloor \frac{s}{2} \rfloor} L(s-2j) \label{Lorredrep}
\end{align}
of $\mf{so}(d-1,1)$. 
To isolate each irrep, we introduce the following traceless projection operators
\begin{align}
\Delta_{(s,s-2j)} = \prod_{0\leq k \leq \lfloor s/2 \rfloor}^{k\neq j}\frac{(\mb{M}^2-\gamma_{s-2k})}{(\gamma_{s-2j}  - \gamma_{s-2k})}~,
\end{align}
where $j$ is an integer satisfying $0\leq j \leq \lfloor s/2 \rfloor$.\footnote{Computation of the traceless projectors via the Brauer algebra was recently described in \cite{Bulgakova2022bsk}. } To facilitate the derivation of their projector properties, we first provide a proof for the decomposition of a traceful field $\Phi_{a(s)}$ into irreducible (traceless) parts. 

By making repetitive use of identity \eqref{M2red}, one may show that for traceful $\Phi_{a(s)}$
\begin{align}
\eta^{bc}\Delta_{(s,s)}\Phi_{bca(s-2)} =0~.
\end{align}
Moreover, one may show that the orthogonal complement $\Delta^{\text{c}}_{(s,s)}= \mathds{1} - \Delta_{(s,s)}$ satisfies
\begin{align}
\Delta^{\text{c}}_{(s,s)} \Phi_{a(s)} = \eta_{a(2)}\Phi_{a(s-2)} ~,
\end{align}
for some traceful tensor $\Phi_{a(s-2)}$. This means that $\Phi_{a(s)}$ can be decomposed into irreducible parts (w.r.t. $\mf{so}(d-1,1)$) according to 
\begin{align}
\Phi_{a(s)} = \phi_{a(s)} +\eta_{a(2)}\phi_{a(s-2)} + \cdots +
\begin{cases} 
(\eta_{a(2)})^{(s-1)/2}\phi_a & ~~s \text{~odd}\\
(\eta_{a(2)})^{s/2}\phi & ~~s \text{~even}
\end{cases}
~~=~~ \sum_{j=0}^{\lfloor s/2 \rfloor} (\eta_{a(2)})^j\phi_{a(s-2j)} \label{LorDecomp}
\end{align}
for some $\phi_{a(s-2j)}$ which are traceless. In fact, by utilising the identity
\begin{align}
\big(\eta^{b(2)}\big)^j\eta_{(b_1b_2}\cdots\eta_{b_{2j-1}b_{2j}}\Phi_{a_1\cdots a_k)}= \bigg [  \prod_{l=1}^{j}\frac{\big(\mb{M}^2-\gamma_{k+2l}\big)}{(k+2l-1)(k+2l)} \bigg ] \Phi_{a(k)}
\end{align}
which holds for arbitrary $j$ and $k$ and traceful $\Phi_{a(k)}$, and which may be proved by induction on $j$. We note that the parameter $l$ which labels the product appears only in the numerical coefficient and not in the corresponding tensorial structure. This occurs frequently throughout this paper. To clarify this and avoid confusion, we will always encase the product symbol and the corresponding numeric factors in square brackets.

One may recover each $\phi_{a(s-2j)}$ in \eqref{LorDecomp} in terms of the original field $\Phi_{a(s)}$
\begin{align}
\phi_{a(s-2j)}= \frac{s!}{(s-2j)!} \bigg [ \prod_{l=1}^{j}\frac{1}{(\gamma_{s-2j}-\gamma_{s-2j+2l})} \bigg ] \Delta_{(s,s-2j)}\Phi_{a(s-2j)b(j)}{}^{b(j)}~.\label{trextract}
\end{align}
By using this decomposition in conjunction with \eqref{M2irred}, one may establish the following properties. Independent of the rank of symmetric tensors on which they act, the operators $\Delta_{(s,s-2j)}$ form a partition of identity, 
\begin{align}
\mathds{1} = \sum_{j=0}^{\lfloor s/2 \rfloor} \Delta_{(s,s-2j)}~. \label{traceres}
\end{align}
When acting on symmetric tensors $\Phi_{a(l)}$ with $0\leq l \leq s$, they are orthogonal projectors,
\begin{align}
\Delta_{(s,s-2j)} \Delta_{(s,s-2k)}= \delta_{jk} \Delta_{(s,s-2j)}~.
\end{align}
The operator $\Delta_{(s,s-2j)}$ acts as the identity on the space of traceless rank-$(s-2j)$ fields
\begin{align}
0=\phi_{a(s-2j-2)b}{}^{b} \qquad \implies \qquad \Delta_{(s,s-2j)}\phi_{a(s-2j)}=\phi_{a(s-2j)}~.
\end{align}
For fixed integers $s$ and $j$ the operator $\Delta_{(s,s-2j)}$ satisfies 
\begin{subequations}\label{traceanihil}
	\begin{align}
	\Delta_{(s,s-2j)}\Phi_{a(l)} &= (\eta_{a(2)})^{j-\frac{1}{2}(s-l)}\phi_{a(s-2j)} \qquad\qquad\qquad s-2j\leq l\leq s~,\\
	\Delta_{(s,s-2j)}\Phi_{a(l)} &= 0 \qquad\qquad\qquad\qquad\qquad\qquad\qquad ~~~~~\phantom{...} 0\leq l < s-2j~,
	\end{align}
\end{subequations}
for traceless $\phi_{a(s-2j)}$ given by \eqref{trextract}. Here $s$ and $l$ must have the same parity (odd or even). Thus when acting on fields $\Phi_{a(l)}$ with $l\leq s$, the operator $\Delta_{(s,s-2j)}$ projects onto fields with vanishing higher trace which are eigentensors of $\mb{M}^2$
\begin{subequations}
	\begin{align}
	0&=(\eta^{b(2)})^{j-\frac{1}{2}(s-l)+1}\Delta_{(s,s-2j)}\Phi_{b(2j-s+l+2)a(s-2j-2)} ~,\\
	0&=\big(\mb{M}^2-\gamma_{s-2j}\big)\Delta_{(s,s-2j)}\Phi_{a(l)}~.
	\end{align}
\end{subequations}
Projector $\Delta_{(s,s-2j)}$ selects the irrep $L(s-2j)$ from the reducible representation \eqref{Lorredrep}.

We may now focus our attention solely on symmetric traceless fields, denoted generally by $\phi_{a(s)}$, since we may always use the decomposition \eqref{LorDecomp} if the starting point is not a traceless field. 
Consequently, for the remainder of this paper we deal only with traceless symmetric fields unless stated otherwise. 

\subsubsection{Transverse projectors} \label{secProjMd}

On a symmetric and traceful field $\Phi_{a(s)}$ the spin operator $\mathbb{W}^2$ acts according to the rule
\begin{align}
\mathbb{W}^2 \Phi_{a(s)} &= \lambda_{(s,d)}\Box \Phi_{a(s)} -s(2s+d-4)\pa_{a}\pa^{b}\Phi_{b a(s-1)}+s(s-1)\eta_{a(2)}\pa^b\pa^c\Phi_{bc a(s-2)}~\non\\
&\phantom{=}-s(s-1)\big(\eta_{a(2)}\Box-\pa_a\pa_a\big)\Phi_{a(s-2)b}{}^{b}~. \label{W2red}
\end{align}
It follows that any TT field is an eigentensor of $\mb{W}^2$ 
\begin{align}
0 = \big(\mb{W}^2-\lambda_{(s,d)}\Box\big)\phi^{\perp}_{a(s)}~,\qquad \lambda_{(s,d)}:=s(s+d-3)~. \label{lambdadeff}
\end{align}
For the remainder of this paper we omit the label $d$ from $\lambda_{(s,d)}$ and instead use $\lambda_s$.

We are now in a position to define the spin-$(s,j)$ projectors via the prescription \eqref{linalg}:\footnote{In this paper we always leave the quadratic Casimir operator $\Box$ (or $\mb{Q}$ in the case of (A)dS$_d$) unresolved in the projectors, never replacing it with $m^2$, and in this sense everything is off-shell.}  
\begin{align}
\Pi_{(s,j)}^{\perp} = \prod_{0\leq k \leq s}^{k \neq j} \frac{\big(\mb{W}^2-\lambda_k\Box\big)}{(\lambda_j-\lambda_k)\Box}~. \label{sjproj}
\end{align}
For $d=4$ they coincide with the projectors provided in \cite{AuriliaUmezawa1967, AuriliaUmezawa1969}.
For $j=s$ they reduce to 
\begin{align}
\Pi_{(s,s)}^{\perp} = \prod_{k=0}^{s-1} \frac{\big(\mb{W}^2-\lambda_k\Box\big)}{(\lambda_s -\lambda_k)\Box} = \frac{(s + d - 4)!}{s! (2s+d-4)!}\frac{1}{\Box^s}\prod_{k=0}^{s-1}\big(\mb{W}^2-\lambda_k\Box\big)~,
\end{align}
and prove to coincide with the TT projectors derived in \cite{Fproj, Se} when restricted to $\mathscr{V}_{(s)}$.
In appendix \ref{AppExp} we derive explicit expressions for $\Pi_{(s,j)}^{\perp}$ when acting on $\phi_{a(s)}$. 

In order to prove the projector properties of $\Pi_{(s,j)}^{\perp}$, we need to establish an analogue of the irreducible decomposition. To this end, repetitive use of \eqref{W2red} allows one to show that 
\begin{align}
\pa^b\Pi_{(s,s)}^{\perp}\phi_{ba(s-1)} = 0~.
\end{align}
Next, one may show that its orthogonal complement $\Pi_{(s,s)}^{\parallel} = \mathds{1} - \Pi_{(s,s)}^{\perp}$ satisfies
\begin{align}
\Pi_{(s,s)}^{\parallel}\phi_{a(s)} = \pa_{\{a}\phi_{a(s-1)\}}~,
\end{align}
for some $\phi_{a(s-1)}$.
From these two facts follows the irreducible (w.r.t. $\mf{iso}(d-1,1)$) decomposition 
\begin{align}
\phi_{a(s)} = \phi_{a(s)}^{\perp} + \pa_{\{a}\phi^{\perp}_{a(s-1)\}}+\cdots + \pa_{\{a_1}\cdots\pa_{a_s\}}\phi^{\perp}~, \label{poindecomp}
\end{align}
for some TT fields $\phi^{\perp}_{a(j)}$  with $1\leq j \leq s$ and scalar $\phi^{\perp}$ whose explicit forms are given in \eqref{divextract}.

Above and below we use the notation $\{\cdots\}$ to denote the symmetric traceless part of a tensor, defined in this case as
\begin{align}
\pa_{\{a}\phi_{a(s-1)\}} = \pa_a \phi_{a(s-1)} -\frac{s-1}{2s+d-4}\eta_{a(2)}\pa^{b}\phi_{ba(s-2)}~. \label{tracesym1}
\end{align}
We define  traceless symmetrisation across multiple traceful indices as a nested operation,
\begin{align}
\pa_{\{a_1}\pa_{a_2}\cdots\pa_{a_j}\phi_{a_{j+1}\cdots a_{s}\}}\equiv \pa_{\{a_1}\pa_{\{a_2}\cdots\pa_{\{a_j}\phi_{a_{j+1}\cdots a_{s}\}\cdots\}\}}~.\label{nested}
\end{align}
For consistency the $\{\cdots\}$ should be opened from outside to inside, which allows us to make repetitive use of \eqref{tracesym1}. 
Using this definition one may prove the following useful identity
\begin{align}
(\pa^b)^j\pa_{\{b_1}\cdots\pa_{b_j}\phi_{a_1\cdots a_k\}}=(-1)^j\Big[\prod_{l=1}^{j}\frac{\big(\mb{W}^2-\lambda_{k+l}\Box\big)}{(k+l)(2k+2l+d-4)}\Big]\phi_{a(k)} ~,\label{magicid}
\end{align}
which holds for arbitrary $j$ and $k$ and traceless $\phi_{a(k)}$, and may be proved by induction on $j$.
If in addition $\phi_{a(k)}$ is transverse, $\phi_{a(k)}=\phi_{a(k)}^{\perp}$, then this reduces to 
\begin{align}
(\pa^b)^j\pa_{\{b_1}\cdots\pa_{b_j}\phi^{\perp}_{a_1\cdots a_k\}}=\Box^j\Big[\prod_{l=1}^{j}\frac{l(2k+l+d-3)}{(k+l)(2k+2l+d-4)}\Big]\phi^{\perp}_{a(k)}~.
\end{align}
The latter allows one to recover each $\phi^{\perp}_{a(j)}$ in \eqref{poindecomp} in terms of the original $\phi_{a(s)}$
\begin{align}
\phi^{\perp}_{a(j)}=\Box^{j-s} \bigg [ \prod_{k=1}^{s-j}\frac{(j+k)(2j+2k+d-4)}{k(2j+k+d-3)} \bigg ]\Pi_{(s,j)}^{\perp}(\pa^{b})^{s-j}\phi_{a(j)b(s-j)}~.\label{divextract}
\end{align}
See appendix \ref{AppExp} for a more explicit expression for $\phi^{\perp}_{a(j)}$.

Having obtained the irreducible decomposition \eqref{poindecomp}, we now state the properties of the operators $\Pi_{(s,j)}^{\perp}$, which follow directly from \eqref{poindecomp}. To this end, we recall the space $\mathscr{V}_{(l)}$ of rank-$l$ symmetric traceless tensor fields. Independent of $l$, the operators form a partition of identity
\begin{align}
\mathds{1} = \sum_{j=0}^s\Pi_{(s,j)}^{\perp}~. \label{didres}
\end{align}
On any $\mathscr{V}_{(l)}$ with $0\leq l \leq s$ they are orthogonal projectors
\begin{align}
\Pi^{\perp}_{(s,j)}\Pi^{\perp}_{(s,k)}=\delta_{jk}\Pi^{\perp}_{(s,k)}~,\qquad \text{on } \mathscr{V}_{(l)} \text{ with } 0\leq l\leq s~. \label{orthog2}
\end{align}
The operator $\Pi^{\perp}_{(s,j)}$ acts as the identity on the space of transverse rank-$j$ fields
\begin{align}
0=\pa^b\phi^{\perp}_{ba(j-1)} \qquad \implies \qquad \Pi^{\perp}_{(s,j)}\phi^{\perp}_{a(j)}=\phi^{\perp}_{a(j)}~. \label{actid2}
\end{align}
Given three integers $s,j,k$ satisfying the inequalities $0\leq j\leq s-k$ and $0\leq k \leq s$, the following projectors are equivalent
\begin{align}
\Pi^{\perp}_{(s,k)} = \Pi^{\perp}_{(s-j,k)}~\qquad \text{on } \mathscr{V}_{(k)} \label{equivalence}~,
\end{align}
when restricted to the space $\mathscr{V}_{(k)}$.

For fixed integers $s$ and $j$ satisfying $j\leq s$, an important property of $\Pi^{\perp}_{(s,j)}$ is
\begin{subequations} \label{Longi2}
	\begin{align}
	\Pi^{\perp}_{(s,j)}\phi_{a(l)}&=\pa_{\{a_1}\cdots\pa_{a_{l-j}}\phi^{\perp}_{a_{l-j+1}\cdots a_l\}} \qquad & &j\leq l \leq s~, \label{Longi2a}\\
	\Pi^{\perp}_{(s,j)}\phi_{a(l)}&=0 \phantom{(\pa_{\a\ad})^{l-j}S^{\perp}_{\a(j)\ad(j)}(\phi)} & &0\leq l \leq j-1~,  \label{Longi2b}
	\end{align} 
\end{subequations}
where $\phi^{\perp}_{a(j)}$ is given by \eqref{divextract}. From \eqref{Longi2} it follows that $\Pi^{\perp}_{(s,j)}\phi_{a(l)}$ is an eigentensor of $\mb{W}^2$,
\begin{align}
0=\big(\mb{W}^2-\lambda_{j}\Box\big)\Pi^{\perp}_{(s,j)}\phi_{a(l)}\qquad\qquad ~~~~~~~0\leq l \leq s~, \label{did55}
\end{align}
and has vanishing $(l-j+1)$ divergence
	\begin{align}
	0&=(\pa^b)^{l-j+1}\Pi^{\perp}_{(s,j)}\phi_{b(l-j+1)a(j-1)}\qquad\qquad 0 \leq l \leq s~.\label{did54}
	\end{align}
Actually, one can prove the following stronger version of identity \eqref{did54}
\begin{align}
0=(\pa^b)^k\Pi_{(s,j)}^{\perp}\phi_{b(k)a(l-k)}~,\label{did53}
\end{align}
for arbitrary positive integers $s,j,k,l$ satisfying the inequalities $l-k+1\leq j\leq s$ and $l\geq k$ (in particular this identity allows for $\phi_{a(l)}$ with $l\geq s$).
From \eqref{W2red} it follows that the projectors $\Pi^{\perp}_{(s,j)}$ are symmetric in the sense that
\begin{align}
\int \rd^d x ~ \Psi^{a(s)} \P^{\perp}_{(s,j)} \Phi_{a(s)} = \int \rd^d x ~ \Phi^{a(s)} \P^{\perp}_{(s,j)} \Psi_{a(s)} ~,
\end{align}
where the symmetric and traceful fields $\Psi_{a(s)} $ and $\Phi_{a(s)} $ are well-behaved.

We emphasise that $\Pi^{\perp}_{(s,j)}$ selects the UIR $D(m, j)$ from the reducible representation \eqref{decomp1} which $\phi_{a(s)}$ furnishes. 
In the introduction we posed the question: what constraints can be placed on $\phi_{a(s)}$ in order to isolate the UIR $D(m ,j)$ from the decomposition \eqref{decomp1}? We see now that the answer is to impose the constraint 
\begin{align}
\phi_{a(s)} = \Pi_{(s,j)}^{\perp}\phi_{a(s)}\qquad \Longleftrightarrow \qquad \phi_{a(s)}&=\pa_{\{a_1}\cdots\pa_{a_{s-j}}\phi^{\perp}_{a_{s-j+1}\cdots a_s\}}~,
\end{align}
which in turn implies $0=(\pa^b)^{s-j+1}\phi_{b(s-j+1)a(j-1)}$. In the case $j=s$ this corresponds to the well-known result that $D(m ,s)$ is furnished by a TT field.

Finally, we note that \eqref{traceres} and \eqref{didres} may be used to decompose a symmetric traceful tensor $\Phi_{a(s)}$ into irreducible (i.e. TT) parts as follows
\begin{align}
\Phi_{a(s)} = \sum_{k=0}^{\lfloor s/2 \rfloor}\sum_{j=0}^s\Pi_{(s,j,k)}^{\text{TT}} \Phi_{a(s)} = \sum_{k=0}^{\lfloor s/2 \rfloor}\sum_{j=0}^{s-2k}\Pi_{(s,j,k)}^{\text{TT}} \Phi_{a(s)} = \sum_{j=0}^s \sum_{k=0}^{\lfloor \frac{s-j}{2} \rfloor}\Pi_{(s,j,k)}^{\text{TT}} \Phi_{a(s)}~,\label{TTdecomp}
\end{align}
where we have used \eqref{traceanihil} and \eqref{Longi2b}. 
 Hence $\Phi_{a(s)}$  describes the reducible representation 
\begin{align}
D(m ,s)\oplus D(m , s-1)\oplus 2D(m , s-2)\oplus\cdots = \bigoplus_{j=0}^{s}\big(\big\lfloor \frac{s-j}{2} \big\rfloor +1\big)D(m , j)~,\label{bigdecomp}
\end{align}
when $\Phi_{a(s)}$ satisfies the KG equation \eqref{KG1}. The notation $\ell D(m , j)$ means that the representation $D(m , j)$ appears with multiplicity $\ell$.\footnote{In the case when $s=2$ this reproduces the well known (see e.g. \cite{Barnes, Rivers, vN73}) decomposition of a symmetric rank-2 field $g_{ab}$ into a spin-2, spin-1 and two spin-0 modes $g_{ab} = g_{ab}^{\perp}+\pa_{\{a}A_{b\}}^{\perp}+\eta_{ab}\sigma +\pa_a\pa_b\varphi$.}
In eq. \eqref{TTdecomp} we have defined the projector 
\begin{align}
\Pi^{\text{TT}}_{(s,j,k)}:=\Pi_{(s,j)}^{\perp} \Delta_{(s,s-2k)}~.
\end{align}
Given two fixed integers $s$ and $j$ with $0\leq j \leq s$, this projector 
 selects one of the $(\lfloor \frac{s-j}{2} \rfloor +1)$ copies of the UIR $D(m , j)$, indexed by $k$ with $0\leq k \leq \lfloor (s-j)/2 \rfloor $, from \eqref{bigdecomp} when acting on $\Phi_{a(s)}$. This may be seen from the property
\begin{align}
\Pi^{\text{TT}}_{(s,j,k)} \Phi_{a(s)} = (\eta_{a(2)})^{k}(\pa_{\{a})^{s-2k-j}\phi^{\perp}_{a(j)\}}~,
\end{align}
which follows from \eqref{traceanihil} and \eqref{Longi2a}. Here  $\pa_{\{a_1}\cdots\pa_{a_j}\psi_{a_{j+1}\cdots a_{j+k}\}}\equiv (\pa_{\{a})^j\psi_{a(k)\}} $.

\subsection{Spin-$(s,j)$ projection operators in (A)dS$_d$} \label{secProjAdSd}
In the field representation the AdS$_d$ and dS$_d$ isometry algebras $\mf{so}(d-1,2)$ and $\mf{so}(d,1)$ respectively are generated by the Lorentz generators $M_{ab}$ and the covariant derivative $\mc{D}_a$,
\begin{align}
\mc{D}_a=e_a{}^m\pa_m+\frac{1}{2}\omega_a{}^{bc}M_{bc}~.
\end{align}
Here $e_a{}^m$ is the inverse vielbein and $\omega_{abc}$ is the (torsion-free) Lorentz connection.  The commutation relations of $M_{ab}$ and $\mc{D}_a$ may be obtained from \eqref{PoinAl} by replacing $\pa_a \mapsto \mc{D}_a$, except $\mc{D}_a$ no longer commutes with itself\footnote{We use the convention $[\mc{D}_a,\mc{D}_b]=\frac{1}{2}R_{ab}{}^{cd}M_{cd}$ which yields $[\mc{D}_a,\mc{D}_b]V_c=R_{abcd}V^d$.}
\begin{align}
\big[\mc{D}_a,\mc{D}_b\big]=-\Omega M_{ab}~.
\end{align}
Here $\Omega$ is related the Riemann and scalar curvatures via 
\begin{align}
R_{abcd} = -\Omega(\eta_{ac}\eta_{bd}-\eta_{ad}\eta_{bc}) \qquad \implies \qquad R =-d(d-1)\Omega~,
\end{align} 
and to the radius $\ell$ of AdS$_d$ via $\Omega = \ell^{-2}$, or of dS$_d$ via $\Omega = -\ell^{-2}$.   

The mass and spin Casimir operators of the Poincar\'{e} algebra were central to the analysis and derivation of the spin-$(s,j)$ projectors in M$^d$. The quadratic and quartic Casimir operators  of $\mf{so}(d-1,2)$ and $\mf{so}(d,1)$ will play an equally important role when constructing the spin-$(s,j)$ projectors in (A)dS$_d$. In both AdS$_d$ and dS$_d$ they are given by
\begin{subequations}
	\begin{align}
	\mb{Q} := &~\Box-\frac{1}{2}\Omega \mb{M}^2~, \label{DAQuadCasimir}\\
	\mb{W}^2 := &-\frac{1}{2}\big(\mb{Q}+\frac{1}{4}(d-2)\Omega\big)\mb{M}^2+\mc{D}^{(a}\mc{D}^{b)}M_{a}{}^{c}M_{bc} \non
	\\ &  - \frac{1}{8} \Omega \big( \mb{M}^2\mb{M}^2+ 2 M^{ab}M^{cd}M_{ac}M_{bd}\big)~,
	\end{align}
\end{subequations}
where $\Box = \mc{D}^a\mc{D}_a$. They have been built in such a way that they commute with $M_{ab}$ and $\mc{D}_a$
\begin{align}
\big[\mc{D}_a,\mb{Q}\big]=0~,\qquad\qquad \big[\mc{D}_a,\mb{W}^2\big] =0~. \label{ValCas}
\end{align}
The proof that $\mb{W}^2$ is indeed a Casimir operator is non-trivial. 
To the best of our knowledge, this is the first time it has been presented in terms of the (A)dS$_d$ covariant derivative $\mc{D}_a$ for arbitrary $d$ (see \cite{HKP} for $d=3,4$ in two component spinor notation).

\subsubsection{Field theoretic representations of $\mf{so}(d-1,2)$ and $\mf{so}(d,1)$}\label{secAdSrep}

In this section we summarise aspects of the unitary irreducible representations of the Lie algebra $\mf{so}(d-1,2)$ of the AdS$_d$ isometry group, and in particular its field theoretic realisations. Our discussion is by no means complete or mathematically rigorous, as we only present the results relevant to our analysis. 
The UIRs of $\mf{so}(d-1,2)$ have been studied in detail by many authors, see e.g. \cite{Vasiliev04, Ponomarev22} and references therein.
For a discussion on the UIRs of the dS$_d$ isometry algebra $\mf{so}(d,1)$ we refer the reader to \cite{Basile16} and references therein.  At the end of this section we briefly comment on the field representations of $\mf{so}(d,1)$.

The UIRs of $\mf{so}(d-1,2)$ are labelled according to the UIRs of its maximal compact subalgebra $\mf{so}(2)\oplus\mf{so}(d-1)$. The UIRs of  $\mf{so}(2)$ are characterised by a real number $E_0$ known as the minimal energy. Generally, the UIRs of $\mf{so}(d-1)$ are labelled by Young diagrams of varying shapes and sizes which, in the field theoretic context, are realised on tensors of a mixed symmetry type. However, as mentioned in the introduction, in this paper we are concerned only with representations characterised by single row Young diagrams, or in other words carrying a single integer spin $s$. Thus the UIRs of $\mf{so}(d-1,2)$ that we are interested in are labelled by the minimal energy $E_0$ and the spin $s$, and are denoted $\mf{D}(E_0,s)$.

At the field theoretic level, the representation $\mf{D}(E_0,s)$ can be furnished by the tensor field $\phi_{a(s)}^{\perp}$ lying on the mass-shell,
\bsubeq \label{DAOnShell}
\begin{align}
0 &= ( \mb{Q} - \r^2)\phi^{\perp}_{a(s)}~,  \label{DAMassShell}
\end{align}
that also satisfies the transverse and traceless constraints
\begin{align} 
0 &= \cD^b \f_{b a(s-1)}^{\perp}~, \label{AdSTrans}\\
0&= \f^{\perp}_{a(s-2)b}{}^{b} ~. \label{AdSTrace}
\end{align}
\esubeq
The parameter $\rho$ is known as the pseudo-mass, and is related to the minimal energy through
\be
\r^2 = \big [E_0 \big ( E_0 - d+1 \big ) +s \big ( s +d -3 \big )  \big ] \Omega~.\label{rhoE0}
\ee
The prefix `pseudo' is due to the fact that the massless case does not correspond to $\rho=0$. For this one usually works with the physical mass $m$, related to $\rho$ via
\be \label{DAPsuedoPhysicalRelation}
m^2 = \r^2 - \t_{(1,s)} \Omega~, \qquad \t_{(1,s)} := 2(s-1)(s+d-3)~.
\ee
It has the advantages that $m^2=0$ corresponds to a massless field and the unitarity bound is
\begin{align}
m^2\geq 0 \qquad \implies \qquad \rho^2\geq \tau_{(1,s)}~, \label{Ubound}
\end{align}
for $s\geq1$. Despite this, we will continue to use $\rho$ in place of $m$ because $\rho$ is the mass that appears in the mass-shell equation \eqref{DAMassShell}. Due to the relationship \eqref{rhoE0} we will also use $\mf{D}(\rho^2,s)$ in place of $\mf{D}(E_0,s)$. 
A traceless symmetric field $\phi_{a(s)}$ satisfying the mass-shell equation \eqref{DAMassShell}  furnishes the reducible representation
\begin{align}
\mf{ D}(\r^2 , s) ~\oplus~ \mf{D}(\r^2 , s-1) ~\oplus~ \cdots ~\oplus~ \mf{D}(\r^2, 0) ~. \label{decomp2}
\end{align}

A massive representation with pseudo-mass $\rho$ and spin $s$ refers to any representation $\mf{D}(\rho^2, s)$ satisfying \eqref{Ubound}, the latter ensuring unitarity.    At the special mass values $\rho^2=\rho_{(t,s)}^2$ with
\begin{subequations}\label{PMvals}
\begin{align}
\r^2_{(t,s)} &= \t_{(t,s)} \Omega~, \qquad 1 \leq t \leq s~,\\[6pt]
\t_{(t,s)} = 2(s-1)&(s+d-3) - (t-1)(2s+d-t-4) ~,
\end{align}
\end{subequations}
the representation $\mf{D}(\rho^2,s)$ shortens. This shortening signifies the presence of a gauge redundancy in the field theory. 
Indeed, a field $\phi_{a(s)}^{\perp}$ satisfying the equations 
\begin{subequations} \label{dAdSPMOnShell}
\begin{align} 
0 &= ( \mb{Q} - \r_{(t,s)}^2)\phi^{\perp}_{a(s)}~,\\
0 &= \cD^b \f_{b a(s-1)}^{\perp}~, \\
0 &= \f^{\perp}_{a(s-2)b}{}^{b} ~,
\end{align}
\end{subequations}
is defined modulo the depth-$t$ gauge transformations
\begin{align}
\d \f^{\perp}_{a(s)} = \cD_{ \{ a_1} \cdots \cD_{  a_t} \x^{\perp}_{ a_{t+1}  \ldots a_s \} }~,
\end{align}
where the gauge parameter $\x^{\perp}_{ a (s-t)}$ is also constrained to satisfy 
\begin{subequations}
\begin{align}
0 &= \big ( \mb{Q} - \rho^2_{(t,s)}\big ) \x^{\perp}_{a(s-t)} ~,  \label{dAdMassShellPM}\\
0 &=\cD^b \x^{\perp}_{b a(s-t-1)} ~, \\
0 &= \x^{\perp}_{a(s-t-2)b}{}^{b} ~.
\end{align}
\end{subequations}
This can be proved using \eqref{magicId2} with $j=1$.
The field $\phi_{a(s)}^{\perp}$ is called a depth-$t$ partially massless (PM) field (see e.g. \cite{DeserN1, Higuchi, DW2, Zinoviev, Metsaev06} for earlier works), and the representation it furnishes is denoted $\mf{P}(t,s)$. For $t\geq 2$ the pseudo-masses \eqref{PMvals} violate the unitarity bound \eqref{Ubound} and $\mf{P}(t,s)$ is non-unitary. 
In the massless case $t=1$ the unitarity bound is saturated and $\mf{P}(1,s)$  is unitary.

Instead of using $\mb{Q}$  and $\rho$ in the mass-shell equation \eqref{DAMassShell}, it is more common for practitioners to use the AdS$_d$ d'Alembert operator $\Box$ and $m$. To make contact with this tradition one can use \eqref{DAQuadCasimir} and \eqref{DAPsuedoPhysicalRelation} to obtain 
\begin{align}
0 = \big ( \Box - \big [ (s-2)(s+d-3)-s \big ] \Omega -m^2 \big ) \f^{\perp}_{a(s)}~.
\end{align}
The partially-massless modes appear at the physical masses $m^2=m^2_{(t,s)}$ with
\be
m^2_{(t,s)} = -(t-1)(2s+d-t-4) \Omega ~, \qquad 1 \leq t \leq s~,
\ee
which makes their violation of \eqref{Ubound} manifest. 

To conclude this section we note that a very similar discussion takes place for field theoretic representations of the dS$_d$ isometry algebra $\mf{so}(d,1)$. The main difference in this case is that the unitarity bound \eqref{Ubound} is replaced with the Higuchi bound \cite{Higuchi}
\begin{align}
m^2 \leq ~ (s-1)(s+d-4)|\Omega|\qquad \implies \qquad \rho^2 \leq (s-1)(s+d-2)|\Omega|~,
\end{align}
and the partially-massless representations are unitary. We remind the reader that our conventions are such that  $\Omega=-\ell^{-2}<0$ in dS$_d$.

\subsubsection{Transverse projectors} \label{secAdSTrans}

On a symmetric and traceful field $\Phi_{a(s)}$ which is otherwise unconstrained, the spin operator $\mathbb{W}^2$ acts according to the rule
\begin{align} \label{DAQuartCasTracefulSymField}
\mb{W}^2 \Phi_{a(s)} = & ~ \lambda_{(s,d)} \big ( \mb{Q} -\tau_{(s,s)} \Omega \big ) \Phi_{a(s)}    \non \\
&-s(2s+d-4)\cD_a \cD^b \Phi_{a(s-1)b}+ s(s-1)\eta_{a(2)} \cD^{b}\cD^{c}\Phi_{ a(s-2)bc} \non \\
&-s(s-1) \big [ (\mb{Q} -\alpha_{(s,d)} \Omega) \eta_{a(2)}-\mc{D}_a\mc{D}_a \big ]  \Phi_{ a(s-2)b}{}^{b} \non \\
&-s(s-1)(s-2)(s-3) \Omega \eta_{a(2)} \eta_{a(2)} \Phi_{a(s-4)bc}{}^{bc} ~, 
\end{align}
where $\alpha_{(s,d)}=2s^2+2sd -10s -3d+11$. It follows that TT fields $\phi^{\perp}_{a(s)}$ are eigentensors of $\mb{W}^2$
\begin{subequations}
\begin{align}
0=\big(\mb{W}^2& -\L_{(s,d)} \big)\phi^{\perp}_{a(s)}~,\label{eigent}\\[6pt]
\L_{(s,d)}  := \lambda_{(s,d)} \big ( \mb{Q} -\t_{(s,s)} \Omega &\big )~, \qquad \t_{(s,s)}=(s-1)(s+d-2) ~,
\end{align}
\end{subequations}
where $\lambda_{(s,d)}$ is defined in \eqref{lambdadeff}. We hereby omit the label $d$ and instead use $\Lambda_{s}$ and $\lambda_{s}$.

We now have the ingredients necessary to construct the spin-$(s,j)$ projection operators in (A)dS$_d$, which are given by
\begin{align} \label{dAdSprojector}
\P^{\perp}_{(s,j)} = \prod_{0 \leq k \leq s}^{k \neq j} \frac{( \mb{W}^2 - \L_{k})}{(\L_{j} - \L_{k})}~. 
\end{align}
In section \ref{secPoles} we comment on the poles of these projectors. 
As was the case in $\text{M}^d$, an efficient way to deduce the properties of $\P^{\perp}_{(s,j)}$ is to make use of the irreducible decomposition of an unconstrained field $\f_{a(s)}$, which we now turn to deriving. 
First we study the projector with $j=s$, which
was first computed in (A)dS$_4$ in \cite{KPproj} and its representation in terms of Casimir operators was derived later in \cite{Hutchings2023}. In (A)dS$_3$ it was constructed recently in \cite{HKP}.
 By making using of the identity \eqref{DAQuartCasTracefulSymField} reiteratively, it can be shown that  
\begin{align}
\cD^b \P^{\perp}_{(s,s)}\f_{b a(s-1)} = 0~.
\end{align}
Next, one may show that $\P^{\parallel}_{(s,s)}\f_{a(s)}$, where  $\P^{\parallel}_{(s,s)} = \mathds{1} - \P^{\perp}_{(s,s)}$, is pure gauge
\begin{align}
\P^{\parallel}_{(s,s)}  \f_{a(s)} = \cD_{\{ a }\f_{a(s-1) \} }~,
\end{align}
for some (non-local) $\f_{a(s-1)}$. It follows from these two results that the irreducible decomposition of $\f_{a(s)}$ is
\begin{align}
\phi_{a(s)} = \phi_{a(s)}^{\perp} + \cD_{\{a}\phi^{\perp}_{a(s-1)\}}+\cdots + \cD_{\{a_1}\cdots\cD_{a_s\}}\phi~, \label{dAdSpoindecomp}
\end{align}
for some (non-local) TT fields $\phi^{\perp}_{a(j)}$ with $1\leq j \leq s$ and scalar $\phi^{\perp}$ whose specific form are given below. The notation $\{\cdots\}$ used to denote the symmetric traceless part of a tensor is defined in a similar way as in $\text{M}^d$, see equation \eqref{tracesym1}. 

Given two positive integers $j\geq 0$ and $k\geq 0$, and traceless $\phi_{a(k)}$, one may show that the following identity holds
\begin{align}
(\cD^b)^j\cD_{\{b_1}\cdots\cD_{b_j}\phi_{a_1\cdots a_k\}}=(-1)^j\Big[\prod_{l=1}^{j}\frac{\big(\mb{W}^2-\L_{k+l}\big)}{(k+l)(2k+2l+d-4)}\Big]\phi_{a(k)} \label{magicId2}
\end{align}
via induction on $j$. If in addition $\phi_{a(k)}$ is transverse, $\phi_{a(k)}=\phi_{a(k)}^{\perp}$, then this reduces to 
\begin{align}
(\cD^b)^j\cD_{\{b_1}\cdots\cD_{b_j}\phi^{\perp}_{a_1\cdots a_k\}}=\bigg [ \prod_{l=1}^{j}\frac{l(2k+l+d-3)}{(k+l)(2k+2l+d-4)}\big(\mb{Q}-\rho^2_{(l,k+l)}\big) \bigg ] \phi^{\perp}_{a(k)}~, \label{magicId2b}
\end{align}
where we have used \eqref{eigent} and \eqref{EnterPM}. This useful identity allows one to recover each $\phi^{\perp}_{a(j)}$ in \eqref{dAdSpoindecomp} in terms of the original field $\phi_{a(s)}$
\begin{align}
\f^{\perp}_{a(j)} = \bigg [ \prod_{k=1}^{s-j} \frac{(j+k)(2j+2k+d-4)}{k(2j+k+d-3)}\big( \mb{Q} - \rho^2_{(k,j+k)} \big)^{-1} \bigg ] \P^{\perp}_{(s,j)} (\cD^b)^{s-j} \f_{a(j) b(s-j)}~. \label{ch}
\end{align}
Here $0\leq j \leq s$ and $\rho^2_{(t,s)}$ are the PM values \eqref{PMvals}, see section \ref{secPoles} for more details. 

Having obtained the irreducible decomposition \eqref{dAdSpoindecomp}, we may now state the properties of the operators $\Pi_{(s,j)}^{\perp}$, which are direct generalisations of those in M$^d$. They resolve the identity
\begin{align}
\mathds{1} = \sum_{j=0}^s\Pi_{(s,j)}^{\perp}~, \label{dAdSidres}
\end{align}
on $\mathscr{V}_{(l)}$ with $l$ arbitrary. On any $\mathscr{V}_{(l)}$ with $0\leq l \leq s$ they are orthogonal projectors
\begin{align}
\Pi^{\perp}_{(s,j)}\Pi^{\perp}_{(s,k)}=\delta_{jk}\Pi^{\perp}_{(s,k)}~,\qquad \text{on } \mathscr{V}_{(l)} \text{ with } 0\leq l\leq s~. \label{dAsSorthog2}
\end{align}
The operator $\Pi^{\perp}_{(s,j)}$ acts as the identity on the space of TT rank-$j$ fields
\begin{align}
0=\cD^b\phi^{\perp}_{ba(j-1)} \qquad \implies \qquad \Pi^{\perp}_{(s,j)}\phi^{\perp}_{a(j)}=\phi^{\perp}_{a(j)}~. \label{dAdSactid2}
\end{align}
Given three integers $s,j,k$ satisfying the inequalities $0\leq j\leq s-k$ and $0\leq k \leq s$, the following projectors are equivalent
\begin{align}
\Pi^{\perp}_{(s,k)} = \Pi^{\perp}_{(s-j,k)}~\qquad \text{on } \mathscr{V}_{(k)} 
\end{align}
when restricted to the space $\mathscr{V}_{(k)}$.

For fixed integers $s$ and $j$ satisfying $j\leq s$, an important property of $\Pi^{\perp}_{(s,j)}$ is
\begin{subequations} \label{dAdSLongi2}
	\begin{align}
	\Pi^{\perp}_{(s,j)}\phi_{a(l)}&=\cD_{\{a_1}\cdots\cD_{a_{l-j}}\phi^{\perp}_{a_{l-j+1}\cdots a_l\}} \qquad & &j\leq l \leq s~, \label{dAdSLongi2a}\\
	\Pi^{\perp}_{(s,j)}\phi_{a(l)}&=0 \phantom{(\pa_{\a\ad})^{l-j}S^{\perp}_{\a(j)\ad(j)}(\phi)} & &0\leq l \leq j-1~,  \label{dAdSLongi2b}
	\end{align} 
\end{subequations}
where $\phi^{\perp}_{a(j)}$ is given by \eqref{ch}. From \eqref{dAdSLongi2a} it follows that $\Pi^{\perp}_{(s,j)}\phi_{a(l)}$ is an eigentensor of $\mb{W}^2$,
\begin{align}
0&=\big(\mb{W}^2-\L_{j}\big)\Pi^{\perp}_{(s,j)}\phi_{a(l)}~, \label{dAdSid55}
\end{align}
and has vanishing $(l-j+1)$ divergence
\begin{align}
0&=(\cD^b)^{l-j+1}\Pi^{\perp}_{(s,j)}\phi_{b(l-j+1)a(j-1)}~.\label{dAdSid54}
\end{align}
 A stronger version of identity \eqref{dAdSid54} is 
\begin{align}
0=(\mc{D}^b)^k\Pi_{(s,j)}^{\perp}\phi_{b(k)a(l-k)}~,\label{dAdSdid53}
\end{align}
for arbitrary positive integers $s,j,k,l$ satisfying the inequalities $l-k+1\leq j\leq s$ and $l\geq k$ (in particular this allows $l\geq s$).
We emphasise that $\Pi^{\perp}_{(s,j)}$ selects the UIR $\mf{D}(\r^2 ,j)$ from the reducible representation \eqref{decomp2} which  a field $\phi_{a(s)}$ on the mass-shell furnishes.
From \eqref{W2red} it follows that the projectors $\Pi^{\perp}_{(s,j)}$ are symmetric in the sense that
\begin{align}
\int \rd^d x \, e \,  \Psi^{a(s)} \P^{\perp}_{(s,j)} \Phi_{a(s)} = \int \rd^d x \, e \, \Phi^{a(s)} \P^{\perp}_{(s,j)} \Psi_{a(s)} ~,\qquad e^{-1}:= \text{det}(e_a{}^{m})~,
\end{align}
where the fields $\Psi_{a(s)} $ and $\Phi_{a(s)} $ are unconstrained and well-behaved.

\subsubsection{Poles of the projectors}\label{secPoles}

As mentioned above, the projectors $\P^{\perp}_{(s,s)}$ have been constructed within the following contexts: (i) (A)dS$_4$ \cite{KPproj} (see also \cite{Hutchings2023}); (ii) (A)dS$_3$ \cite{HKP}; (iii) $\mc{N}=1$ AdS$_4$ superspace \cite{BHKP}; and (iv) $\mc{N}=1$ AdS$_3$ superspace \cite{HKP}.  In all cases it was observed that the special masses corresponding to partially massless (super)fields of all depths $1\leq t \leq s$ appeared in the poles of $\P^{\perp}_{(s,s)}$. 
In this section we confirm that this pattern persists in all dimensions, and uncover the analogous pattern for the projectors $\P_{(s,j)}^{\perp}$ with $j<s$.

To this end, we point out the following identity for $\Lambda_j=\lambda_j(\mb{Q}-\tau_{(s,s)}\Omega)$
\begin{align}
\Lambda_j-\Lambda_k = (j-k)(j+k+d-3)\big(\mb{Q}-\rho^2_{(j-k,j)}\big)~,\label{EnterPM}
\end{align}
which may be used to put the projectors into the form
\begin{align}
\P^{\perp}_{(s,j)} = \prod_{0 \leq k \leq s}^{k \neq j} \frac{1}{(j-k)(j+k+d-3)}\frac{( \mb{W}^2 - \L_{k})}{(\mb{Q}-\rho^2_{(j-k,j)})}~.
\end{align}
Using $\rho^2_{(t,s)}=\rho^2_{(-t,s-t)}$ it is not difficult to see that
\begin{align}
 \prod_{0 \leq k \leq s}^{k \neq j}(\mb{Q}-\rho^2_{(j-k,j)})=
 \prod_{m=1}^{j} \big ( \mb{Q} - \rho^2_{(m,j)}\big ) \prod_{n=1}^{s-j} \big ( \mb{Q} - \rho^2_{(n,n+j)} \big )~. \label{Poles}
\end{align}
Here $\rho^2_{(t,s)}$ are the masses associated with the partially (and strictly) massless irreps \eqref{PMvals}.

In the case $j=s$ we see that the mass-values $\rho_{(t,s)}$ appearing within the poles of $\Pi_{(s,s)}^{\perp}$ are those associated with a spin-$s$ field of all possible depths $1\leq t \leq s$. Another way of interpreting this is that projector $\Pi_{(s,s)}^{\perp}$ is aware of all the possible mass values for which the representation $\mf{D}(\rho^2,s)$ shortens and a gauge symmetry appears. This confirms that the pattern observed earlier in $d=4$ and $d=3$ persists in all dimensions. 

In the case $0\leq j\leq s-1$ we observe that the spin-$(s,j)$ projector $\Pi_{(s,j)}^{\perp}$, which selects the spin-$j$ UIR $\mf{D}(\rho^2,j)$, possesses two distinct classes of poles. The first occur at $ \rho_{(t,j)} $ for $1 \leq t \leq j$, which are the masses associated with a spin-$j$ partially massless field with all possible depths $1\leq t \leq j$. Thus $\Pi_{(s,j)}^{\perp}$ is aware of all the possible mass values for which the representation $\mf{D}(\rho^2,j)$ shortens. This is to be expected and is in agreement with the pattern for $j=s$ noticed above. The second class of poles occur at $\rho_{(t,t+j)}$ for $1 \leq t \leq s-j$. These are the mass values associated with a spin-$k$ depth-$(k-j)$ partially massless field with $j+1\leq k \leq s$. This is unexpected as it does not conform to the pattern observed previously, and the projector $\Pi_{(s,j)}^{\perp}$ is privy to short representations  carrying spin not equal to the one it selects (i.e. $\mf{D}(\rho^2,j)$). It would be interesting to better understand the significance of this second class of poles.


\section{Gauge-invariant actions} \label{Sec4Actions}
In this section, we make use of the spin-$(s,j)$ projection operators computed in section \ref{SpinProjectionOperators} to systematically derive second-order-derivative linearised actions for rank-$s$ gauge fields in  M$^d$ which propagate (at least) a massless spin-$s$ mode. We also construct analogous actions in (A)dS$_d$ in addition to all PM actions. However, in (A)dS$_d$ there is a much greater freedom to construct actions due to the presence of the mass-like curvature parameter $\Omega$, and our analysis is consequently not exhaustive. We also elaborate on some interesting higher-derivative actions which do not require compensating fields to achieve gauge invariance.

We emphasise that the approach used below to construct actions in this paper is tailored towards \textit{gauge} theories. Thus, this method cannot be used to produce non-gauge theories such as Singh-Hagen's massive spin-$s$ model \cite{SH}. However, it should be possible to adapt our approach to reproduce  the gauge-invariant St\"uckelberg reformulation of the Singh-Hagen model. The latter was first derived for spins $s=2$ and $s=3$ by Zinoviev \cite{Zin1985}, and later extended to arbitrary integer spin in \cite{Klishevich1997pd} by Klishevich  and Zinoviev.


\subsection{Gauge-invariant actions in $\text{M}^d$} \label{secActMd}

The most general action that is quadratic in both spacetime derivatives and the rank-$s$ traceless  field $\phi_{a(s)}$, and which does not contain a dimensionful parameter, is
\begin{subequations}
	\begin{align}
	\mc{S}_{\text{gen}}^{(s)}[\phi]=\int \text{d}^dx \,\big(a_1\phi^{a(s)}\Box \phi_{a(s)}+a_2\phi^{a(s)}\pa_{a}\pa^{b}\phi_{ba(s-1)}\big)~.\label{dGenActa}
	\end{align}
		The action \eqref{dGenActa} may be recast into the following form
	\begin{align}
	\mc{S}_{\text{gen}}^{(s)}[\phi]=\int \text{d}^dx ~ \phi^{a(s)}\big(b_1\Box + b_2 \mb{W}^2\big)\phi_{a(s)}~.\label{dGenActb}
	\end{align}
	Inserting the partition of identity \eqref{didres} and using \eqref{did55}, one arrives at the equivalent expression
	\begin{align}
	\mc{S}_{\text{gen}}^{(s)}[\phi] = \int \text{d}^dx~ \phi^{a(s)}\Box\sum_{j=0}^{s}c_j\Pi^{\perp}_{(s,j)}\phi_{a(s)}~, \label{dGenActc}
	\end{align}
\end{subequations}
which is local despite its appearance. Here the constants $a_i,b_i$ and $c_i$ are related via
\begin{subequations}\label{constants66}
\begin{align}
&a_1=b_1+\lambda_s b_2~,\qquad\qquad\qquad \phantom{..} a_2 = -s(2s+d-4)b_2~,\\
&b_1=a_1+\frac{(s+d-3)}{(2s+d-4)}a_2~,\qquad b_2=-\frac{1}{s(2s+d-4)}a_2~,\\
&c_j=b_1+\lambda_j b_2 = a_1+\frac{\lambda_s-\lambda_j}{s(2s+d-4)}a_2~.
\end{align}
\end{subequations}

We wish to construct an action which describes a massless spin-$s$ field. The projector $\Pi^{\perp}_{(s,s)}$ selects the spin-$s$ part from the field $\phi_{a(s)}$ and  must be present in such an action, thus we require $c_s\neq 0$ and consequently $a_1\neq 0$. This allows us to hereby set $c_s=a_1=1$. Now, from property \eqref{Longi2b} it is clear that $\Pi^{\perp}_{(s,s)}\phi_{a(s)}$ is invariant under the gauge transformations 
\begin{align}
\delta \phi_{a(s)}=\pa_{\{a}\xi_{a(s-1)\}}=\pa_a\xi_{a(s-1)}-\frac{s-1}{2s+d-4}\eta_{a(2)}\pa^b\xi_{ba(s-2)}~,  \label{dmasslessgt}
\end{align}
for traceless $\xi_{a(s-1)}$.
However, the presence of the other projectors in \eqref{dGenActc} breaks this gauge symmetry.
In particular, for the gauge variation of \eqref{dGenActc} under \eqref{dmasslessgt} to vanish, we require
\begin{align}
0= c_j\,\Box\,\Pi^{\perp}_{(s,j)}\xi_{a(s-1)}~,\qquad 0\leq j\leq s-1~.\label{GIcondM}
\end{align}
For future reference we point out that 
$\Pi^{\perp}_{(s,j)}\xi_{a(s-1)}$ contains the following number of divergences
\begin{align}
\Pi^{\perp}_{(s,j)}\xi_{a(s-1)}~\sim~ (\pa^{b})^{s-j-1}\xi_{a(j)b(s-j-1)}~,\qquad 0\leq j\leq s-1~, \label{GPstructM}
\end{align}
 at the very least, see eq. \eqref{divstructure} with $s\rightarrow s-1$. 
We now proceed to systematically search for actions which can be made gauge invariant (perhaps by introducing compensating fields), by setting the coefficients $c_j$ of projectors in \eqref{dGenActc} to zero. It is clear that we have the freedom to set only a single $c_j$ to zero at a time. In section \ref{secAllprojMd} we consider the case when all $c_j\neq 0$.


\subsubsection{Projector $\Pi_{(s,s-1)}^{\perp}$ absent; massless Fronsdal action} \label{secFronsMd}

We first consider the case when the projector $\Pi^{\perp}_{(s,s-1)}$  is absent in \eqref{dGenActc}, i.e.  $c_{s-1}=0$. This fixes $b_1$ and $b_2$ in \eqref{dGenActb} (recall we also chose the normalisation $c_s=1$),  
\begin{align}
\mc{S}_{\text{gen}}^{(s)}[\phi]\Big|_{c_{s-1}=0}=\frac{1}{\lambda_s-\lambda_{s-1}}\int \text{d}^dx\, \phi^{a(s)}\big(\mb{W}^2-\lambda_{s-1}\Box \big)\phi_{a(s)}~,\label{dFronsActkin}
\end{align}
which is the version of the action we prefer to work with for computations.

With $c_{s-1} = 0$ it is clear that for gauge invariance  we need \eqref{GIcondM} to hold for $0\leq j \leq s-2$. Inspecting \eqref{GPstructM} we see that this can be achieved if the gauge parameter is transverse, $0=\pa^b\xi_{ba(s-2)}$.
This constraint may be arrived at by introducing a compensator $\chi_{a(s-2)}$ which transforms according to
\begin{align}
\delta \chi_{a(s-2)} = \pa^{b}\xi_{ba(s-2)}~. \label{dcompgt1}
\end{align}
Imposing the gauge condition $\chi_{a(s-2)}=0$ yields the desired constraint. 

To obtain an action that is invariant under \eqref{dmasslessgt} and \eqref{dcompgt1}, we must supplement \eqref{dFronsActkin} with a $\phi\chi$ and $\chi\chi$ sector. The result of this routine procedure, 
\begin{align} \label{dFMFronsdalAction}
\mc{S}_{\text{Fronsdal}}^{(s)}[\phi,\chi]=\frac{1}{\lambda_s-\lambda_{s-1}}\int \text{d}^dx\, &\Big\{\phi^{a(s)}\big(\mb{W}^2-\lambda_{s-1}\Box\big)\phi_{a(s)}
+2(s-1)(2s+d-6)\phi^{a(s)}\pa_a\pa_a\chi_{a(s-2)}  \non\\
&+\frac{(s-1)(2s+d-6)}{s(2s+d-4)}\chi^{a(s-2)}\big(\mb{W}^2-\lambda_s\Box\big)\chi_{a(s-2)}\Big\}~,
\end{align}
is nothing but Fronsdal's formulation for a massless spin-$s$ field \cite{Fronsdal1} expressed using Casimir operators. To make contact with the literature, we introduce the doubly traceless field $\Phi_{a(s)}$ transforming according to 
\begin{align}
\delta\Phi_{a(s)}=\pa_a\xi_{a(s-1)}~,\qquad \Phi_{a(s-4)bc}{}^{bc}=0~.
\end{align}
The two traceless fields $\phi_{a(s)}$ and $\chi_{a(s-2)}$ may be unified via
\begin{align}
\Phi_{a(s)} = \phi_{a(s)}+\frac{s-1}{2s+d-4}\eta_{a(2)}\chi_{a(s-2)}~,
\end{align}
where we have made the identifications 
\begin{align}
\phi_{a(s)}= \Phi_{\{a(s)\}}=\Phi_{a(s)}-\frac{s(s-1)}{2(2s+d-4)}\eta_{a(2)}\Phi_{a(s-2)b}{}^{b}~,\qquad \chi_{a(s-2)}=\frac{s}{2}\Phi_{a(s-2)b}{}^{b}~.
\end{align}
If we substitute these into the action \eqref{dFMFronsdalAction} then we find
\begin{align} \label{dFMFronsdalActionFS}
\mc{S}^{(s)}_{\text{Fronsdal}}[\Phi] = \int \rd^d x ~ \Phi^{a(s)}\Big( \mc{R}_{a(s)}(\Phi)-\frac{1}{4}s(s-1)\eta_{a(2)}\mc{R}_{a(s-2)b}{}^{b}(\Phi)\Big)~,
\end{align}
where $\mc{R}_{a(s)}(\Phi)=\Box\Phi_{a(s)}-s\pa_a\pa^b\Phi_{a(s)b}+\frac{1}{2}s(s-1)\pa_a\pa_a\Phi_{a(s-2)b}{}^{b}$ is the rank-$s$ Fronsdal tensor. 

We note that action \eqref{dFronsActkin} (see also \eqref{dAdSFronsActkin}) coincides with the `Maxwell-like' action derived by Campleoni and Francia in \cite{Campoleoni2012th}.\footnote{The analysis of \cite{Campoleoni2012th} was not limited to traceless symmetric fields, but also included generalisations to: (i) traceful symmetric fields; (ii) mixed-symmetric fields; and (iii) (A)dS$_d$ space. } It was assumed in  \cite{Campoleoni2012th} that the gauge parameter is transverse from the on-set, thus it was not necessary to introduce the compensator $\chi_{a(s-2)}$. 

\subsubsection{Projector $\Pi_{(s,s-t)}^{\perp}$ absent; depth-$t$}\label{secNewDepthMd}

We can generalise the analysis presented in the previous subsection. Let us fix two integers $s$ and $t$ satisfying $1\leq t \leq s$. We consider the case when the projector $\Pi_{(s,s-t)}^{\perp}$  is absent in \eqref{dGenActc}, i.e. when $c_{s-t}=0$ and hence 
\begin{align}
\mc{S}_{\text{gen}}^{(s)}[\phi]\Big|_{c_{s-t}=0}=\frac{1}{\lambda_s-\lambda_{s-t}}\int \text{d}^dx \, \phi^{a(s)}\big(\mb{W}^2-\lambda_{s-t}\Box\big)\phi_{a(s)}~.\label{dFronsActkin3}
\end{align}
It is convenient to split the conditions for gauge invariance \eqref{GIcondM} into the following sets
\begin{subequations}
\begin{align}
0&=c_k\,\Box\,\Pi^{\perp}_{(s,k)}\xi_{a(s-1)}~\qquad s-t+1\leq k\leq s-1~,\label{repairjob1M}\\
0&=c_k\,\Box\,\Pi^{\perp}_{(s,k)}\xi_{a(s-1)}~\qquad ~~~~~~~~~~0\leq k\leq s-t-1~.\label{repairjob2M}
\end{align}
\end{subequations}
To satisfy both sets simultaneously we require \eqref{repairjob1M} to hold identically and \eqref{repairjob2M} to be reached by the introduction of a compensator. 
From the identity \eqref{Longi2b} we see that  \eqref{repairjob1M} is satisfied identically if $\xi_{a(s-1)}=\pa_{\{a_1}\cdots \pa_{a_{t-1}}\xi_{a_{t}\cdots a_{s-1}\}}$ for unconstrained $\xi_{a(s-t)}$. 
With this the gauge transformations \eqref{dmasslessgt} become
\begin{align}
\delta \phi_{a(s)}=\pa_{\{a_1}\cdots\pa_{a_{t}}\xi_{a_{t+1}\cdots a_{s}\}}~,  \label{ddepthxgt}
\end{align}
and are said to have depth-$t$.

To arrive at conditions \eqref{repairjob2M} we introduce a compensator $\chi_{a(s-t-1)}$ transforming as 
\begin{align}
\delta \chi_{a(s-t-1)}= \pa^{b}\xi_{ba(s-t-1)}~.\label{dcompgtx}
\end{align}
The gauge condition $\chi_{a(s-t-1)}=0$, which implies $0=\pa^b\xi_{ba(s-t-1)}$, yields the constraints \eqref{repairjob2M}.  To confirm this we observe that upon substituting 
$\xi_{a(s-1)}=\pa_{\{a_1}\cdots \pa_{a_{t-1}}\xi_{a_{t}\cdots a_{s-1}\}}$ into \eqref{GPstructM} we find that $\Pi^{\perp}_{(s,k)}\xi_{a(s-1)}\sim (\pa^b)^{s-t-k}\xi_{b(s-t-k)a(k)}$ for $0\leq k \leq s-t-1$.

It remains to determine the coupling between $\phi_{a(s)}$ and $\chi_{a(s-t-1)}$, and the kinetic sector for $\chi_{a(s-t-1)}$, which must accompany \eqref{dFronsActkin3} for invariance under \eqref{ddepthxgt} and \eqref{dcompgtx}. After performing this routine procedure, one arrives at the final action 
\begin{align} \label{dFMDepthSJAction}
\mc{S}^{(t,s)}_{\text{massless}}[\phi,\chi] &= \frac{1}{\lambda_s-\lambda_{s-t}}\int \text{d}^dx\, \Big\{ \phi^{a(s)}\big(\mb{W}^2-\lambda_{s-t}\Box\big)\phi_{a(s)}\non\\
&+(s-t)(2s-2t+d-4)\Big[2\phi^{a(s)}(\pa_{a})^{t+1}\chi_{a(s-t-1)}  \non \\
&+\chi^{a(s-t-1)}\prod_{k=1}^{t}\frac{1}{(s-k+1)(2s-2k+d-2)}\big (\mb{W}^2-\l_{s-k+1}\Box \big )\chi_{a(s-t-1)}\Big]\Big\}~.  
\end{align}
Note that  the $\chi\chi$ sector of the above action is higher-derivative, having order $2t$. To compute the gauge variation of the compensator sector the identity \eqref{magicid} proves useful. 

In the $t=1$ case, the action  \eqref{dFMDepthSJAction} coincides with the Fronsdal action \eqref{dFMFronsdalAction}. Notice that in the maximal depth case ($t=s$), the compensator $\chi_{a(s-t-1)}$  is ill-defined, however, this issue does not plague the action \eqref{dFMDepthSJAction} as the $\chi_{a(s-t-1)}$ sector is accompanied by a factor of $s-t$. In this case, the compensators are absent in the action, leaving only the dynamical $\phi_{a(s)}$ sector
\begin{align}  \label{dFMDepthSAction}
\mc{S}^{(s,s)}_{\text{massless}}[\phi] &= \frac{1}{\lambda_s}\int \text{d}^dx \, \phi^{a(s)} \mb{W}^2 \phi_{a(s)}~.
\end{align}
In $d=4$ this coincides with the conformally-invariant action for a maximal depth-$s$ conformal spin-$s$ field.
Such actions without compensators will be the topic of focus in section \ref{secNoCompMd}.

It is instructive to analyse the on-shell content of the action \eqref{dFMDepthSJAction}. First, we note that after gauging away the compensator $\chi_{a(s-t-1)}$, 
the equations of motion are
\begin{subequations}
\begin{align}
0&=(\pa^b)^{t+1}\phi_{b(t+1)a(s-t-1)}~,\label{EoM1}\\
0&=\big (\mb{W}^2-\lambda_{s-t}\Box \big )\phi_{a(s)}~, \label{EoM2}
\end{align}
\end{subequations}
with residual gauge symmetry 
\begin{align}
\delta \phi_{a(s)}=\pa_{\{a_1}\cdots\pa_{a_{t}}\rho^{\perp}_{a_{t+1}\cdots a_{s}\}}~,  \qquad 0=\pa^b\rho^{\perp}_{ba(s-t-1)}~.\label{resgs}
\end{align}
In the maximal-depth case $s=t$ there is no compensator and hence only \eqref{EoM2} is present.

The irreducible decomposition of $\phi_{a(s)}$ is given by 
\begin{align}
\phi_{a(s)}=\phi^{\perp}_{a(s)}+\pa_{\{a}\phi^{\perp}_{a(s-1)\}}+\cdots + \pa_{\{a_1}\cdots\pa_{a_{t}}\phi^{\perp}_{a_{t+1}\cdots a_{s}\}}~,
\end{align}
where the tail is cut off due to \eqref{EoM1}. We observe that the residual gauge symmetry \eqref{resgs} is precisely enough to gauge away the last term in this decomposition,
\begin{align}
\phi_{a(s)}=\phi^{\perp}_{a(s)}+\pa_{\{a}\phi^{\perp}_{a(s-1)\}}+\cdots + \pa_{\{a_1}\cdots\pa_{a_{t-1}}\phi^{\perp}_{a_{t}\cdots a_{s}\}}~.\label{eomdecomp}
\end{align}
This is equivalent to imposing the gauge condition 
\begin{align}
0=(\pa^b)^{t}\phi_{b(t)a(s-t)}~. \label{gc1}
\end{align}
The residual gauge symmetry is \eqref{resgs} with the extra condition $0=\Box^{t}\rho_{a(s-t)}$.

Since the factor $(\mb{W}^2-\lambda_{s-t}\Box)$ appears in all projectors $\Pi_{(s,k)}^{\perp}$ except for $k=s-t$, we see that $\Pi_{(s,k)}^{\perp}\phi_{a(s)}=0$ for $k\neq s-t$ with $0\leq k\leq s$ due to the equation of motion \eqref{EoM2}.  This would seem to indicate that no modes of $\phi_{a(s)}$ propagate. However, on the space of constrained fields $\phi_{a(s)}$ satisfying \eqref{gc1} with $2 \leq t \leq s$, which we denote by $\mathscr{V}_{(s,t)}^{\perp}$, there exists another set of projectors.\footnote{Relevant projectors for $t=0$ are $\Pi_{(s,k)}^{\perp}$ whilst for $t=1$ the field is irreducible and projectors are unnecessary.} Given two integers $s$ and $t$, which characterise the rank of the field $\phi_{a(s)}$ and the constraint \eqref{gc1} respectively, the new set of projectors are defined as 
\begin{align}
\hat{\mathcal{P}}^{\perp}_{(s,t,j)}=\prod_{s-t+1\leq k \leq s}^{k\neq j}\frac{(\mb{W}^2-\l_{k}\Box)}{(\l_j-\l_k)\Box}~,
\end{align}
with $2\leq t\leq s$ and $s-t+1 \leq j \leq s$. When acting on $\mathscr{V}_{(s,t)}^{\perp}$ with $s$ and $t$ fixed, they satisfy the usual projector properties
\begin{align}
\mathds{1}=\sum_{j=s-t+1}^s\hat{\mathcal{P}}^{\perp}_{(s,t,j)}~, \qquad \hat{\mathcal{P}}^{\perp}_{(s,t,j)}\hat{\mathcal{P}}^{\perp}_{(s,t,l)}=\delta_{jl}\hat{\mathcal{P}}^{\perp}_{(s,t,j)}~,
\end{align}
where $s-t+1 \leq l \leq s$. In addition projector $\hat{\mathcal{P}}^{\perp}_{(s,t,j)}$ satisfies the property 
\begin{align}
\hat{\mathcal{P}}^{\perp}_{(s,t,j)}\phi_{a(s)}=\pa_{\{a_1}\cdots\pa_{a_{s-j}}\phi^{\perp}_{a_{s-j+1}\cdots a_s\}}~,
\end{align} 
which has the following consequences 
\begin{subequations}
\begin{align}
0&=(\pa^b)^{s-j+1}\hat{\mathcal{P}}^{\perp}_{(s,t,j)}\phi_{b(s-j+1)a(j-1)}~,\\
0&=\big(\mb{W}^2-\lambda_j\Box\big)\hat{\mathcal{P}}^{\perp}_{(s,t,j)}\phi_{a(s)}~.\label{npprop2}
\end{align}
\end{subequations}
It follows that $\phi_{a(s)}\in \mathscr{V}_{(s,t)}^{\perp}$ has irreducible decomposition \eqref{eomdecomp} and $\hat{\mathcal{P}}^{\perp}_{(s,t,j)}\phi_{a(s)}$ carries spin $j$.

Resuming the on-shell analysis, we note that $\hat{\mathcal{P}}^{\perp}_{(s,t,j)}\phi_{a(s)}\neq 0$ since the factor $(\mb{W}^2-\lambda_{s-t}\Box)$ appearing in \eqref{EoM2} is not present in $\hat{\mathcal{P}}^{\perp}_{(s,t,j)}$. Thus the projectors $\hat{\mathcal{P}}^{\perp}_{(s,t,j)}$ are not plagued by the issue noted above for $\Pi^{\perp}_{(s,j)}$. Applying $\hat{\mathcal{P}}^{\perp}_{(s,t,j)}$ to \eqref{EoM2} we find 
\begin{align}
0=\Box \big(\hat{\mathcal{P}}^{\perp}_{(s,t,j)}\phi_{a(s)}\big)\qquad\qquad s-t+1\leq j \leq s~,
\end{align} 
by virtue of \eqref{npprop2}. We see that the spin-$j$ modes of $\phi_{a(s)}$ with $s-t+1\leq j \leq s$ are propagating and massless, whilst those with $j\leq s-t$ are non-propagating. Thus, on-shell, the model \eqref{dFMDepthSJAction} describes the reducible representation
\begin{align}
D(0,s)~\oplus~ D(0,s-1)~\oplus~\cdots~\oplus ~ D(0,s-t+1)~ \oplus (\cdots)
\end{align}
of the Poincar\'{e} algebra. The second ellipsis signifies the potential presence of extra (possibly ghost-like) states which may originate from the higher-derivative nature of the compensator sector. A more careful analysis, which is outside the scope of the current paper, should be conducted to establish their presence.

\subsubsection{All projectors present}\label{secAllprojMd}
Let us now consider the case when all the projectors appearing in \eqref{dGenActc} are present, i.e. when $c_j \neq 0$ for $0 \leq j \leq s$. The resulting  action takes the general form
\begin{align} \label{dActComp}
\mc{S}_{\text{gen}}^{(s)}[\phi] \Big |_{c_j \neq 0}=\int \text{d}^dx ~ \phi^{a(s)}\big(b_2 \mb{W}^2 +  (1 - \l_s b_2)\Box \big)\phi_{a(s)}~,
\end{align}
where the constant $b_2$ remains arbitrary. We note that $c_s$ is still normalised to be  $c_s = 1$.

In accordance with \eqref{GIcondM}, there are numerous ways to ensure that the action \eqref{dActComp} is gauge invariant. To see this, we again split the conditions \eqref{GIcondM} for gauge invariance into the two sets
\vspace{-\baselineskip}
\begin{subequations}
	\begin{align}
	0&=c_l\,\Box\,\Pi^{\perp}_{(s,l)}\xi_{a(s-1)}~,\qquad s-j \leq l \leq s-1~, \label{dMGCProjPres1}\\
	0&=c_l\,\Box\,\Pi^{\perp}_{(s,l)}\xi_{a(s-1)}~,\qquad ~~~~~0\leq l \leq s-j-1~. \label{dMGCProjPres2}
	\end{align}
\end{subequations}
It follows from identity \eqref{Longi2b} that  condition \eqref{dMGCProjPres1} can be identically satisfied if $\x_{a(s-1)} = \pa_{ \lbrace a_1} \cdots \pa_{a_j} \x_{a_{j+1} \cdots a_{s-1} \rbrace}$ for unconstrained $\x_{a(s-j-1)} $. The remaining equation \eqref{dMGCProjPres2} can be  arrived at via one of the following three conditions:
\begin{enumerate}
	\item  Given fixed $j$ and $k$ for $0 \leq j \leq s-1$ and $s-j \leq t \leq s$, we require $\x_{a(s-j-1)} $ to satisfy
	\bsubeq \label{dMCondProjPres}
	\be 
	(\mb{W}^2 - \l_{t} \Box) \x_{a(s-j-1)} = 0~. \label{dMAllProjGC2}
	\ee
	\item Given fixed $j$ for $0 \leq j \leq s-1$, we require $\x_{a(s-j-1)} $ to satisfy  
	\be
	\pa^b \x_{a(s-j-2) b} = 0~. \label{dMAllProjGC0}
	\ee
	\item Given fixed $j$ for $0 \leq j \leq s-1$, we require $\x_{a(s-j-1)} $ to satisfy  
	\be
	\Box \x_{a(s-j-1)} = 0~. \label{dMAllProjGC1}
	\ee
	\esubeq
\end{enumerate}
We note that conditions \eqref{dMAllProjGC2} and \eqref{dMAllProjGC0}  are made apparent once the projector $\P^{\perp}_{(s,l)}$ in \eqref{dMGCProjPres2} is evaluated explicitly. Seemingly, it appears that condition \eqref{dMAllProjGC1} immediately satisfies \eqref{dMGCProjPres2}. However, one may notice that this condition gives rise to problems within the poles of the projector in \eqref{dMGCProjPres2} for  $0 \leq j \leq s-2$. The exception to this is the case when the gauge parameter is scalar ($j = s-1$), as the problem of d'Alembert operators appearing in the poles is avoided.
Let us now investigate the actions formulated from each of the cases \eqref{dMCondProjPres} separately. 

In order to arrive at the first condition \eqref{dMAllProjGC2}, it is necessary to introduce the compensator $\c_{a(s-j-1)}$ which transforms in the following fashion 
\be
\d \c_{a(s-j-1)} =  \big (\mb{W}^2 - \l_{t} \Box \big )  \x_{a(s-j-1)} ~,
\ee
where $0 \leq j \leq s-1$ and  $s-j \leq t \leq s$.  Employing the general prescription yields the action 
\begin{align} \label{dMAct2AllProj}
\mc{S} [\phi, \c] &=  \frac{1}{\l_s - \l_t}  \int \text{d}^dx ~ \Big \lbrace \phi^{a(s)}\big(  \mb{W}^2 - \l_t \Box \big)\phi_{a(s)} -2  \f^{a(s)} (\pa_{ a} )^{j+1}  \c_{a(s-j-1) }   \non \\
&+ \prod_{l=1}^{j+1} \frac{1}{(s-l+1)(2s-2l+d-2)} \c^{a(s-j-1)}   \prod_{ s- j \leq k \leq s }^{k \neq t} ( \mb{W}^2 - \l_k \Box)  \c_{a(s-j-1)} \Big \rbrace~. 
\end{align}
We first note that a gauge-invariant action could not be constructed in the case $t=s$, thus the action \eqref{dMAct2AllProj}  is only defined for values of $t$ within the range  $s-j \leq t \leq s-1$. This avoids any inconsistencies with the overall factor of $\frac{1}{\l_s - \l_t}$ appearing in the action above.  

Since the coefficient $b_2$ was fixed to be $\frac{1}{\l_s - \l_t}$ via gauge invariance, we find that the constant $c_l$, which we recall is associated with the projector $\P^{\perp}_{(s,l)}$ in \eqref{dGenActc} for $0 \leq l \leq s$, is given by $c_l =\frac{\l_l - \l_t}{\l_s - \l_t}$. It follows that the projector $\P^{\perp}_{(s,l)}$    is not present in \eqref{dGenActc} for the case $l = t$, hence, the action \eqref{dMAct2AllProj} coincides with a special case of the massless depth-$(s-t)$ action \eqref{dFMDepthSJAction} for $s-j \leq t \leq s-1$, by which we mean that  the corresponding gauge parameter and the compensating field are defined to be $\x_{a(t)} = \pa_{\lbrace a_1} \cdots \pa_{a_{t+j-s+1}} \x_{a_{t+j-s+2} \cdots a_t \rbrace}$ and $\c_{a(t-1)} = \pa_{\lbrace a_1} \cdots \pa_{a_{t+j-s} } \c_{a_{t+j-s +1} \cdots a_{t-1} \rbrace } $, respectively. We observe that our initial assumption that all the projectors in \eqref{dGenActc}  are present has been violated and therefore there are no gauge invariant actions for the case \eqref{dMAllProjGC2} with all $c_l \neq 0$. Finally, it appears that the massless depth-$(s-t)$ actions for the case $0 \leq t \leq s-j-1$ cannot be formulated in the above approach.

In order to arrive at the second condition \eqref{dMAllProjGC0}, it is necessary to introduce the compensating field $\c_{a(s-j-2)}$ which transforms as 
\be
\d  \c_{a(s-j-2)} = \pa^b \x_{b a(s-j-2)}~,
\ee
where $0 \leq j \leq s-1$. Employing the general prescription allows one to generate the following gauge-invariant action in terms of the fields $\f_{a(s)}$ and $\c_{a(s-j-2)} $
\begin{align} 
\mc{S}[\phi,\chi] &= \frac{1}{\lambda_s-\lambda_{s-j-1}}\int \text{d}^dx\, \Big\{ \phi^{a(s)}\big(\mb{W}^2-\lambda_{s-j-1}\Box\big)\phi_{a(s)}\non\\
&+(s-j-1)(2s-2j+d-6)\Big[2\phi^{a(s)}(\pa_{a})^{j+2}\chi_{a(j+2)}  \non \\
&+\chi^{a(s-j-2)}\prod_{k=1}^{j+1}\frac{1}{(s-k+2)(2s-2k+d+4)}\big (\mb{W}^2-\l_{s-k+2}\Box \big )\chi_{a(s-j-2)}\Big]\Big\}~.  
\end{align}
This action coincides with the  massless depth-$(j+1)$ action \eqref{dFMDepthSJAction}, as expected. Once again, for invariance under the gauge transformations \eqref{dMAllProjGC0},  at least one projector must be absent.

As discussed earlier, the final condition \eqref{dMAllProjGC1} is naive since inconsistencies arise in the poles of the projector in \eqref{dMGCProjPres2} for  $0 \leq j \leq s-2$, due to the presence of inverse powers of $\Box$. The only case where this problem does not plague \eqref{dMAllProjGC1} occurs when $j =s-1$ or in other words, when the gauge parameter is scalar. To arrive at this condition, we need to introduce the compensator $\c$ which transforms via $
\d \c = \Box \x $. Constructing the most general action involving the dynamical field $\f_{a(s)}$ and compensating field $\c$ yields 
\begin{align} \label{dMActioallproj}
\mc{S}[\f ,\c] =\int \text{d}^dx ~ \Big \lbrace &\phi^{a(s)}\big( (1 - \l_s b_2)\Box + b_2 \mb{W}^2\big)\phi_{a(s)} - 2 (1- \l_s  b_2) \f^{a(s)} ( \pa_{a} )^s \c \non  \\
&+ (-1)^s (1-\l_s b_2) \prod_{k=1}^{s}\frac{\l_k}{k(2k+d-4)}\c \Box^{s-1} \c \Big \rbrace~,
\end{align}
which is invariant under the gauge transformations
\bsubeq
\begin{align}
\d \f_{a(s)} &= \pa_{ \lbrace a_1}\cdots \pa_{a_s \rbrace} \x ~,\label{spinsdepths} \\
\d \c &= \Box \x~.
\end{align}
\esubeq

The action \eqref{dMActioallproj} describes a novel one-parameter family of gauge-invariant actions, whose on-shell content we now elucidate. In the case $b_2 = \frac{1}{\l_s}$ the compensator decouples\footnote{Up to an overall normalisation the action \eqref{dMActioallproj} with $b_2 = \frac{1}{\l_s}$ coincides with the action \eqref{randact} when $t=s$. In $d=4$ this is the maximal-depth conformal spin-$s$ action.} and the equation of motion is 
\be \label{RedEomScalAct}
0 = \mathbb{W}^2  \f_{a(s)}~.
\ee
The gauge freedom \eqref{spinsdepths} allows one to impose the gauge condition $0=(\pa ^b)^s \f_{b(s)}$. Since the factor $(\mb{W}^2-\lambda_0\Box)=\mathbb{W}^2$ is present in all projectors $\P^{\perp}_{(s,j)}$ except $\P^{\perp}_{(s,0)}$, we find that only $\P^{\perp}_{(s,0)}\phi_{a(s)}$ is non-zero. Instead of $\P^{\perp}_{(s,j)}$ we need to consider the projectors $\hat{\mathcal{P}}^{\perp}_{(s,t,j)}$ with $t=s$ and hence $1\leq j \leq s$. Applying $\hat{\mathcal{P}}^{\perp}_{(s,s,j)}$ to \eqref{RedEomScalAct} yields $0=\Box(\hat{\mathcal{P}}^{\perp}_{(s,s,j)}\phi_{a(s)})$. 
Thus the spin-$j$ modes with $1 \leq j \leq s$ encoded in the field $\f_{a(s)}$ are propagating and massless. Thus, when $b_2 = \frac{1}{\l_s}$, the action \eqref{dMActioallproj} describes the following reducible representation of the Poincar\'{e} algebra
\begin{align} \label{allbutscalar}
D(0,s)~\oplus~ D(0,s-1)~\oplus~\cdots~\oplus ~ D(0,1)~.
\end{align}

 Next we consider the case when $b_2 \neq \frac{1}{\l_s}$. After gauging away the compensator $\c$, the equations of motion of \eqref{dMActioallproj} reduce to
\bsubeq \label{EoMScalAct}
\begin{align} 
0 &= \big [ b_2\mathbb{W}^2 + \big (1-\l_s b_2 \big ) \Box \big ] \f_{a(s)}~, \label{EoMScalActa}\\
0&= (\pa ^b)^s \f_{b(s)}~,\label{EoMScalActb}
\end{align}
\esubeq
with residual gauge symmetry 
\be
\d \f_{a(s)} = \pa_{ \lbrace a_1}\cdots \pa_{a_s \rbrace} \x ~, \qquad \Box \x = 0~. 
\ee
Equation of motion \eqref{EoMScalActb} implies that the spin-$0$ mode of $\f_{a(s)}$ is non-propagating. If $b_2\neq \frac{1}{\l_s -\l_j }$  then \eqref{EoMScalActa} does not yield the equation $0=(\mathbb{W}^2-\lambda_j\Box)\phi_{a(s)}$, and hence we may freely use $\P^{\perp}_{(s,j)}$. Applying $\P^{\perp}_{(s,j)}$ to \eqref{EoMScalActa} with $1 \leq j \leq s$ yields
\be \label{EoM4}
0 = \Box  \P^{\perp}_{(s,j)} \f_{a(s)}~,
\ee
and we once again arrive at \eqref{allbutscalar}.
 
  One may notice that the remaining case $b_2 = \frac{1}{\l_s -\l_j }$ with $1 \leq j \leq s-1$ (the case $j=0$ implies $b_2=\frac{1}{\lambda_s}$), implies $c_{s-j} = 0 $. Thus, the resulting actions coincide with special cases of Fronsdal's spin-$s$ action \eqref{dFMFronsdalAction} in the case $j = 1$, and the new depth-$j$ actions \eqref{dFMDepthSJAction} when $2 \leq j \leq s-1$. They are special cases in the sense that they coincide upon specifying that the compensator $\c_{a(s-t-1)}$ and gauge parameter $\x_{a(s-t)}$ take the forms $\c_{a(s-t-1)} = \pa_{ \lbrace a_1 } \cdots \pa_{a_{s-t-1} \rbrace } \c$ and $\x_{a(s-t)} = \pa_{ \lbrace a_1 } \cdots \pa_{a_{s-t} \rbrace } \x $, respectively.

\subsubsection{Actions without compensators} \label{secNoCompMd}

The properties of the spin-$(s,j)$ projectors allow for the immediate construction of gauge-invariant actions which consist of only the traceless dynamical field $\phi_{a(s)}$, and no compensating fields, at the expense of being higher-derivative. For example, given two fixed integers $s$ and $t$ with $1\leq t \leq s$, one can define the order-$2s$ action 
\begin{align} \label{dFMActionNoComp1}
\mc{S}[\phi] = \int \text{d}^dx \, \phi^{a(s)} \Box^s \Pi^{\perp}_{(s,s-t+1)}\phi_{a(s)}~,
\end{align}
or the order-$2(s-t+1)$ action 
\begin{align}
\mc{S}[\phi] = \int \text{d}^dx \, \phi^{a(s)} \Box^{s-t+1} \Pi^{\perp}_{(s-t+1,s-t+1)}\phi_{a(s)}~.\label{randact}
\end{align}
Both of these actions are invariant under the depth-$t$ gauge transformations 
\begin{align}
\delta \phi_{a(s)}= \pa_{\{a_1}\cdots\pa_{a_{t}}\xi_{a_{t+1}\cdots a_{s}\}}~, \qquad 0=\xi_{a(s-t-2)b}{}^{b}~.
\end{align}
The actions \eqref{dFMActionNoComp1} and \eqref{randact}   coincide for $t=1$.  In the case $t=s$, the action \eqref{randact} coincides (up to an overall factor) with the depth-$s$ action \eqref{dFMDepthSAction}.
We emphasise that for $t>1$, actions \eqref{dFMActionNoComp1} and \eqref{randact} are not equivalent through the relations \eqref{equivalence}. 
Gauge invariance of action \eqref{dFMActionNoComp1} is manifest due to the projector properties \eqref{Longi2b} and \eqref{did54}, whilst gauge invariance of \eqref{randact} is manifest due to \eqref{Longi2b} and \eqref{did53}.

\subsubsection{Generalised conformal higher-spin actions} \label{secGenCHSMd}

The exponents of the operator $\Box$ in actions \eqref{dFMActionNoComp1} and \eqref{randact} are the smallest numbers which cancel the non-localities $\sim \frac{1}{\Box^k}$ with $k\leq s$ that are present in $\Pi_{(s,j)}^{\perp}$. However, inserting $\mc{A}\Box^{s}$ with $\mc{A}$ an arbitrary operator will achieve this.
By choosing an appropriate $\mc{A}$ and amending \eqref{randact} one may construct conformally invariant actions, a.k.a. conformal higher-spin (CHS) actions. Any local and parity invariant action involving $\Pi^{\perp}_{(s-t+1,s-t+1)}$ for $1\leq t \leq s$ must take the form 
\begin{align}
\mc{S}^{(t,s)}[\phi] = \int \text{d}^dx \, \phi_{(t)}^{a(s)}\Box^{s-t+1} \Pi^{\perp}_{(s-t+1,s-t+1)} \mc{A}_{a(s),}{}^{b(s)}\phi^{(t)}_{b(s)}~.\label{dgenCHS1}
\end{align}
Here $\mc{A}_{a(s), b(s)}$ is some local operator constructed from $\pa_a$ and $\eta_{ab}$ which is independently traceless in its two groups of indices (i.e. $\mc{A}_{a(s)}(\phi) \equiv \mc{A}_{a(s),}{}^{b(s)}\phi_{b(s)}$ is traceless and symmetric). The latter requirement ensures that \eqref{dgenCHS1} is invariant under the gauge transformations
\begin{align}
\delta \phi^{(t)}_{a(s)}= \pa_{\{a_1}\cdots\pa_{a_{t}}\xi_{a_{t+1}\cdots a_{s}\}}~,\qquad 0=\xi^{(t)}_{a(s-t-2)b}{}^{b}~.\label{gt555}
\end{align}
If $\phi^{(t)}_{a(s)}$ is considered to be a primary field with conformal weight $t-s+1$,\footnote{This weight is required for consistency with \eqref{gt555} if $\xi^{(t)}_{a(s-t)}$ is a primary field with weight $t$ less than $\phi^{(t)}_{a(s)}$. } then in order for \eqref{dgenCHS1} to be scale invariant we require that $\mc{A}$ has conformal weight $d-4$. In other words it is of order $[\mc{A}]:=d-4$ in derivatives. To progress our analysis we must consider the cases of odd and even $d$ separately.

\textbf{Conformal actions in even dimensions $d \geq 4$}

In even dimensions $[\mc{A}]=d-4$ is even and hence $\mc{A}$ should consist of products of second order operators. The most general operator of this type yields the action
\begin{align}
\mc{S}^{(t,s)}_{\text{CHS}}[\phi] = \int \text{d}^dx \, \phi_{(t)}^{a(s)} \prod_{k=1}^{\frac{1}{2}(d-4)}(\alpha_k\mb{W}^2+\beta_k\Box) \Box^{s-t+1} \Pi^{\perp}_{(s-t+1,s-t+1)}\phi^{(t)}_{a(s)}~,\label{evendgenCHS}
\end{align}
for some undetermined constants $\alpha_k$ and $\beta_k$. In the case $t=1$ the field $\Pi^{\perp}_{(s-t+1,s-t+1)}\phi^{(t)}_{a(s)}$ is irreducible and $\mb{W}^2\sim \lambda_s\Box$, hence the action reduces to
\begin{align}
\mc{S}^{(1,s)}_{\text{CHS}}[\phi] \propto \int \text{d}^dx \, \phi^{a(s)} \Box^{s+\frac{1}{2}(d-4)} \Pi^{\perp}_{(s,s)}\phi_{a(s)}~.
\end{align}
 This may be shown to coincide with the action for standard minimal depth conformal fields introduced in \cite{FT} for $d=4$ and in \cite{Se} for generic even $d$. 
 
 In $d=4$ we must have $\mc{A}=\mathds{1}$ for conformal invariance, and the action reduces to
 \begin{align}
 \mc{S}^{(t,s)}_{\text{CHS}}[\phi] = \int \text{d}^4x \, \phi_{(t)}^{a(s)} \Box^{s-t+1} \Pi^{\perp}_{(s-t+1,s-t+1)}\phi_{a(s)}^{(t)}~.
 \end{align}
 This gives rise to a new realisation for the generalised spin-$s$ depth-$t$ Bach tensor,
\begin{align}
\mc{B}_{a(s)}^{(t)} (\phi) = \Box^{s-t+1}\Pi_{(s-t+1, s-t+1)}^{\perp}\phi^{(t)}_{a(s)}~, 
\end{align}
whose properties of partial-conservation, tracelessness and gauge-invariance are manifest.  
  
In the case $t\geq 2$, the constants $\alpha_k$ and $\beta_k$ should be uniquely determined by requiring the action to be invariant under special conformal transformations. Equivalently, in the framework of conformal space (see e.g. \cite{PondsThesis}), the Lagrangian should be constrained to be primary, i.e. annihilated by the special conformal generator $K_a$. We leave this task for future work. Once complete, the resulting action should coincide with the generalised conformal higher-spin actions given by Vasiliev \cite{Vasiliev2009} (see also \cite{Marnelius}).

\textbf{Conformal actions in odd dimensions $d \geq 5$}

For odd $d$ the number $[\mc{A}]=d-4$ is odd and it is not possible to construct an operator with an odd number of $\pa_a$ which has index structure $\mc{A}_{a(s),b(s)}$. 
One possible way out is to sacrifice locality and use the operator $\sqrt{\Box}$, whereupon the most general form of the action \eqref{dgenCHS1} is 
\begin{align}
\mc{S}^{(t,s)}_{\text{CHS}}[\phi] = \int \text{d}^dx \, \phi_{(t)}^{a(s)} \sqrt{\Box}\prod_{k=1}^{\frac{1}{2}(d-5)}(\alpha_k\mb{W}^2-\beta_k\Box)\Box^{s-t+1} \Pi^{\perp}_{(s-t+1,s-t+1)}\phi^{(t)}_{a(s)}~.\label{odddgenCHS2}
\end{align}
 The non-local operator $\sqrt{\Box}$ requires a careful definition (see e.g. \cite{FracLap} and references therein) which is outside the scope of this paper. In the minimal depth case $t=1$ this reduces to 
\begin{align}
\mc{S}^{(1,s)}_{\text{CHS}}[\phi] \propto \int \text{d}^dx \, \phi^{a(s)} \sqrt{\Box}\Box^{s+\frac{1}{2}(d-5)} \Pi^{\perp}_{(s,s)}\phi_{a(s)}~.
\end{align}
In the case $j\geq 2$, the coefficients $\alpha_k$ and $\beta_k$ should be determined by imposing conformal invariance. Whether this is possible is unclear since to the best of our knowledge the behaviour of $\sqrt{\Box}$ under conformal transformations is uncharted territory. 

One other possibility is to sacrifice invariance under parity transformations by inserting the operator $\mb{H} :=\varepsilon^{ba_1\cdots a_{d-1}}\pa_bM_{a_1a_2}\cdots M_{a_{d-2}a_{d-1}}$ instead of $\sqrt{\Box}$. However, it may be shown that for $d\geq 5$,  $\mb{H}$ annihilates totally symmetric fields, $\mb{H}\Phi_{a(s)}=0$ (this is not true in $d=3$, which we exploit below). One would therefore need to reformulate the problem using conformal fields possessing mixed symmetry, which are discussed in \cite{Vasiliev2009}, and derive the relevant spin projectors.

\textbf{Conformal actions in three dimensions $d=3$}

For conformal invariance in $d=3$ we require $[\mc{A}] =d-4=-1$ and hence $\mc{A}$ is non-local. An appropriate operator is provided by $\mc{A}=\frac{1}{\mb{W}}$, where $\mb{W}=-\frac{1}{2}\ve_{abc}\pa^aM^{bc}$ ($\mb{W}$ is proportional to $\mb{H}$ introduced above) is the Pauli-Lubanski pseudo scalar. 
With this choice the order-$(2s-2t+1)$ action \eqref{dgenCHS1} becomes  
\begin{align} \label{TMGenCHS}
\cS^{(t,s)}_{\text{CHS}}[\phi] = \int \rd^3x ~ \phi_{(t)}^{a(s)} \frac{1}{\mb{W}}\Box^{s-t+1} \P^{\perp}_{(s-t+1,s-t+1)} \phi^{(t)}_{a(s)}~,
\end{align}
for $1\leq t \leq s$, which is invariant under the higher-depth gauge symmetries \eqref{gt555}.
Though dividing by $\mb{W}$ may look dangerous and unusual, it is harmless as its affect is to cancel one of the factors of $\mb{W}$ in $\P^{\perp}_{(s-t+1,s-t+1)}$, which always contains a factor $(\mb{W}^2-\lambda_0\Box)=\mb{W}^2=\mb{W}\mb{W}$. 

It can be shown that action \eqref{TMGenCHS} is equivalent to the spin-$s$ depth-$t$ conformal action 
\be
\cS^{(t,s)}_{\text{CHS}}[\phi] = \frac{1}{2(s-j)}\int \rd^3x ~ \f_{(t)}^{\a(2s)} \mc{C}^{(t)}_{\a(2s)} (\f)~, \label{genCHS12123}
\ee
which is formulated in terms of the spin-$s$ depth-$t$ Cotton tensor
\begin{align}
\mc{C}^{(t)}_{\a(2s)}(\phi) := \frac{1}{2^{2s-2t+1}} \sum_{k=t-1}^{s-1} \binom{2s}{2k+1} \binom{k}{t-1} \Box^{k-t+1} \big ( \pa_{\a}{}^\b \big )^{2s-2k-1} \f^{(t)}_{\b(2s-2k-1)\a(2k+1)}~. \label{Cottonspinor}
\end{align}
Here we have used two-component notation (see e.g. \cite{KuzenkoPonds2019}), which is the setting in which \eqref{Cottonspinor} was originally derived in \cite{KuzenkoPonds2019}.\footnote{See \cite{PopeTownsend1989} and \cite{Kuzenko16} for the bosonic and fermionic Cotton tensors with $t=1$ (minimal depth), respectively.}
This leads to a new realisation of the spin-$s$ depth-$t$ Cotton tensor (which is equivalent to \eqref{Cottonspinor} up to an overall normalisation) 
\begin{align}
\mc{C}^{(t)}_{a(s)}(\phi) = \frac{1}{\mb{W}}\Box^{s-t+1} \P^{\perp}_{(s-t+1,s-t+1)} \phi_{a(s)}^{(t)}=\frac{1}{\lambda_{s-t+1}}\mb{W}\Box^{s-t} \P^{\perp}_{(s-t+1,s-t+1)} \phi_{a(s)}^{(t)}~.\label{Cotton}
\end{align}
We emphasise that despite its appearance this is a local tensor due to the appearance of a $\mb{W}^2$ factor in the projector.
This version of the generalised Cotton tensor is far simpler than \eqref{Cottonspinor}. Moreover, the properties of partial-conservation,
\begin{align}
0=(\pa^b)^{t}\mc{C}^{(t)}_{b(t)a(s-t)}(\phi)~, \label{parcon}
\end{align}
invariance under the gauge transformations \eqref{gt555},
\begin{align}
0=\mc{C}_{a(s)}^{(t)}(\delta \phi)~,
\end{align}
and tracelessness follow from the projector properties \eqref{Longi2a} and \eqref{did54}. It is a non-trivial problem to establish these properties when working with the form \eqref{Cottonspinor}. 
The generalised spin-$s$ depth-$t$ CHS action is naturally factorised into multiple second-order differential operators, and a single first-order one. 

Finally, we note that the $d=3$ minimal-depth ($t=1$) actions \eqref{dgenCHS1} with the choices $\mc{A}=\mathds{1}$ and $\mc{A}=\frac{1}{\mb{W}^2}$ were investigated in \cite{Bergshoeff11} and \cite{Dalmazi21} respectively. These actions are not conformal. 

\subsubsection{Topologically massive higher-depth actions in $d=3$}\label{secTMHS}

Topologically massive models for higher-spin gauge fields refer to higher-spin extensions \cite{Bergshoeff11, KP18, Dalmazi21} of any of the following linearised massive spin-2 actions (i) topologically massive gravity \cite{DJT1,DJT2}; (ii) new topologically massive gravity \cite{Bergshoeff09, Dalmazi09}; or (iii) new massive gravity \cite{Bergshoeff092}. 
Topologically massive integer spin-$s$ models were obtained in (i) \cite{KP18} by coupling the massless Fronsdal action \cite{Fronsdal1} to the conformal higher-spin action \cite{PopeTownsend1989}; (ii) \cite{Bergshoeff11} by constructing a self-dual action using  the higher-spin Cotton tensor; and (iii) \cite{Dalmazi21} via a similar method to \cite{Bergshoeff11} but using a tensor closely related to the Cotton tensor.\footnote{In \cite{Boulanger2014vya} all possible field equations describing topologically massive theories in AdS$_3$ in terms of tensor(-spinor) fields were found, including their `critical' versions. It would be interesting to investigate the significance of the latter within the framework developed here.}
We refer the reader to \cite{Dalmazi21} for a more detailed account on the nuances between them. Below we provide higher-depth extensions of these models and perform an initial\footnote{As mentioned previously, since these actions are higher-derivative a more careful analysis should investigate the presence or absence of ghosts. } on-shell analysis.

The spin-$s$ depth-$t$ topologically massive higher-spin action we propose is
\begin{align}
\mc{S}_{\text{TM}}^{(t,s)}[\phi,\chi] = \mc{S}_{\text{CHS}}^{(t,s)}[\phi] + \a m^{2s-2t-1}\mc{S}^{(t,s)}_{\text{massless}}[\phi,\chi] ~,\label{tma1}
\end{align}
where $m$ is a positive mass-parameter and $\alpha$ is a real constant. Action $\mc{S}_{\text{CHS}}^{(t,s)}$ is defined in \eqref{TMGenCHS} whilst $\mc{S}^{(t,s)}_{\text{massless}}$ is given in \eqref{dFMDepthSJAction}.  Action $\mc{S}_{\text{TM}}^{(t,s)}$ enjoys the gauge symmetry
\begin{subequations}
\begin{align}
\delta \phi_{a(s)} &= \pa_{\{a_1}\cdots\pa_{a_{t}}\xi_{a_{t+1}\cdots a_{s}\}}~,\\
\delta \chi_{a(s-t-1)} &= \pa^b\xi_{ba(s-t-1)}~,
\end{align}
\end{subequations}
with $\xi_{a(s-t)}$ traceless. After gauging away the compensating field $\chi_{a(s-t-1)}$ the equations of motion resulting from this action are
\begin{subequations}
\begin{align}
0&=(\pa^b)^{t+1}\phi_{b(t+1)a(s-t-1)}~,\label{3EoM1}\\[6pt]
0&=\mc{C}^{(t)}_{a(s)}(\phi)+\frac{\alpha m^{2s-2t-1}}{\lambda_{s}-\lambda_{s-t}}(\mb{W}^2-\lambda_{s-t}\Box)\phi_{a(s)}~, \label{3EoM2}
\end{align}
\end{subequations}
where $\mc{C}^{(t)}_{a(s)}(\phi)$ is the generalised Cotton tensor \eqref{Cotton}.  After entering the gauge $0=(\pa^b)^t\phi_{b(t)a(s-t)}$ we see that all spin-$k$ modes with $0\leq k \leq s-t$ are non-propagating, $\Pi_{(s,k)}^{\perp}\phi_{a(s)}=0$. Analysis of \eqref{3EoM2} reveals that amongst the non-trivial solutions to \eqref{3EoM2}  are propagating massive spin-$k$ modes with $s-t+1\leq k \leq s$ with distinct masses $m_k$ (which are proportional to $m$) that depend on $k,s,t$ and $\a$  in a complicated way. The signs of their helicities depend on the sign of the constant $\alpha$.  

The spin-$s$ depth-$t$ new topologically massive higher-spin action  we propose is
\begin{align}
\mc{S}^{(t,s)}_{\text{NTM}}[\phi_{(s)}] = \int \text{d}^3x\,  \phi_{(t)}^{a(s)}\big(\mb{W}-\sigma s m\big)\mc{C}_{a(s)}^{(t)}(\phi) ~, \label{tma2}
\end{align}
where $m$ is a positive mass-parameter and $\sigma=\pm 1$. The resulting equation of motion is
\begin{align}
0 = \big(\mb{W}-\sigma s m\big)\mc{C}_{a(s)}^{(t)}(\phi)~. 
\end{align}  
For $t>1$ this does not lead to a Klein-Gordon equation on $\mc{C}_{a(s)}^{(t)}(\phi)$ since the latter is only partially conserved \eqref{parcon}.  Amongst the non-trivial solutions to this equation are the massive spin-$k$ modes with $s-t+1\leq k \leq s$ with mass $m$ and helicity $\sigma k$.
  
The spin-$s$ depth-$t$ new massive higher-spin action we propose is  
\begin{align}
\mc{S}^{(t,s)}_{\text{NM}}[\phi_{(s)}] = \int \text{d}^3x\,  \phi_{(t)}^{a(s)}\big(\mb{W}-\sigma s m\big)\mc{E}_{a(s)}^{(t)}(\phi)~, \label{tma3}
\end{align}
where we have defined the local (even though it does not appear to be) tensor 
\begin{align}
\mc{E}_{a(s)}^{(t)}(\phi):=\frac{1}{\mb{W}}\mc{C}_{a(s)}^{(t)}(\phi) =\frac{1}{\lambda_{s-t+1}}\Box^{s-t} \P^{\perp}_{(s-t+1,s-t+1)} \phi_{a(s)}^{(t)}~,
\end{align}
was first introduced in \cite{Dalmazi21} for $t=1$. This action also has propagating massive spin-$k$ modes with $s-t+1\leq k \leq s$ with mass $m$ and helicity $\sigma k$.

\subsection{Gauge-invariant actions in (A)dS$_d$}\label{secActAdSd}

The most general action in (A)dS$_d$ that is quadratic in the covariant derivative $\cD_a$ and in the rank-$s$ traceless field $\phi_{a(s)}$ is given by
\begin{subequations}
	\begin{align}
	\mc{S}_{\text{gen}}^{(s)}[\phi]=\int \text{d}^dx ~e~ \big(a_1\phi^{a(s)}\Box \phi_{a(s)}+a_2\phi^{a(s)}\cD_{a}\cD^{b}\phi_{ba(s-1)} + a_3 \Omega \f^{a(s)} \f_{a(s)}\big)~.\label{dAdSGenActa}
	\end{align}
Note there is an additional term proportional to the (A)dS$_d$ curvature $\Omega$ as compared to \eqref{dGenActa}. This action can be recast into the following form
	\begin{align}
	\mc{S}_{\text{gen}}^{(s)}[\phi]=\int \text{d}^dx ~ e ~ \phi^{a(s)}\big(b_1\mb Q + b_2 \mb{W}^2 + b_3 \Omega \big)\phi_{a(s)}~.\label{dAdSGenActb}
	\end{align}
Inserting the resolution of identity \eqref{dAdSidres} and using \eqref{dAdSid55}, we find
	\begin{align}
	\mc{S}_{\text{gen}}^{(s)}[\phi] = \int \text{d}^dx ~ e ~ \phi^{a(s)}\sum_{j=0}^{s} \big ( c_j \mb{Q} +d_j \Omega \big )\Pi^{\perp}_{(s,j)}\phi_{a(s)}~. \label{dAdSGenActc}
	\end{align}
\end{subequations}
The constants $a_i,b_i,c_i$ are all related to each other via the same equations \eqref{constants66} as in M$^d$, and $d_j$ is related to them through
\begin{align}
d_j := b_3 - \l_j \t_{(j,j)} b_2 = a_3 - s(s+d-2)a_1 - \frac{\lambda_s \t_{(s,s)} -\lambda_j \t_{(j,j)}}{s(2s+d-4)}a_2~.
\end{align}
Recall $\lambda_s = s(s+d-3)$ and $\tau_{(s,s)}=(s-1)(s+d-2)$.

We wish to construct an action with a propagating massless spin-$s$ field on the equations of motion, and for this we require $c_s\neq 0$ and consequently $a_1\neq 0$. This allows us to hereby set $c_s=a_1=1$. 
The requirement of having a propagating spin-$s$ mode does not put a restriction on the constant $d_s$, so we leave it arbitrary for now. 
From property \eqref{dAdSLongi2b} it is clear that $\Pi^{\perp}_{(s,s)}\phi_{a(s)}$ is invariant under the gauge transformations 
\begin{align}
\delta \phi_{a(s)}=\cD_{\{a}\xi_{a(s-1)\}}= \cD_a\xi_{a(s-1)}-\frac{s-1}{2s+d-4}\eta_{a(2)}\cD^b\xi_{ba(s-2)}~,  \label{dAdSmasslessgt}
\end{align}
for traceless $\xi_{a(s-1)}$.
The presence of the other projectors breaks this gauge symmetry. 
In particular, in order for the gauge variation of \eqref{dAdSGenActc} under \eqref{dAdSmasslessgt} to vanish, we require
\begin{align}
0=( c_j\mb{Q} +d_j \Omega \big )\Pi^{\perp}_{(s,j)}\xi_{a(s-1)} \qquad 0\leq j\leq s-1~.\label{GIcond}
\end{align}
For future reference we point out that $\Pi^{\perp}_{(s,j)}\xi_{a(s-1)}$ contains the following number of divergences
\begin{align}
\Pi^{\perp}_{(s,j)}\xi_{a(s-1)}~\sim~ (\cD^{b})^{s-j-1}\xi_{a(j)b(s-j-1)}~,\qquad 0\leq j\leq s-1~, \label{GPstruct}
\end{align}
where we have used  \eqref{ch} and \eqref{dAdSLongi2a}. 
As we did in M$^d$, we now proceed to search for actions which can be made gauge invariant by setting various coefficients $c_j$ and $d_j$ to zero. 

Before doing this, we note that in contrast to M$^d$, after setting $c_s=a_1=1$ in the (A)dS$_d$ action \eqref{dAdSGenActa}, there are two free parameters as opposed to one (due to the third term in \eqref{dAdSGenActa} not being present in \eqref{dGenActa}). 
In the next two sections, we will use this freedom to set a single pair $(c_j,d_j)$ to zero at a time, meaning that the corresponding projector $\Pi_{(s,j)}^{\perp}$ will be completely absent. This will provide generalisations of the models derived in sections \ref{secFronsMd} and \ref{secNewDepthMd} to (A)dS$_d$. 
However, rather than setting a pair $(c_j,d_j)$ to zero, we could equally well set one $c_j=0$ and another $d_k=0$, or set two different $c_j$ to zero etc.\footnote{It turns out that most of these are special cases of the models derived in sections \ref{secFronsAdSd} and \ref{secNewDepthAdSd}. By special case we mean the scenario when a combination of the relevant gauge parameter or auxiliary field is specified to be the symmetric trace-free gradient of a lower-rank field. } 
This is tantamount to having all projectors present in \eqref{dAdSGenActc}. 
We do not exhaust all of the different possibilities in this work. However, in section \ref{secPM} we do consider the case $c_{s-1}=0$ and fix the remaining parameter in such a way that $d_{s-1}\neq 0$, and this leads to the partially-massless actions.

\subsubsection{Projector $\Pi_{(s,s-1)}^{\perp}$ absent; massless Fronsdal action} \label{secFronsAdSd}

We first consider the case when $\Pi^{\perp}_{(s,s-1)}$  is absent in \eqref{dAdSGenActc}, i.e. $c_{s-1}=d_{s-1} = 0$ which fixes the constants $b_1,b_2,b_3$ and the action \eqref{dAdSGenActb} is 
\begin{align}
\mc{S}_{\text{gen}}^{(s)}[\phi]\Big|_{c_{s-1}=0}=\frac{1}{\lambda_s-\lambda_{s-1}}\int \text{d}^dx ~ e~  &\phi^{a(s)}\big(\mb{W}^2-\L_{s-1} \big)\phi_{a(s)}~. \label{dAdSFronsActkin}
\end{align}
With $c_{s-1}=d_{s-1} = 0$ it is clear that for gauge invariance  we need \eqref{GIcond} to hold for $0\leq j \leq s-2$. By inspection of \eqref{GPstruct}, we see that this can be achieved if the gauge parameter were to be transverse, $0=\mc{D}^b\xi_{ba(s-2)}$.\footnote{Alternatively we could try requiring, say, $0=\mc{D}^b\mc{D}^b\xi_{b(2)a(s-3)}$ to ensure \eqref{GIcond} for $0\leq j \leq s-3$ are satisfied and $0=(\mb{Q}+\frac{d_{s-2}}{c_{s-2}}\O)\xi_{a(s-1)}=(\mb{Q}-\rho^2_{(1,s-1)})\xi_{a(s-1)}$ to satisfy \eqref{GIcond} for $j=s-2$. In this case one would have to introduce two compensators to achieve these constraints, and on dimensional grounds they cannot be used to achieve gauge invariance. Another way to see this wouldn't work is by observing that the factor  $(\mb{Q}-\rho_{(1,s-1)}^2)$ resides within the poles (see \eqref{Poles}) of $\Pi_{(s,s-2)}^{\perp}$ in \eqref{GIcond} with $j=s-2$.}
This constraint may be arrived at by introducing a compensator $\chi_{a(s-2)}$ which transforms according to
\begin{align}
\delta \chi_{a(s-2)} = \cD^{b}\xi_{ba(s-2)}~. \label{dAdScompgt1}
\end{align}
Imposing the gauge condition $\chi_{a(s-2)}=0$ yields the desired constraint.

To obtain an action that is invariant under \eqref{dAdSmasslessgt} and \eqref{dAdScompgt1}, we must supplement \eqref{dAdSFronsActkin} with a $\phi\chi$ and $\chi\chi$ sector. The result of this procedure yields
\begin{align} \label{dAdSFMFronsdalAction}
\mc{S}_{\text{Fronsdal}}^{(s)}[\phi,\chi]=\frac{1}{\lambda_s-\lambda_{s-1}}\int \text{d}^dx ~ e~  &\Big\{\phi^{a(s)}\big(\mb{W}^2-\L_{s-1} \big)\phi_{a(s)}
+2(s-1)(2s+d-6)\phi^{a(s)}\cD_a\cD_a\chi_{a(s-2)}  \non\\
&+\frac{(s-1)(2s+d-6)}{s(2s+d-4)}\chi^{a(s-2)}\big(\mb{W}^2-\L_s\big)\chi_{a(s-2)}\Big\}~.
\end{align}
This action can be shown to coincide with Fronsdal's formulation for a massless spin-$s$ field in (A)dS$_d$, which was first computed by Fronsdal \cite{Fronsdal5} for the case $d =4$, and later extended to arbitrary $d$ in\cite{Buchbinder2001bs}. To show this one can combine $\phi_{a(s)}$ and $\chi_{a(s-2)}$ into a doubly traceless field, in the same way as was done in section \ref{secFronsMd}.

\subsubsection{Projector $\Pi_{(s,s-t)}^{\perp}$ absent; depth-$t$} \label{secNewDepthAdSd}

Let us fix two integers $s$ and $t$ satisfying $1\leq t \leq s$. We now consider the general case when the projector $\Pi_{(s,s-t)}^{\perp}$  is absent in \eqref{dAdSGenActc}, i.e. when $c_{s-t}=d_{s-t}=0$  and hence
\begin{align}
\mc{S}_{\text{gen}}^{(s)}[\phi]\Big|_{c_{s-t}=0}=\frac{1}{\lambda_s-\lambda_{s-t}}\int \text{d}^dx \, e \, \phi^{a(s)}\big(\mb{W}^2-\L_{s-t} \big)\phi_{a(s)}~.\label{dAdSFronsActkin3}
\end{align}
It is convenient to split the conditions for gauge invariance \eqref{GIcond} into the following sets
\begin{subequations}
\begin{align}
0&=( c_k\mb{Q} +d_k \Omega \big )\Pi^{\perp}_{(s,k)}\xi_{a(s-1)}~,\qquad s-t+1\leq k\leq s-1~,\label{repairjob1}\\
0&=( c_k\mb{Q} +d_k \Omega \big )\Pi^{\perp}_{(s,k)}\xi_{a(s-1)}~,\qquad ~~~~~0\leq k\leq s-t-1~.\label{repairjob2}
\end{align}
\end{subequations}
To satisfy both sets simultaneously we require \eqref{repairjob1} to hold identically and introduce a compensator to reach \eqref{repairjob2}. 
From the identity \eqref{dAdSLongi2b} we see that  \eqref{repairjob1} is identically satisfied if $\xi_{a(s-1)}=\cD_{\{a_1}\cdots \cD_{a_{t-1}}\xi_{a_{t}\cdots a_{s-1}\}}$ with $\xi_{a(s-t)}$ unconstrained. 
With this the gauge transformations \eqref{dAdSmasslessgt} become depth-$t$ ones
\begin{align}
\delta \phi_{a(s)}=\cD_{\{a_1}\cdots\cD_{a_{t}}\xi_{a_{t+1}\cdots a_{s}\}}~.  \label{dAdSdepthxgt}
\end{align}

To arrive at conditions \eqref{repairjob2} we introduce a compensator $\chi_{a(s-t-1)}$ transforming as 
\begin{align}
\delta \chi_{a(s-t-1)}= \cD^{b}\xi_{ba(s-t-1)}~.\label{dAdScompgtx}
\end{align}
The gauge condition $\chi_{a(s-t-1)}=0$, which implies $0=\mc{D}^b\xi_{ba(s-t-1)}$, yields the constraints \eqref{repairjob2}.  To confirm this we observe that upon substituting 
$\xi_{a(s-1)}=\cD_{\{a_1}\cdots \cD_{a_{t-1}}\xi_{a_{t}\cdots a_{s-1}\}}$ into \eqref{GPstruct} we find that $\Pi^{\perp}_{(s,k)}\xi_{a(s-1)}\sim (\mc{D}^b)^{s-t-k}\xi_{b(s-t-k)a(k)}$ for $0\leq k \leq s-t-1$.   

It remains to determine the coupling between $\phi_{a(s)}$ and $\chi_{a(s-t-1)}$, and the kinetic sector for $\chi_{a(s-t-1)}$, which must accompany \eqref{dAdSFronsActkin3} for invariance under \eqref{dAdSdepthxgt} and \eqref{dAdScompgtx}. After performing this routine procedure, one arrives at the final action 
\begin{align} \label{dAdSFMDepthSJAction}
\mc{S}^{(t,s)}_{\text{massless}}[\phi,\chi] &=\frac{1}{\lambda_s-\lambda_{s-t}}\int \text{d}^dx ~ e \, \Big\{ \phi^{a(s)}\big(\mb{W}^2-\L_{s-t} \big)\phi_{a(s)}\non\\
&+(s-t)(2s-2t+d-4)\Big[ 2\phi^{a(s)}(\cD_{a})^{t+1}\chi_{a(s-t-1)}  \non \\
&+\chi^{a(s-t-1)}\prod_{k=1}^{t}\frac{1}{(s-k+1)(2s-2k+d-2)}\big (\mb{W}^2-\L_{s-k+1}  \big )\chi_{a(s-t-1)}\Big]\Big\}~.  
\end{align}
The identity \eqref{magicId2} is indispensable for this calculation. Note that the $\chi\chi$ sector of the above action is higher-derivative, having order $2t$. In the $t=1$ case, the action  \eqref{dAdSFMDepthSJAction} coincides with the Fronsdal action \eqref{dAdSFMFronsdalAction}. In the $t=s$ case, $\chi_{a(s-t-1)}$ is not present or necessary, and in $d=4$ \eqref{dAdSFMDepthSJAction} coincides with the action for a depth-$s$ conformal spin-$s$ field in (A)dS$_4$.

It is instructive to analyse the on-shell content of the action \eqref{dAdSFMDepthSJAction}. First, we note that after gauging away the compensator $\chi_{a(s-t-1)}$, 
the equations of motion are
\begin{subequations}
\begin{align}
0&=(\mc{D}^b)^{t+1}\phi_{b(t+1)a(s-t-1)}~,\label{AEoM1}\\
0&=(\mb{W}^2-\Lambda_{s-t})\phi_{a(s)}~, \label{AEoM2}
\end{align}
\end{subequations}
with residual gauge symmetry 
\begin{align}
\delta \phi_{a(s)}=\mc{D}_{\{a_1}\cdots\mc{D}_{a_{t}}\xi^{\perp}_{a_{t+1}\cdots a_{s}\}}~,  \qquad 0=\mc{D}^b\xi^{\perp}_{ba(s-t-1)}~.\label{Aresgs}
\end{align}
In the maximal-depth case $s=t$ there is no compensator and hence only \eqref{AEoM2} is present. 

For all values of $t$ we may impose the gauge
\begin{align}
0=(\mc{D}^b)^{t}\phi_{b(t)a(s-t-1)}~,\label{Agc1}
\end{align}
whereupon the irreducible decomposition of $\phi_{a(s)}$ is
\begin{align}
\phi_{a(s)}=\phi^{\perp}_{a(s)}+\mathcal{D}_{\{a}\phi^{\perp}_{a(s-1)\}}+\cdots + \mathcal{D}_{\{a_1}\cdots\mathcal{D}_{a_{t-1}}\phi^{\perp}_{a_{t}\cdots a_{s}\}}~.\label{Aeomdecomp}
\end{align}
The residual gauge symmetry is \eqref{Aresgs} with the extra condition
\begin{align}
 0 = \prod_{k=s-t+1}^{s}\big(\mb{Q}-\rho^2_{(k-s+t,k)}\big)\xi_{a(s-t)}^{\perp}~,
\end{align}
 which follows from varying \eqref{Agc1}, and using \eqref{magicId2} and \eqref{EnterPM}. The parameters $\rho_{(k-s+t,k)}$ are all the possible partial masses \eqref{PMvals} that  have $\xi_{a(s-t)}^{\perp}$ as the corresponding gauge parameter, such that its gauge field has rank $s-t+1\leq k\leq s$.

As was the case for Minkowski spacetime, the equation of motion \eqref{AEoM1} leads us to consider an alternative set of projectors $\hat{\mathcal{P}}^{\perp}_{(s,t,j)}$. The latter are defined to act on the space $\mathscr{V}_{(s,t)}^{\perp}$ of constrained fields $\phi_{a(s)}$ satisfying \eqref{Agc1} and take the form
\begin{align}
\hat{\mathcal{P}}^{\perp}_{(s,t,j)}=\prod_{s-t+1\leq k \leq s}^{k\neq j}\frac{ ( \mb{W}^2-\Lambda_{k}  )}{ ( \L_j-\L_k  )}~,
\end{align}
with $2\leq t\leq s$ and $s-t+1 \leq j \leq s$. 
They satisfy the (A)dS$_d$ analogues of the properties listed in section \ref{secNewDepthMd}.
It follows that $\phi_{a(s)}\in \mathscr{V}_{(s,t)}^{\perp}$ has irreducible decomposition \eqref{Aeomdecomp} and $\hat{\mathcal{P}}^{\perp}_{(s,t,j)}\phi_{a(s)}$ carries spin $j$.

Resuming the on-shell analysis, we note that $\hat{\mathcal{P}}^{\perp}_{(s,t,j)}\phi_{a(s)}\neq 0$ since the factor $(\mb{W}^2-\Lambda_{s-t})$ appearing in \eqref{AEoM2} is not present in $\hat{\mathcal{P}}^{\perp}_{(s,t,j)}$. Applying $\hat{\mathcal{P}}^{\perp}_{(s,t,j)}$ to \eqref{AEoM2} we find 
\begin{align}
0=\big(\mb{Q}-\rho^2_{(j-s+t,j)}\big) \big( \hat{\mathcal{P}}^{\perp}_{(s,t,j)}\phi_{a(s)}\big) ~ \qquad\qquad s-t+1\leq j \leq s~,
\end{align}
where we have used $0=(\mathbb{W}^2-\L_j)\hat{\mathcal{P}}^{\perp}_{(s,t,j)}\phi_{a(s)}$ and \eqref{EnterPM}.
We see that the spin-$j$ modes of $\phi_{a(s)}$ with $s-t+1\leq j \leq s$ are propagating and are partially-massless with depth-$(j-s+t)$. These modes cannot be gauged away. Thus, on-shell this model describes the reducible representation
\begin{align}
\mf{P}(t,s) ~\oplus ~\mf{P}(t-1,s-1)~\oplus~ \cdots ~\oplus~ \mf{P}(1,s-t+1)\oplus (\cdots) \label{redPM}
\end{align}
of the (A)dS$_d$ algebra (see section \ref{secAdSrep} for the notation). Though this model describes partially-massless modes, it does not coincide with the partially-massless model originally put forward by Zinoviev \cite{Zinoviev}. The latter describes the single irreducible representation $\mf{P}(t,s)$, and does not include the tail in \eqref{redPM}. We will investigate this model next.

\subsubsection{Spin-$s$ depth-$t$ partially massless action} \label{secPM}

We wish to employ the prescription developed in the previous sections to reproduce the spin-$s$ depth-$t$ partially massless action of Zinoviev \cite{Zinoviev} (or an equivalent version of it). 
To arrive at the latter, the author of \cite{Zinoviev} begins by constructing a model for the mass-$m$ spin-$s$ field within a St\"{u}ckelberg-like framework.
Initially, the spectrum of fields used for this model is the set $\{\Phi_{a(s)},\Phi_{a(s-1)},\dots, \Phi\}$ of symmetric and doubly traceless fields, however there exist $s$ critical values of $m$,\footnote{These critical values are related to the partial masses \eqref{PMvals}. } labelled by $t$ with $1 \leq t \leq s$,
whereupon the bottom $(s-t+1)$ fields decouple. The resulting action (after ignoring the decoupled fields) describes a spin-$s$ depth-$t$ PM field and is formulated using the set  of fields $\{\Phi_{a(s)},\Phi_{a(s-1)},\dots, \Phi_{a(s-t+1)}\}$.   The action consists of three distinct sectors:
(i) the diagonal sum of second-order massless Fronsdal actions \eqref{dFMFronsdalActionFS} for each $\Phi_{a(k)}$, which have been minimally uplifted
to (A)dS$_d$; (ii)  a sum of non-diagonal first-order terms which couple the $\Phi_{a(k)}$; and 
(iii) a sum of (non-)diagonal zeroth-order curvature terms.

 Below we work with symmetric and traceless fields, and we accordingly point out that the above set of doubly traceless fields decomposes into the following set of traceless fields
\be
\{\Phi_{(s)},\Phi_{(s-1)},\dots, \Phi_{(s-t+1)}\} \rightarrow \{\phi_{(s)},\tilde{\phi}_{(s-1)},\dots, \tilde{\phi}_{(s-t+1)}\}\cup \{\chi_{(s-2)},\chi_{(s-3)},\dots, \chi_{(s-t-1)}\}~.
\ee
In Zinoviev's formulation all of the fields $\tilde{\phi}_{(k)}$ with $s-t+1\leq k \leq s-1$, which correspond to the symmetric traceless part of $\Phi_{a(k)}$, are St\"{u}ckelberg fields. This means they may be gauged away off-shell (without constraining any of the remaining fields or gauge parameters), after which we are left with the set $\{\phi_{(s)},\chi_{(s-2)},\chi_{(s-3)},\dots, \chi_{(s-t-1)}\}$. Upon fixing this gauge, the residual gauge symmetry goes from the first-order gauge transformations in \cite{Zinoviev}, to the higher-order ones that are characteristic of a PM field (see \eqref{gtPM}).  The former spectrum and gauge transformations are the ones that naturally appear in the model we now derive from first principles.

In the spirit of \cite{Zinoviev}, we are lead to investigate the case $c_{s-1} = 0$ and $d_{s-1} \neq 0$ of action \eqref{dAdSGenActc}, since such conditions generated Fronsdal's massless spin-$s$ action in M$^d$ (see section \ref{secFronsMd}).\footnote{If we allowed $d_{s-1}=0$ in addition to $c_{s-1}=0$ then we would simply arrive at the actions in section \ref{secNewDepthAdSd}.} The corresponding action is thus given by
\begin{align} \label{dAdSGenPMAct}
\cS^{(s)}_{\text{gen}}[\f] \Big|_{c_{s-1} = 0} = \frac{1}{\l_{s} - \l_{s-1}} \int \rd^d x \, e \,  \f^{a(s)}\big(\mb{W}^2-\lambda_{s-1}\mb{Q} + (\l_s - \l_{s-1})b_3 \O \big)\f_{a(s)}~.
\end{align}
We note that the constants $b_1$ and $b_2$ are fixed in accordance with \eqref{constants66}, however, in contrast to the Fronsdal case \ref{secFronsAdSd}, the constant $b_3$ is now arbitrary (since we are not yet fixing any $d_{j}$). 

It is convenient to split the conditions for gauge invariance \eqref{GIcond} into the following sets
\bsubeq \label{dAdSGCPM}
\begin{align}
0&=( c_j\mb{Q} +d_j \Omega \big )\Pi^{\perp}_{(s,j)}\xi_{a(s-1)}~,\qquad &s-t+1 \leq j\leq s-1~, \label{dAdSGCPM3}\\
0&=( c_{s-t} \mb{Q} +d_{s-t} \Omega \big )\Pi^{\perp}_{(s,s-t)}\xi_{a(s-1)}~, &  \label{dAdSGCPM2}\\
0&=( c_j\mb{Q} +d_j \Omega \big )\Pi^{\perp}_{(s,j)}\xi_{a(s-1)}~,\qquad &0\leq j\leq s-t -1~.  \label{dAdSGCPM1}
\end{align}
\esubeq
We see that  \eqref{dAdSGCPM3} is satisfied identically if the gauge parameter $\x_{a(s-1)}$ takes the form $\x_{a(s-1)} = \cD_{\lbrace a_1 } \cdots \cD_{a_{t-1}} \x_{a_{t} \cdots a_{s-1} \rbrace }$, where $2 \leq t \leq s$ and $\x_{a(s-t)}$ is unconstrained.  With this, the gauge transformation \eqref{dAdSmasslessgt} becomes a depth-$t$ transformation
\begin{align}
\d \f_{a(s)} = \cD_{\lbrace a_1 } \cdots \cD_{a_{t}} \x_{a_{t+1} \cdots a_{s} \rbrace }~.
\end{align}

Next, we observe that the condition \eqref{dAdSGCPM1} can be achieved  if the gauge parameter is transverse, $\cD^b \x_{b a(s-t-1)} = 0$. This condition is arrived upon if one introduces the compensator $\chi_{a(s-t-1)}$ which transforms as
\begin{align} \label{dAdSGT2}
\d \chi_{a(s-t-1)} = ( \sqrt{\O} )^{t-1} \cD^b \x_{b a(s-t-1)} ~.
\end{align}
The factor $( \sqrt{\O} )^{t-1} $ was introduced in order for the kinetic sector of $\chi_{a(s-t-1)}$ to be second-order.

It remains to satisfy the final condition \eqref{dAdSGCPM2}. This can be accomplished by requiring that the gauge parameter $\x_{a(s-1)}$  satisfies the second-order differential equation
\be \label{dAdSPMCond2}
0 = \big ( \mb{W}^2 +f_{s-t} \mb{Q} + g_{s-t} \O \big ) \x_{a(s-t)}~,
\ee
where the constants $f_{s-t}$ and $g_{s-t}$ are chosen such that \eqref{dAdSPMCond2} coincides with the condition 
\begin{align}
0 =( c_{s-t} \mb{Q} +d_{s-t} \Omega \big ) \x^{\perp}_{a(s-t)}~,\label{redres1}
\end{align}
when  $\xi_{a(s-t)}$ is transverse (i.e. when one gauges away $\chi_{a(s-t-1)}$ on-shell), which in turn ensures \eqref{dAdSGCPM2} is satisfied. 
Fixing the coefficients in this way, one finds that
\begin{subequations}
\begin{align} 
f_{s-t} &=  A_{s-t} \big ( \l_{s-t} -\l_{s-1} \big ) - \l_{s-t}~,\\
g_{s-t} &= A_{s-t} \Big ( \big ( \l_s - \l_{s-1} \big ) b_3 - \l_{s-t} \t_{(s-t,s-t)} \Big ) + \l_{s-t} \t_{(s-t,s-t)}~,
\end{align}
\end{subequations}
where $A_{s-t}$ is an arbitrary constant. Thus off-shell, we can arrive at the condition  \eqref{dAdSPMCond2} by introducing the additional compensator  $\chi_{a(s-t)}$ which transforms in the following fashion
\begin{align} \label{dAdSGT3}
\d \chi_{a(s-t)} = ( \sqrt{\O} )^{t-2}  \Big ( \mb{W}^2 +f_{s-t} \mb{Q} + g_{s-t} \O \Big ) \x_{a(s-t)}~.
\end{align}
Again, we supplement the gauge transformation with the dimensionful factor   $( \sqrt{\O} )^{t-2}$ to ensure that the kinetic sector of $\chi_{a(s-t)} $ is second-order. 

We anticipate that the action we are constructing will describe a spin-$s$ depth-$t$ partially massless field on-shell. In accordance with this, it is required that on-shell, the gauge parameter satisfies the condition $0 = (\mb{Q} - \t_{(t,s)} \O) \x^{\perp}_{a(s-t)} $ (cf. \eqref{dAdMassShellPM}). Thus, upon fixing the gauge $\chi_{a(s-t-1)} = \chi_{a(s-t)} =0$, we see that $b_3$ must be chosen to be
\begin{align} \label{AdSb3}
b_3 = \frac{1}{\l_s - \l_{s-1}} \Big [ \big ( \l_{s-1} - \l_{s-t} \big ) \t_{(t,s)} + \l_{s-t} \t_{(s-t,s-t)} \Big ]~,
\end{align}
so that the residual gauge symmetry \eqref{redres1} reduces to $0 = (\mb{Q} - \t_{(t,s)} \O) \x^{\perp}_{a(s-t)} = 0$.

At this stage of our analysis, we are unable to construct a gauge-invariant action from the fields $\{\f_{a(s)}, \chi_{a(s-t-1)}, \chi_{a(s-t)}\}$ that is second-order. This is because $\f_{a(s)}$ cannot couple to $\chi_{a(s-t-1)}$ or $\chi_{a(s-t)}$ on dimensional grounds. To address this problem, we need to introduce a tower of compensating fields $\chi_{a(k)}$, with $s-t+1 \leq k \leq s-2$, which transform via
\begin{align}
\d \chi_{a(k)}  = ( \sqrt{\O} )^{s-k-2} \Big ( \mb{W}^2 +f_k \mb{Q} + g_k \O \Big ) \cD_{\lbrace a_1} \cdots \cD_{ a_{k-s+t}}  \x_{a_{k-s+t+1} \cdots a_k \rbrace}~.
\end{align}
For consistency, the constants $f_k$ and $g_k$ must be such that when $\xi_{a(s-t)}$ is transverse we have $( \mb{W}^2 +f_k \mb{Q} + g_k \O  )\xi_{a(s-t)}^{\perp} \propto (\mb{Q}- \t_{(t,s)} \O) \x^{\perp}_{a(s-t)} $. Consequently, we arrive at the expressions
\begin{subequations}
\begin{align} \label{dAdSPMC}
f_{k} &=  A_{k} \big ( \l_{s-t} -\l_{s-1} \big ) - \l_{s-t}~,\\
g_{k} &=  A_{k} \Big ( \big ( \l_s - \l_{s-1} \big ) b_3 - \l_{s-t} \t_{(s-t,s-t)} \Big ) + \l_{s-t} \t_{(s-t,s-t)}~,
\end{align}
\end{subequations}
where the constant $A_k$ embodies the (undetermined) coefficient of proportionality. Therefore, in summary we have the following fields and gauge transformations
\begin{subequations}\label{gtPM}
\begin{align}
\d \f_{a(s)} &= \cD_{\lbrace a_1 } \cdots \cD_{a_{t}} \x_{a_{t+1} \cdots a_{s} \rbrace }~,\label{gtPM1}\\
\d \chi_{a(k)}  &= ( \sqrt{\O} )^{s-k-2} \Big ( \mb{W}^2 +f_k \mb{Q} + g_k \O \Big ) \cD_{\lbrace a_1} \cdots \cD_{ a_{k-s+t}}  \x_{a_{k-s+t+1} \cdots a_k \rbrace}~,\label{gtPM2}\\
\d \chi_{a(s-t-1)} &= ( \sqrt{\O} )^{t-1} \cD^b \x_{b a(s-t-1)} ~,\label{gtPM3}
\end{align}
\end{subequations}
with $s-t \leq k\leq s-2$, $f_k$ and $g_k$ given by \eqref{dAdSPMC}, and $b_3$ given by \eqref{AdSb3}.

We now have all the ingredients necessary to construct the following  most general action 
\begin{align} \label{dAdSPMAction}
\cS^{(s,t)}_{\text{PM}} [\f,\chi]= \frac{1}{\l_s - \l_{s-1}} &\int \rd^d x ~e~ \Big \lbrace \f^{a(s)} \Big ( \mb{W}^2 - \l_{s-1} \mb{Q} + \big ( [\l_{s-1} - \l_{s-t} ] \t_{(t,s)} + \l_{s-t} \t_{(s-t,s-t)} \big ) \O \Big )\f_{a(s)} \non ~ \\
& + \D \f^{a(s)} \cD_a \cD_a \chi_{a(s-2)} + \sum_{l = s -t -1}^{s-2} \U_l \,\chi^{a(l)} \big ( \mb{W}^2  +  \a_l \mb{Q} + \b_l \O \big ) \chi_{a(l)}~ \non \\
& +\sum_{l = s-t}^{s-2} \Q_l \, \sqrt{\O } \, \chi^{a(l)}\cD_a \chi_{a(l-1)} + \sum_{l = s-t+1}^{s-2} \G_l \,\chi^{a(l)}\cD_a \cD_a \chi_{a(l-2)} \Big \rbrace ~,
\end{align}
where the coefficients are fixed via gauge invariance to be
\begin{align}
\D &= -2~, \qquad \a_l = - \l_{l+2}~, \qquad \G_l = 0~, \qquad A_k = \frac{\l_{s-t} - \l_{k+1}}{\l_{s-t} - \l_{s-1}}~, \\
\b_l &= \frac{1}{\l_{l+1} - \l_l} \Big [  (\l_{l+2} - \l_{l+1}  )  ( \l_{s-t} - \l_l  ) \t_{(t,s)} +  (\l_{l+1} - \l_{l+2}  ) \l_{s-t} \t_{(s-t,s-t)}  \non\\
& \hspace{2.35cm}+  (\l_{l+2} - \l_{l}  ) \l_{l+1} \t_{(l+1,l+1)}  \Big ]~, \non \\
\Q_m &= 2(-1)^{s-m}\frac{ m!}{s! (2m+d-2)}  \prod_{n = m+1}^{s-1} \big (2n+d-4 \big)^{-1} \times \non \\
& \hspace{2.7cm}\times\Big ( \big (\l_{n} - \l_{s-t} \big )\t_{(t,s)} +\l_{s-t} \t_{(s-t,s-t)} - \l_n \t_{(n,n)} \Big )~, \non \\
\Q_{s-t} &=  2(-1)^{t+1}\frac{   (s-t)!}{ s!} (t-1) (s-t) (2s+d-t-2) (2s -2t+d-4)  \times \non  \\
& \hspace{1.5cm}\times\prod_{k=1}^{t-1} \Big (2s+d-2k-4 \Big)^{-1} \prod_{n = s-t+2}^{s-1} \Big ( \big (\l_{n} - \l_{s-t} \big )\t_{(t,s)} +\l_{s-t} \t_{(s-t,s-t)} - \l_n \t_{(n,n)} \Big )~, \non \\
\U_p &= \frac{ (-1)^{s-p}  p!}{(2p+d)(2p+d-2)s!}  \prod_{n = p+2}^{s-1} \big (2n+d-4 \big)^{-1} \times \non \\
&\hspace{2.7cm}\times\Big ( \big (\l_{n} - \l_{s-t} \big )\t_{(t,s)} +\l_{s-t} \t_{(s-t,s-t)} - \l_n \t_{(n,n)} \Big )~, \non  \\
\U_{s-t-1} &= (-1)^{t+1} \frac{(s-t)!}{s!} (s-t) (t-1) (2s-2t+d-4)  (2 s +d - t-2) \times \non  \\
& \hspace{1.5cm}\times\prod_{k=1}^{t-1} \Big (2s+d-2k-4 \Big)^{-1} \prod_{n = s-t+2}^{s-1} \Big ( \big (\l_{n} - \l_{s-t} \big )\t_{(t,s)} +\l_{s-t} \t_{(s-t,s-t)} - \l_n \t_{(n,n)} \Big )~, \non 
\end{align}
where $ s-t+1 \leq m \leq s-2$ and $s-t \leq p \leq s-2$.

Next, we need to perform an on-shell analysis to verify that the model does indeed describe a spin-$s$ depth-$t$ partially massless field. The equations of motion corresponding to \eqref{dAdSPMAction} are 
\bsubeq \label{dAdSEoM}
\begin{align}
0 &= \Big ( \mb{W}^2 - \l_{s-1} \mb{Q} +\big [ ( \l_{s-1} - \l_{s-t} ) \t_{(t,s)} + \l_{s-t} \t_{(s-t,s-t)} \big ] \O \Big ) \f_{a(s)} - \cD_a \cD_a \chi_{a(s-2)}~, \\
0 &= - 2\cD^b \cD^c \f_{bca(s-2)} + 2\U_{s-2} \big ( \mb{W}^2 +\a_{s-2} \mb{Q} + \b_{s-2} \O \big ) \chi_{a(s-2)} + \Q_{s-2}  \sqrt{\O} \, \cD_a \chi_{a(s-3)}~, \\
0 &= 2 \U_k \big ( \mb{W}^2 +\a_{k} \mb{Q} + \b_{k} \O \big ) \chi_{a(k)} - \Q_{k+1} \sqrt{\O} \, \cD^b \chi_{ba(k)} + \Q_k \sqrt{\O} \, \cD_a \chi_{a(k-1)}~, \\
0 &= 2 \U_{s-t-1} \big ( \mb{W}^2 +\a_{s-t-1} \mb{Q} + \b_{s-t-1} \O \big ) \chi_{a(s-t-1)} - \Q_{s-t} \sqrt{\O} \, \cD^b \chi_{ba(s-t-1)}~,
\end{align}
\esubeq
where $s-t \leq k \leq s-3$. 

Making use of the gauge symmetry allows us to gauge away the fields $\chi_{a(s-t-1)}$ and $\chi_{a(s-t)}$
\begin{align} \label{dAdSGaugeCon}
\chi_{a(s-t-1)} = 0~, \qquad \chi_{a(s-t)} = 0~,
\end{align}
for $2 \leq t \leq s-1$. Note that $\chi_{a(s-t)} $ is the only field that can be gauged away in  the  case $t=s$.
As a consequence of \eqref{dAdSGaugeCon}, the residual gauge symmetry  is \eqref{gtPM1} with $\xi_{a(s-t)}$ satisfying 
\begin{align} \label{dAdSRGS}
0 = \cD^b \x_{b a(s-t-1)} ~,  \qquad 0 = \big ( \mb{Q}  -  \t_{(t,s)} \O \big )   \x_{a(s-t)}~.
\end{align}
It follows from \eqref{dAdSRGS} that the transformations for $\chi_{a(k)} $ with $s-t+1 \leq k \leq s-2$ trivialise
\begin{align}
\d \chi_{a(k)}  = 0~.
\end{align}
This is an indication that these fields must vanish on-shell.

The equations of motion \eqref{dAdSEoM} reduce to the following form in the gauge \eqref{dAdSGaugeCon}
\bsubeq
\begin{align}
0 &= \Big ( \mb{W}^2 - \l_{s-1} \mb{Q} +\big [ ( \l_{s-1} - \l_{s-t} ) \t_{(t,s)} + \l_{s-t} \t_{(s-t,s-t)} \big ] \O  \Big ) \f_{a(s)} - \cD_a \cD_a \chi_{a(s-2)}~, \label{PMActEoM1}\\
0 &= - 2\cD^b \cD^c \phi_{bca(s-2)} + 2\U_{s-2} \big ( \mb{W}^2 +\a_{s-2} \mb{Q} + \b_{s-2} \O  \big ) \chi_{a(s-2)} + \Q_{s-2}  \sqrt{\O} \, \cD_a \chi_{a(s-3)}~, \label{PMActEoM2}\\
0 &= 2 \U_k \big ( \mb{W}^2 +\a_{k} \mb{Q} + \b_{k} \O \big ) \chi_{a(k)} - \Q_{k+1} \sqrt{\O} \, \cD^b \chi_{ba(k)} + \Q_k \sqrt{\O} \, \cD_a \chi_{a(k-1)}~, \\
0 &= 2 \U_{s-t+1} \big ( \mb{W}^2 +\a_{s-t+1} \mb{Q} + \b_{s-t+1} \O \big ) \chi_{a(s-t+ 1)} - \Q_{s-t+2} \sqrt{\O } \, \cD^b \chi_{ba(s-t+1)}~,  \\
0 &=  \cD^b \chi_{b a(s-t)}~, 
\end{align}
\esubeq
where $s-t+2 \leq k \leq s-3$. Suppose one is able to show that all the compensators  $\chi_{a(k)} $ vanish on-shell, then it follows from \eqref{PMActEoM1} and \eqref{PMActEoM2} that the dynamical field $\f_{a(s)}$ describes a spin-$s$ depth-$t$ partially massless field, which we recall is defined to satisfy \eqref{dAdSPMOnShell}. This is to be expected since up to an overall factor, the action \eqref{dAdSPMAction}\footnote{Action describing a spin-$s$ depth-$t$ field \cite{Zinoviev} was recently rederived in \cite{Bekaert2023uve} using ambient-space techniques.  } possesses the same spectrum of fields, with the same gauge transformations,  as Zinoviev's  spin-$s$ depth-$t$ model \cite{Zinoviev} once the St\"{u}ckelberg fields are gauged away and the doubly traceless fields are expressed in terms of traceless fields. Showing that all  compensators $\chi_{a(k)} $ indeed vanish on-shell is a tedious task.

Our construction of the spin-$s$ depth-$t$ partially massless action was guided by characteristics of Zinoviev's action \cite{Zinoviev}. In particular, this motivated us to choose the starting conditions $c_{s-1} = 0$ and $d_{s-1} \neq 0$.  It must be noted that the results of this section can be arrived at naturally, and independently, via the prescription developed in this paper. A systematic analysis would involve investigating all allowed combinations of $c_j$ and $d_j$ for $0 \leq j \leq s-1$. Furthermore, our choice of $b_3$ was motivated by the desire to reproduce the PM action. However, this was just a short-cut, and one can fix $b_3$ to be \eqref{AdSb3} via gauge invariance alone.

\subsubsection{Actions without compensators}

In this section we provide comments on some higher-derivative models which do not require compensators for gauge invariance. Conceptually this is very similar to what was done in M$^d$ in section \ref{secNoCompMd}. Given two fixed integers $s$ and $t$ with $1\leq t \leq s$, the (A)dS$_d$ version of the order-$2s$ action \eqref{dFMActionNoComp1} is 
\begin{align} \label{dFMActionNoComp1AdS}
\mc{S}[\phi] = \int \text{d}^dx \, e \, \phi^{a(s)}  \prod_{m=1}^{s-t+1} \big ( \mb{Q} - \rho^2_{(m,s-t+1)}\big ) \prod_{n=1}^{t-1} \big ( \mb{Q} - \rho^2_{(n,n+s-t+1)} \big )\Pi^{\perp}_{(s,s-t+1)}\phi_{a(s)}~,
\end{align}
and of the order-$2(s-t+1)$ action \eqref{randact} is 
\begin{align}
\mc{S}[\phi] = \int \text{d}^dx \, e \,  \phi^{a(s)}  \prod_{m=1}^{s-t+1} \big ( \mb{Q} - \rho^2_{(m,s-t+1)}\big ) \Pi^{\perp}_{(s-t+1,s-t+1)}\phi_{a(s)}~.\label{randactAdS}
\end{align}
Both of these actions are invariant under the depth-$t$ gauge transformations 
\begin{align}
\delta \phi_{a(s)}= \mc{D}_{\{a_1}\cdots\mc{D}_{a_{t}}\xi_{a_{t+1}\cdots a_{s}\}}~, \qquad 0=\xi_{a(s-t-2)b}{}^{b}~. \label{AdSgt}
\end{align}
Gauge invariance of action \eqref{dFMActionNoComp1AdS} is manifest due to the projector properties \eqref{dAdSLongi2b} and \eqref{dAdSid54}, whilst gauge invariance of \eqref{randactAdS} is manifest due to \eqref{dAdSLongi2b} and \eqref{dAdSdid53}.
The actions \eqref{randactAdS} and \eqref{dFMActionNoComp1AdS} coincide for $t=1$.  In the case $t=s$, the action \eqref{randactAdS} coincides (up to an overall factor) with the depth-$s$ action \eqref{dAdSFMDepthSJAction}.

\subsubsection{Generalised conformal higher-spin actions and their factorisation}

In this section we will only investigate the case of even $d$ and $d=3$. We will focus our discussion on the factorisation of the generalised CHS actions. 

\textbf{Conformal actions in even dimensions $d \geq 4$}

In (A)dS$_d$ with $d$ even, the appropriate extension of \eqref{evendgenCHS} is
\begin{align}
\mc{S}^{(t,s)}_{\text{CHS}}[\phi] = \int \text{d}^dx \, e \, \phi_{(t)}^{a(s)}& \prod_{k=1}^{\frac{1}{2}(d-4)}(\alpha_k\mb{W}^2+\beta_k\mb{Q}+\gamma_k\Omega) \non\\
\times  &\prod_{l=1}^{s-t+1} \big ( \mb{Q} - \rho^2_{(l,s-t+1)}\big )\Pi^{\perp}_{(s-t+1,s-t+1)}\phi^{(t)}_{a(s)}~,\label{evendgenCHSAdS}
\end{align}
for some constants $\alpha_k,\beta_k$ and $\gamma_k$ determined by requiring invariance under rigid special conformal transformations of the background ($\a_k$ and $\b_k$ will coincide with those in \eqref{evendgenCHS}).

In the case $t=1$ the field $\Pi^{\perp}_{(s-t+1,s-t+1)}\phi^{(t)}_{a(s)}$ is irreducible and hence $\mb{W}^2\sim \Lambda_s$ on it, simplifying \eqref{evendgenCHSAdS} somewhat.
Furthermore, off-shell it is always possible to enter the TT gauge for $t=1$, $\phi_{a(s)}\equiv \phi_{a(s)}^{\perp}$, whereupon \eqref{evendgenCHSAdS} becomes 
\begin{align}
\mc{S}^{(1,s)}_{\text{CHS}}[\phi] = \int \text{d}^dx \,e\, \phi_{\perp}^{a(s)}& \prod_{k=1}^{\frac{1}{2}(d-4)}\big((\a_k\l_s+\b_k)\mb{Q}+(\gamma_k-\a_k\t_{(s,s)}\l_s)\Omega\big)\prod_{l=1}^{s} \big ( \mb{Q} - \rho^2_{(l,s)}\big )\phi^{\perp}_{a(s)}~.\label{HiHelloGoodbye}
\end{align}
It is well known \cite{Tseytlin13, Metsaev14, Nutma14} that for $t=1$ the CHS action factorises into Klein-Gordon type operators, with the mass values corresponding to spin-$s$ PM masses with depths $1,\dots,s$, in agreement with \eqref{HiHelloGoodbye}. In addition, a discrete set of $\frac12 (d-4)$ massive points appear for $d>4$, which would correspond to the first set of factors in \eqref{HiHelloGoodbye}. 
 This information could be used to easily fix $\alpha_k,\beta_k,\gamma_k$ for $t=1$. 

For $t>1$ an analogous result does not exist because it is not possible to enter the TT gauge and so the action does not factorise into products of KG operators. However, action \eqref{HiHelloGoodbye} is expressed in terms of second order operators of the type $\sim(\mb{W}^2+\mb{Q}+\Omega)$, which is the next best thing. This was first shown in \cite{Grigoriev18} within a completely different framework. Comparison with this work could provide an alternative way to fix $\a_k,\b_k,\g_k$ for $t>1$. 

Despite the lack of factorisation for $t>1$, we can arrive at the following alternative result. If we impose the gauge $0=(\mc{D}^b)^{t}\phi_{b(t)a(s-t)}$ the irreducible decomposition of $\phi_{a(s)}$ becomes
\begin{align}
\phi_{a(s)} = \phi_{a(s)}^{\perp} + \cD_{\{a}\phi^{\perp}_{a(s-1)\}}+\cdots + \cD_{\{a_1}\cdots\cD_{a_{t-1}}\phi^{\perp}_{a_{t}\cdots a_s\}}~,
\end{align}
From this and the properties detailed in section \ref{secAdSTrans} it follows that
\begin{align}
\prod_{l=1}^{s-t+1} \big ( \mb{Q} - \rho^2_{(l,s-t+1)}\big )\Pi^{\perp}_{(s-t+1,s-t+1)}\phi_{a(s)} = \sum_{j=0}^{t-1}&\prod_{k=0}^{s-t}\frac{(s-k-j)(s+k-j+d-3)}{(s-k-t+1)(s+k-t+d-2)}\times\non\\
& \hspace{-0.5cm}\times \big(\mb{Q}-\rho^2_{(s-k-j,s-j)}\big)\mc{D}_{\{a_1}\cdots\mc{D}_{a_{j}}\phi^{\perp}_{a_{j+1}\cdots a_s\}}~.
\end{align}
Inserting this into \eqref{evendgenCHSAdS} and using \eqref{magicId2b} we obtain
\begin{align}
\mc{S}^{(t,s)}_{\text{CHS}}[\phi] = \int \text{d}^dx \, e \, \sum_{j=0}^{t-1}\phi_{\perp}^{a(s-j)}&\Big\{(-1)^j\Gamma_{j} \prod_{m=1}^{\frac{1}{2}(d-4)}\Big[(\a_m\l_{s-j}+\b_m)\mb{Q}+(\gamma_m-\a_m\t_{(s-j,s-j)}\l_{s-j})\Omega\Big]\times \non\\
\times &~\prod_{k=0}^{s-t}\big(\mb{Q}-\rho^2_{(s-k-j,s-j)}\big)\prod_{l=1}^{j} \big(\mb{Q} - \rho^2_{(l,s-j+l)}\big)\Big\}\phi^{\perp}_{a(s-j)}~,
\end{align}
where we have defined the constants $\Gamma_j:=\prod_{k=0}^{s-t}\frac{(s-k-j)(s+k-j+d-3)}{(s-k-t+1)(s+k-t+d-2)}\prod_{l=1}^{j}\frac{l(2s-2j+l+d-3)}{(s+l-j)(2s+2l-2j+d-4)}$.
We see that the action decomposes into a diagonal sum of  spin-$k$ modes with $s-t+1\leq k \leq s$ such that the kinetic operator for each mode consists of products of KG type operators with various PM masses, and a discrete set of massive points.

 In $d=4$ the action \eqref{evendgenCHSAdS} reduces to
 \begin{align}
 \mc{S}^{(t,s)}_{\text{CHS}}[\phi] = \int \text{d}^4x \,e\,  \phi_{(t)}^{a(s)}\prod_{l=1}^{s-t+1} \big ( \mb{Q} - \rho^2_{(l,s-t+1)}\big )\Pi^{\perp}_{(s-t+1,s-t+1)}\phi^{(t)}_{a(s)}~.\label{4genCHSAdS}
 \end{align}
 This gives rise to a new realisation for the generalised spin-$s$ depth-$t$ Bach tensor,
\begin{align}
\mc{B}_{a(s)}^{(t)} (\phi) =\prod_{l=1}^{s-t+1} \big ( \mb{Q} - \rho^2_{(l,s-t+1)}\big )\Pi^{\perp}_{(s-t+1,s-t+1)}\phi^{(t)}_{a(s)}~,
\end{align}
whose properties of partial-conservation, tracelessness and gauge-invariance are manifest.

\textbf{Conformal actions in three dimensions $d=3$}

The appropriate extension of the $d=3$ generalised CHS action \eqref{TMGenCHS} to (A)dS$_3$ is 
\begin{align} \label{TMGenCHSAdS}
\cS^{(t,s)}_{\text{CHS}}[\phi] = \int \rd^3x \,e\, \phi_{(t)}^{a(s)} \frac{1}{\mb{W}}\prod_{m=1}^{s-t+1} \big ( \mb{Q} - \rho^2_{(m,s-t+1)}\big ) \P^{\perp}_{(s-t+1,s-t+1)} \phi^{(t)}_{a(s)}~,
\end{align}
for $1\leq t \leq s$. Here $\mb{W}=-\frac{1}{2}\ve_{abc}\mc{D}^aM^{bc}$ is a quadratic Casimir of the (A)dS$_3$ isometry algebra, $0=\big[\mc{D}_a,\mb{W}\big]$, and the relation $\mb{W}^2 = \mb{W}\mb{W}$ holds with $\mb{W}^2$ defined in \eqref{DAQuadCasimir}. It is invariant under the higher-depth gauge transformations \eqref{AdSgt}. 

Action \eqref{TMGenCHSAdS} allows us to propose an expression for the spin-$s$ depth-$t$ Cotton tensor 
\begin{align}
\mc{C}_{a(s)}^{(t)}(\phi) &= \frac{1}{\mb{W}}\prod_{m=1}^{s-t+1} \big ( \mb{Q} - \rho^2_{(m,s-t+1)}\big ) \P^{\perp}_{(s-t+1,s-t+1)} \phi^{(t)}_{a(s)}~\non\\
&=\frac{1}{\lambda_{s-t+1}}\mb{W}\prod_{m=1}^{s-t} \big ( \mb{Q} - \rho^2_{(m,s-t+1)}\big ) \P^{\perp}_{(s-t+1,s-t+1)} \phi^{(t)}_{a(s)}~.
\end{align}
It is manifestly gauge invariant,
\begin{align}
\delta \phi_{a(s)}= \mc{D}_{\{a_1}\cdots\mc{D}_{a_{t}}\xi_{a_{t+1}\cdots a_{s}\}}\qquad \implies \qquad \mc{C}_{a(s)}^{(t)}(\delta\phi)=0~,
\end{align}
and partially conserved,
\begin{align}
0 = (\mc{D}^b)^t\mc{C}_{b(t)a(s-t)}^{(t)}(\phi)~.
\end{align}
It is also a primary field with conformal weight $s-t+2$.
For $t=1$ this was derived in \cite{KP21} (see also \cite{HKP}). For $t>1$ this is the first time this tensor has appeared in (A)dS$_3$. Using this field strength it is very straight-forward to construct topologically massive actions with higher depth and spin in (A)dS$_3$, one only has to follow the same recipe used in section \ref{secTMHS} for M$^3$.

\section{$\mc{N}=1$ superspin projection operators in M$^{4|4}$ }\label{secSUSY}

In this section we revisit the problem of building superprojectors for the four-dimensional  $\mc{N}=1$ Poincar\'{e} superalgebra $\mf{iso}(3,1|4)$. 
As mentioned in the introduction, these were previously investigated in the seminal papers \cite{SalamStrathdee1974,Sokatchev, SG}.
We work in four-dimensional  $\mc{N}=1$ Minkowski superspace M$^{4|4}$ and make use of the two-component spinor framework (see \cite{BK} for our conventions). Dotted and undotted spinor indices denoted by the same symbol are assumed to be symmetrised over, e.g.
\begin{align}
U_{\a(m)} V_{\a(n)} = U_{(\a_1\cdots \a_m} V_{\a_{m+1}\cdots \a_{m+n})} =\frac{1}{(m+n)!}\big(U_{\a_1 \cdots \a_m} V_{\a_{m+1}\cdots \a_{m+n}}+\cdots\big)~. \label{conventionSpin}
\end{align}
 The notation $U_{\a_1\dots\a_m}=U_{(\a_1\dots\a_m)}\equiv U_{\a(m)} $ represents totally symmetric indices.

A massive UIR $G(m,\hat{s})$ of $\mf{iso}(3,1|4)$ is labelled by its mass $m>0$ and superspin $\hat{s}$ which takes positive (half-)integer values. It corresponds to the following reducible representation of the Poincar\'e algebra $\mf{iso}(3,1)$
\begin{align}
G(m,\hat{s}) = D(m,\hat{s}+\frac{1}{2})\oplus D(m,\hat{s}) \oplus D(m,\hat{s}) \oplus  D(m,\hat{s}-\frac{1}{2})~.\label{SusyDecomp}
\end{align}
The last term is not present for superspin $\hat{s}=0$. Once supersymmetry is introduced, the construction of the projectors becomes more complicated than in the previous sections. This is because an unconstrained real superfield ${\F}_{\a(s)\ad(s)}$ on the mass-shell
\begin{align}
0=(\Box-m^2)\Phi_{\a(s)\ad(s)}  \label{smassshell}
\end{align}
furnishes the reducible representation 
\begin{align}
G(m,s+\frac{1}{2})\oplus 2\,G(m,s)\oplus 2\,G(m,s-\frac{1}{2})\oplus \cdots \oplus 2\,G(m,\frac{1}{2})\oplus 2\,G(m,0)~. \label{sreducible}
\end{align}
A $2$ in-front of a UIR indicates that this particular UIR appears with multiplicity 2. As we will see, this degeneracy gives rise to complications when implementing the recipe \eqref{linalg}.  From \eqref{sreducible} it follows that there are a total of $4s+3$ super Poincar\'e UIRs encoded by $\Phi_{\a(s)\ad(s)}$, in contrast to the $s+1$ Poincar\'e UIRs encoded by $\phi_{\a(s)\ad(s)}$ in \eqref{decomp1}. The superprojector(s) which isolate the UIR $G(m,j)$ from a superfield $\Phi_{\a(s)\ad(s)}$, with $j$ an integer or half-integer such that $0\leq j \leq s+\frac{1}{2}$, will sometimes be referred to as a superspin-$(s,j)$ projector.

To derive the decomposition \eqref{sreducible} we will first have to review some basic facts about superfields in M$^{4|4}$. 
For the covariant derivatives $D_A=\big(\pa_a, D_{\a},\DB^{\ad}\big)$  of M$^{4|4}$ we adopt the conventions of \cite{BK}, in which $D_A$ satisfy the anti-commutation relations
\begin{subequations}
	\begin{align}
	\big\{D_{\a},\DB_{\ad}\big\}=&-2\ri \pa_{\a\ad}~,\\
	\big\{D_{\a},D_{\b}\big\}=0~,\qquad &\big\{\DB_{\ad},\DB_{\bd}\big\}=0~.
	\end{align}
\end{subequations}
Many useful identities may be readily obtained from this algebra, such as
\begin{subequations}\label{idSUSY}
	\begin{align}
	D_{\a}D_{\b}&=\phantom{-}\frac{1}{2}\ve_{\a\b}D^2~,\qquad \big[D^2,\DB_{\ad}\big]=-4\ri \pa_{\a\ad}D^{\a}~,\label{idSUSYa}\\
	\DB_{\ad}\DB_{\bd}&=-\frac{1}{2}\ve_{\ad\bd}\DB^2~, \qquad \big[\DB^2,D_{\a}\big]=\phantom{-}4\ri \pa_{\a\ad}\DB^{\ad}~.\label{idSUSYb}
	\end{align}
	Another three important identities we will use on  occasion  are
	\begin{align}
	D^{\g}\bar{D}^2D_{\g} &= \bar{D}_{\gd}D^2\bar{D}^{\gd}~,\label{idSUSYc}\\
	\big\{D^2,\bar{D}^2\big\} &= 16\Box + 2D^{\b}\bar{D}^2D_{\b}~,\label{idSUSYd}\\
	\big[D^2,\bar{D}^2\big] &= -4\ri\pa^{\a\ad}\big[D_{\a},\bar{D}_{\ad}\big] ~,\label{idSUSYe}
	\end{align}
\end{subequations}
where we have denoted $D^2=D^{\a}D_{\a}$ and $\DB^2=\DB_{\ad}\DB^{\ad}$.

\subsection{The superspin operator}

In the superfield representation, the Casimir operators of the $\mc{N}=1$ super Poincar\'e algebra $\mf{iso}(3,1|4)$ must commute with all the generators of $\mf{iso}(3,1|4)$, and also the spinor derivatives. 
We observe that $\Box$ continues to be a Casimir operator. However, the spin operator $\mb{W}^2$ \eqref{spinOp}, which in $d=4$ two-component notation takes the form
\begin{align}
\mb{W}^2 = -\frac12 \mb{W}^{\a\ad}\mb{W}_{\a\ad}~,\qquad \mb{W}_{\a\ad}=\ri\pa_{\a}{}^{\bd}\bar{M}_{\ad\bd}-\ri\pa_{\ad}{}^{\b}M_{\a\b}~,
\end{align} 
 is no longer a Casimir operator as it does not commute with $D_{\a}$ and $\bar{D}_{\ad}$.\footnote{Here $M_{\a\b}$ are the generators of the $d=4$ Lorentz group, which have the commutator $[ M_{\a\b}, M_{\g\d} ] = \ve_{\a(\g}M_{\d)\b}+\ve_{\b(\g}M_{\d)\a}$. They act on the tensor $\phi_{\a(m)\ad(n)}$ via $M_{\g\d}\phi_{\a(m)\ad(n)}=m\ve_{\a(\g}\phi_{\d)\a(m-1)\ad(n)}$. Similar remarks hold regarding the complex conjugate $\bar{M}_{\ad\bd}$ of $M_{\a\b}$.}

The supersymmetric extension of $\mb{W}^2$, known as the superspin operator, is 
\begin{align}
\mb{Z}^2 = \Box\mb{W}^2 + \Box^2\big(B+\frac{3}{4}\big)\mc{P}_{\rLAL}~,
\end{align}
where $\mc{P}_{\rLAL}$ is defined in \eqref{aclProj} and $B=\frac{1}{4\Box}\big[D^{\b},\bar{D}^{\bd}\big]\mb{W}_{\b\bd}$. 
The superspin operator is the final Casimir operator that we will need, and we emphasise that it satisfies the important property
\begin{align}
\big[\mb{Z}^2, D_A\big]=0~.
\end{align}
When acting on a superfield furnishing the UIR $G(m,\hat{s})$ of $\mf{iso}(3,1|4)$, the superspin operator is proportional to the identity, with eigenvalue
\begin{align}
\mb{Z}^2 ~\big|_{G(m,\hat{s})} ~ = ~ \lambda_{\hat{s}}\Box^2~,\qquad\qquad \lambda_{\hat{s}} := \hat{s}(\hat{s}+1)~. \label{superspinIr}
\end{align}
Equation \eqref{superspinIr} can be taken as the definition of $\hat{s}$, and then used to determine the superspin of an irreducible superfield.\footnote{However, it is important to note that even if $\mb{Z}^2$ takes a constant value on a superfield, that superfield is not guaranteed to be irreducible. An example of this comes from the next subsection, where we have  $0=\big(\mb{Z}^2-\lambda_{\frac{1}{2}(p+q)}\Box^2\big)\Psi_{\a(p)\ad(q)}^{\pm}$ and hence $0=\big(\mb{Z}^2-\lambda_{\frac{1}{2}(p+q)}\Box^2\big)\big(\Psi_{\a(p)\ad(q)}^{+}+\Psi_{\a(p)\ad(q)}^{-}\big)$, see \eqref{Zacsup}. But the superfield $\big(\Psi_{\a(p)\ad(q)}^{+}+\Psi_{\a(p)\ad(q)}^{-}\big)$ is not irreducible. \label{CautionFN}}

A practical identity is the action of $\mb{Z}^2$ on an unconstrained superfield $\Psi_{\a(p)\ad(p)}$,
\begin{subequations}\label{practical}
	\begin{align}
	\mb{Z}^2\Psi_{\a(p)\ad(p)} = &\big[\lambda_{p+\frac12}+\frac{3}{4}\big(\mc{P}_{\rLAL}-1\big)\big]\Box^2\Psi_{\a(p)\ad(p)}
	+ \frac{\ri}{2}p\Box \Big(
	-2\ri p\pa_{\a\ad}\pa^{\b\bd}\Psi_{\a(p-1)\b\ad(p-1)\bd}\non\\
	&+\pa_{\a\gd}\bar{D}^{\gd}D^{\b}\Psi_{\b\a(p-1)\ad(p)}+\pa_{\g\ad}D^{\g}\bar{D}^{\bd}\Psi_{\a(p)\ad(p-1)\bd}\Big)~, \label{practicala}
	\end{align}
	which is a special case of its action on an unconstrained superfield $\Psi_{\a(p)\ad(q)}$,
	\begin{align}
	\mb{Z}^2\Psi_{\a(p)\ad(q)} = &\big[\lambda_{\frac{1}{2}(p+q+1)}+\frac{3}{4}\big(\mc{P}_{\rLAL}-1\big)\big]\Box^2\Psi_{\a(p)\ad(q)}
	+ \frac{\ri}{4}\Box \Big(
	-4\ri pq\pa_{\a\ad}\pa^{\b\bd}\Psi_{\a(p-1)\b\ad(q-1)\bd}\non\\
	&+(p+q)\pa_{\a\gd}\bar{D}^{\gd}D^{\b}\Psi_{\b\a(p-1)\ad(q)}+(p+q)\pa_{\g\ad}D^{\g}\bar{D}^{\bd}\Psi_{\a(p)\ad(q-1)\bd}\non\\
	&-(p-q)\bar{D}_{\ad}D_{\a}\pa^{\b\bd}\Psi_{\a(p-1)\b\ad(q-1)\bd}+(p-q)\pa_{\a\ad}\bar{D}^{\bd}D^{\b}\Psi_{\a(p-1)\b\ad(q-1)\bd}\Big)~. \label{practicalb}
	\end{align}
\end{subequations}
In the next subsection we specialise these to the case of $\Psi$ satisfying various constraints.

\subsection{Constrained and irreducible superfields}\label{secConstr}

For the purposes of this paper, a constrained superfield is a superfield that is subject to a condition(s) in the form of a differential equation, in any of the derivatives $D_A$, which are not 
wave equations. When a superfield is subject to an appropriate set of constraints (plus the KG equation), it will furnish some irreducible representation of $\mf{iso}(3,1|4)$. Such superfields will be referred to as irreducible superfields.

In the massive case, the standard way to confirm that a superfield is irreducible is to analyse its component content and confirm that it agrees with \eqref{SusyDecomp}. One can also check that $\mb{Z}^2$ is constant on it, bearing in mind the cautionary footnote \ref{CautionFN}. 
We now turn to presenting select constrained and irreducible superfields, though the routine task of performing a  component analysis to show their irreducibility will not be presented.

The first type of constrained superfield that we consider is the \textit{ultra-transverse} superfield $\Psi^{\rT}_{\a(p)\ad(q)}$, for $p>0$ and $q>0$, which is defined to be simultaneously $D$- and $\bar{D}$-transverse\footnote{ In a previous work \cite{BHKP} we called such superfields TLAL.}
\begin{subequations}
	\begin{align}
	0&=\bar{D}^{\bd}\Psi^{\rT}_{\a(p)\ad(q-1)\bd} ~,\qquad 0=D^{\b}\Psi^{\rT}_{\b\a(p-1)\ad(q)}~,\label{TLALa}\\[7pt]
	\implies \qquad 0 = \bar{D}^2\Psi^{\rT}_{\a(p)\ad(q)}&~,\qquad~~~\,\,\, 0= D^2\Psi^{\rT}_{\a(p)\ad(q)}~, \qquad 0=\pa^{\b\bd}\Psi^{\rT}_{\a(p-1)\b\ad(q-1)\bd} ~.~~~~\label{TLALb}
	\end{align}
\end{subequations}
The conditions \eqref{TLALb} are all consequences of the defining constraints \eqref{TLALa}. 
From identity \eqref{practicalb} it follows that $\mb{Z}^2$ is constant on an such a superfield
\begin{align}
0=\big(\mb{Z}^2-\lambda_{\frac{1}{2}(p+q+1)}\Box^2\big)\Psi^{\rT}_{\a(p)\ad(q)}~.
\end{align}
An ultra-transverse superfield $\Psi^{\rT}_{\a(p)\ad(q)}$ is irreducible,
furnishing the UIR $G\big(m , \frac{1}{2}(p+q+1)\big)$. In general we will attach the symbol `T' to a superfield to indicate that it is ultra-transverse. 

When $p>q=0$ and the superfield $\Psi^{\rT}_{\a(p)}$ carries only one type of index, we define ultra-transverse to mean that it satisfies the conditions
\begin{align}
0 = D^{\b}\Psi^{\rT}_{\b\a(p-1)}~,\qquad 0=\bar{D}^2\Psi^{\rT}_{\a(p)} \qquad \implies \qquad 0=D^2\Psi^{\rT}_{\a(p)}~.\label{LTAL}
\end{align}
Such a superfield is irreducible and furnishes the UIR $G\big(m , \frac{1}{2}(p+1)\big)$. 
If on the other hand $q>p=0$, then $\Psi_{\ad(q)}^{\rT}$ is defined to be ultra-transverse if 
\begin{align}
0 = \bar{D}^{\bd}\Psi^{\rT}_{\bd\ad(q-1)}~,\qquad 0=D^2\Psi^{\rT}_{\ad(q)} \qquad \implies \qquad 0=\bar{D}^2\Psi^{\rT}_{\ad(q)}~,\label{ALTL}
\end{align}
and $\Psi_{\ad(q)}$  is irreducible, furnishing the UIR $G\big(m , \frac{1}{2}(q+1)\big)$. For $p=q=0$, we simply define an ultra-transverse superfield $\Psi^{\rT}$ to be simultaneously linear \& anti-linear (LAL)
\begin{align}
0=\bar{D}^2\Psi^{\rT}~,\qquad 0=D^2\Psi^{\rT}~.
\end{align} 
The superfield $\Psi^{\rT}$ is irreducible and furnishes the UIR $G(m , \frac{1}{2})$.

Next we introduce the \textit{ultra-longitudinal}\footnote{This is not to be confused with e.g. a longitudinal linear superfield, which satisfies $\bar{D}_{\ad}\Psi_{\a(p)\ad(q)}=0$.} superfield $\Psi^{\rL}_{\a(p)\ad(q)}$, for $p>0$ and $q>0$, which we define to be a superfield such that it may be expressed in the form 
\begin{subequations}\label{longitudinal}
	\begin{align}
	&\Psi^{\rL}_{\a(p)\ad(q)} = \big[D_{\a},\bar{D}_{\ad}\big] \Omega^{\rT}_{\a(p-1)\ad(q-1)}~,\label{longitudinala}\\[7pt]
	\implies \qquad 0 = \bar{D}^2\Psi^{\rL}_{\a(p)\ad(q)}&~,\qquad~~~\,\,\, 0= D^2\Psi^{\rL}_{\a(p)\ad(q)}~, \qquad 0=\pa^{\b\bd}\Psi^{\rL}_{\a(p-1)\b\ad(q-1)\bd} ~,~~~~\label{longitudinalb}
	\end{align}
\end{subequations}
for some ultra-transverse superfield $\Omega^{\rT}_{\a(p-1)\ad(q-1)}$. The conditions \eqref{longitudinalb} are all consequences of the defining constraint \eqref{longitudinala}. We are not aware of any other constraints that $\Psi^{\rL}_{\a(p)\ad(q)}$ obeys, apart from those related to \eqref{longitudinal}. Since $\mb{Z}^2$ is a Casimir, it passes through the commutator in \eqref{longitudinala} and reaches the ultra-transverse superfield, where it takes the value $\lambda_{\frac{1}{2}(p+q-1)}\Box^2$, 
\begin{align}
0=\big(\mb{Z}^2-\lambda_{\frac{1}{2}(p+q-1)}\Box^2\big)\Psi^{\rL}_{\a(p)\ad(q)}~.
\end{align} 
An ultra-longitudinal superfield $\Psi^{\rL}_{\a(p)\ad(q)}$ is irreducible, furnishing the UIR $G\big(m , \frac{1}{2}(p+q-1)\big)$. In general we will attach the symbol `L' to a superfield to indicate that it is ultra-longitudinal. 

Next we introduce the three projectors \cite{SalamStrathdee1974}
\begin{align}
\mc{P}_{+}=\frac{1}{16\Box}\bar{D}^2D^2~,\qquad \mc{P}_{-}=\frac{1}{16\Box}D^2\bar{D}^2~,\qquad \mc{P}_{\text{LAL}} = -\frac{1}{8\Box}D^{\g}\bar{D}^2D_{\g}~, \label{aclProj}
\end{align}
which are orthogonal and form a partition of identity
\begin{align}
\mc{P}_{i}\mc{P}_{j}=\delta_{ij}\mc{P}_{i}~,\qquad \mathds{1} = \mc{P}_{+} + \mc{P}_{-} + \mc{P}_{\text{LAL}}~, \label{resol2}
\end{align}
where $i=\{+,-,\text{LAL}\}$.
Operators $\mc{P}_{+}$, $\mc{P}_{-}$ and $\mc{P}_{\text{LAL}}$ project onto the constrained spaces of chiral, anti-chiral, and LAL superfields respectively,
\begin{align}
\bar{D}_{\ad}\mc{P}_{+}\Psi=0, \qquad D_{\a}\mc{P}_{-}\Psi=0~,\qquad \bar{D}^2\mc{P}_{\text{LAL}}\Psi=D^2\mc{P}_{\text{LAL}}\Psi=0~,
\end{align} 
where we have suppressed any indices $\Psi$ may carry. One may readily confirm that each of these operators are equal to the identity operator when acting on the space which they project to. 

Imposing any single\footnote{Imposing any combination of them would trivialise the superfield, i.e. force it to vanish. } one of these constraints (chiral, anti-chiral or LAL) on an otherwise unconstrained superfield $\Psi_{\a(p)\ad(q)}$, for generic $p$ and $q$, is not enough to ensure that the result is irreducible. However, let us consider a chiral superfield $\Psi^{+}_{\a(p)\ad(q)}$, or an anti-chiral superfield $\Psi_{\a(p)\ad(q)}^{-}$, with $p>0$ and $q>0$, which is also vector-transverse\footnote{The name `vector-transverse' is used to distinguish from transversality with respect to the spinor derivatives.}
\begin{subequations}
	\begin{align}
	0&=\bar{D}_{\bd}\Psi_{\a(p)\ad(q)}^{+}~,\qquad 0=\pa^{\b\bd}\Psi^{+}_{\a(p-1)\b\ad(q-1)\bd}~,\\
	0&=D_{\b}\Psi_{\a(p)\ad(q)}^{-}~,\qquad 0=\pa^{\b\bd}\Psi^{-}_{\a(p-1)\b\ad(q-1)\bd}~.
	\end{align}
\end{subequations}  
Such superfields will be referred to as being transverse-(anti-)chiral. Using identity \eqref{practicalb}, it is not difficult to convince oneself that $\mb{Z}^2$ is constant on such superfields
\begin{align}
0=\big(\mb{Z}^2-\lambda_{\frac{1}{2}(p+q)}\Box^2\big)\Psi_{\a(p)\ad(q)}^{\pm}~.\label{Zacsup}
\end{align}
This is made manifest by the fact that these superfields may always be expressed in the form
\begin{subequations}
	\begin{align}
	\Psi_{\a(p)\ad(q)}^{+} &= \bar{D}_{\ad}\O^{\rT}_{\a(p)\ad(q-1)}~,\qquad \O^{\rT}_{\a(p)\ad(q-1)} =\phantom{-} \frac{1}{8\Box}\bar{D}^{\bd}D^2\Phi^{+}_{\a(p)\ad(q-1)\bd}~,\\
	\Psi_{\a(p)\ad(q)}^{-} &= D_{\a}\L^{\rT}_{\a(p-1)\ad(q)}~,\qquad \L^{\rT}_{\a(p-1)\ad(q)}=-\frac{1}{8\Box}D^{\b}\bar{D}^2\Phi^{-}_{\b\a(p-1)\ad(q)}~, 
	\end{align}
\end{subequations}
where $\O^{\rT}_{\a(p)\ad(q-1)}$ and $\L^{\rT}_{\a(p-1)\ad(q)}$ are ultra-transverse. Transverse-(anti-)chiral superfields $\Psi_{\a(p)\ad(q)}^{\pm}$, identified by the symbol `$(-)+$', are irreducible and furnish the UIR $G\big(m , \frac12(p+q)\big)$. 

When $q=0$ and $p>0$ then only the chiral or anti-chiral conditions satisfying the KG equation \eqref{smassshell} are needed to furnish the UIR $G(m ,  \frac12 q)$, and not the $\pa$-transverse condition (this obviously cannot be imposed). We will accordingly append the superfield with the relevant $+$ or $-$ symbol, $\Psi^{\pm}_{\a(p)}$. Similar comments apply in the case $p=0$ and $q>0$ and $q=p=0$.
Finally, we point out that an LAL superfield $\Phi^{\rLAL}_{\a(p)\ad(q)}$, which is also vector-transverse, is not irreducible. In the next subsection we will see that it decomposes into the sum of an ultra-transverse and an ultra-longitudinal superfield. 
\subsection{Irreducible content of a transverse superfield } \label{secFPSUSY}

Consider a real superfield $\Phi^{\perp}_{\a(s)\ad(s)}$ which is on the mass-shell and vector-transverse
\begin{subequations}\label{SUSYFP}
	\begin{align}
	0&=(\Box-m^2)\Phi^{\perp}_{\a(s)\ad(s)}~,\label{SUSYFPa}\\
	0&=\pa^{\b\bd}\Phi^{\perp}_{\a(s-1)\b\ad(s-1)\bd}~. \label{SUSYFPb}
	\end{align}
\end{subequations}
We now demonstrate that such a superfield furnishes the reducible representation
\begin{align}
G(m,s+\frac{1}{2})\oplus G(m,s)\oplus G(m,s)\oplus G(m,s-\frac{1}{2})~.\label{SUSYFPrep}
\end{align}

To begin, we trisect $\Phi^{\perp}_{\a(s)\ad(s)}$ using \eqref{resol2} (we use the shorthand $\Psi_{(s)}$ to represent $\Psi_{\a(s)\ad(s)}$),
\begin{align}
\Phi^{\perp}_{(s)}= \Phi_{(s)}^{+}+\Phi^{-}_{(s)}+\Phi^{\text{LAL}}_{(s)}~,\qquad \Phi^{i}_{(s)}:= \mc{P}_{i}\,\Phi_{(s)}^{\perp}~.\label{trisection}
\end{align}
As discussed in the previous section, the chiral and anti-chiral parts are each irreducible,
and together furnish two copies of the UIR $G(m,s)$.  
The (reducible) transverse-LAL part furnishes the remaining two UIRs, to separate them we bisect as follows
\begin{subequations}
	\begin{align}
	&\Phi^{\text{LAL}}_{(s)} = \Phi^{\rT}_{(s)}+\Phi^{\rL}_{(s)}~,\\[7pt]
	\Phi^{\rT}_{(s)} := \frac{(\mb{Z}^2-\lambda_{s-\frac{1}{2}}\Box^2)}{(\lambda_{s+\frac{1}{2}}-\lambda_{s-\frac12})\Box^2}&\Phi^{\text{LAL}}_{(s)}~,\qquad \Phi^{\rL}_{(s)} := \frac{(\mb{Z}^2-\lambda_{s+\frac{1}{2}}\Box^2)}{(\lambda_{s-\frac{1}{2}}-\lambda_{s+\frac12})\Box^2}\Phi^{\text{LAL}}_{(s)}~.
	\end{align}
\end{subequations}
As the notation suggests, the superfield $\Phi^{\rT}_{(s)}$ is ultra-transverse, and thus furnishes the UIR $G(m,s+\frac12)$. On the other-hand, the superfield $\Phi^{\rL}_{(s)}$ is ultra-longitudinal, 
\begin{align}
\Phi^{\rL}_{\a(s)\ad(s)} = \big[D_{\a},\bar{D}_{\ad}\big]\O^{\rT}_{\a(s-1)\ad(s-1)}~,\qquad \O^{\rT}_{\a(s-1)\ad(s-1)}:= \frac{s}{8}\Box\big[D^{\b},\bar{D}^{\bd}\big]\Phi^{\rLAL}_{\b\a(s-1)\bd\a(s-1)}~,
\end{align}
and thus furnishes the UIR $G(m,s-\frac12)$. Both of these facts may be deduced using \eqref{practical}. 

Therefore, in summary, a superfield $\Phi^{\perp}_{(s)}$ satisfying \eqref{SUSYFP} may be dissected into four pieces\footnote{The superfields $\Phi^{\rT}_{(s)}$ and $\Phi^{\rL}_{(s)}$ are real, whilst $\Phi^{+}_{(s)}$ and $\Phi^{-}_{(s)}$ are complex conjugates of each other.   }
\begin{align}
\Phi^{\perp}_{(s)} = \Phi^{\rT}_{(s)} + \Phi^{+}_{(s)} + \Phi^{-}_{(s)} + \Phi^{\rL}_{(s)}~.
\end{align}  
Each piece is irreducible, being ultra-transverse, transverse-chiral, transverse-anti-chiral and ultra-longitudinal respectively.  The individual pieces are related to $\Phi^{\perp}_{(s)}$ through the formulae 
\begin{align}
\Phi^{\rT}_{(s)} &= \frac{(\mb{Z}^2-\lambda_{s-\frac{1}{2}}\Box^2)}{(\lambda_{s+\frac{1}{2}}-\lambda_{s-\frac12})\Box^2}\mc{P}_{\rLAL}\Phi^{\perp}_{(s)}~,\qquad \Phi^{+}_{(s)} =\mc{P}_{+}\Phi^{\perp}_{(s)}~,\non\\[7pt]
\Phi^{\rL}_{(s)} &= \frac{(\mb{Z}^2-\lambda_{s+\frac{1}{2}}\Box^2)}{(\lambda_{s-\frac{1}{2}}-\lambda_{s+\frac12})\Box^2}\mc{P}_{\rLAL}\Phi^{\perp}_{(s)}~,\qquad \Phi^{-}_{(s)} =\mc{P}_{-}\Phi^{\perp}_{(s)}~.
\end{align}
It follows that $\Phi^{\perp}_{(s)}$ furnishes the reducible representation \eqref{SUSYFPrep}.

\subsection{The superprojectors} \label{secSupProj1}

In this work we define a superprojector to be any projector which, when acting on an \textit{unconstrained} superfield satisfying the massive Klein-Gordon equation, extracts a single irreducible part of that superfield, i.e. it furnishes some UIR $G(m,\hat{s})$. 
We are now almost at a stage where we can read off the explicit form of the superprojectors for $\Phi_{\a(s)\ad(s)}$, we just need to relax the vector-transverse constraint \eqref{SUSYFPb}. 

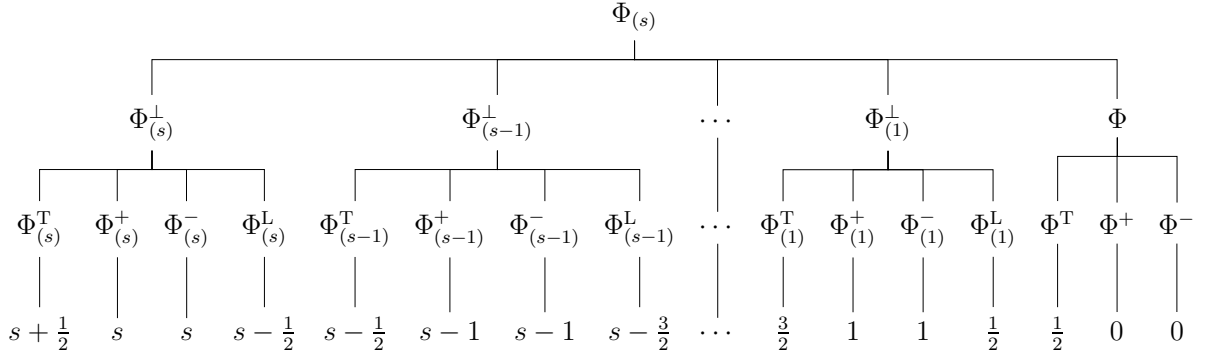
\begin{figure}
	\begin{center}
		\resizebox{0.95\linewidth}{!}{%
			\begin{tikzpicture}
			\tikzset{level distance=45pt}
			\tikzset{edge from parent/.style=
				{draw,
					edge from parent path={(\tikzparentnode.south)
						-- +(0,-8pt)
						-| (\tikzchildnode)}}}
			\Tree[.$\Phi_{(s)}$ [.$\Phi^{\perp}_{(s)}$ [.$\Phi_{(s)}^{\rT}$ [.$s+\frac12$ ] ]
			[.$\Phi_{(s)}^{+}$ [.$s$ ] ]
			[.$\Phi_{(s)}^{-}$ [.$s$ ] ]
			[.$\Phi_{(s)}^{\rL}$ [.$s-\frac12$ ] ]
			]
			[.$\Phi^{\perp}_{(s-1)}$ [.$\Phi_{(s-1)}^{\rT}$ [.$s-\frac12$ ] ]
			[.$\Phi_{(s-1)}^{+}$ [.$s-1$ ] ]
			[.$\Phi_{(s-1)}^{-}$ [.$s-1$ ] ]
			[.$\Phi_{(s-1)}^{\rL}$ [.$s-\frac{3}{2}$ ] ]    
			]
			[.$\cdots$ [.$\cdots$ [.$\cdots$  ] ] ]	
			[.$\Phi^{\perp}_{(1)}$ [.$\Phi_{(1)}^{\rT}$ [.$\frac{3}{2}$ ] ]
			[.$\Phi_{(1)}^{+}$ [.$1$ ] ]
			[.$\Phi_{(1)}^{-}$ [.$1$ ] ]
			[.$\Phi_{(1)}^{\rL}$ [.$\frac12$ ] ]
			] 
			[.$\Phi$ [.$\Phi^{\rT}$ [.$\frac12$ ] ]
			[.$\Phi^{+}$ [.$0$ ] ]
			[.$\Phi^{-}$ [.$0$ ] ]
			]                                    
			]                                     
			\end{tikzpicture}
		}
		\caption{Decomposition of an unconstrained superfield $\Phi_{(s)}$ into irreducible parts. The bottom row provides the superspins of all the constituent superfields, reflecting the pattern in \eqref{sreducible}. }\label{fig1}
	\end{center}
\end{figure}

We already have the tools at our disposal to do this, namely, the spin-$(s,j)$ projectors \eqref{sjproj}.\footnote{ The explicit form of the projectors $\Pi_{(s,j)}^{\perp}$ as defined in \eqref{sjproj} does not depend on whether they act on spinors or tensors since the Lorentz generators appearing in $\mb{W}^2$ are in an unspecified representation.}  
In particular, using the methods of section \ref{secProjMd}, an unconstrained superfield $\Phi_{\a(s)\ad(s)}$ decomposes into vector-transverse parts as
\begin{align}
\Phi_{\a(s)\ad(s)} &= \big(\Pi^{\perp}_{(s,s)} + \Pi^{\perp}_{(s,s-1)}+\cdots+\Pi^{\perp}_{(s,1)} +\Pi^{\perp}_{(s,0)}\big)\Phi_{\a(s)\ad(s)}  \non\\[7pt]
&= \Phi^{\perp}_{\a(s)\ad(s)}+\pa_{\a\ad} \Phi^{\perp}_{\a(s-1)\ad(s-1)}+\cdots+ (\pa_{\a\ad})^{s-1} \Phi^{\perp}_{\a\ad} +(\pa_{\a\ad})^s\Phi\label{vecTsuper}
\end{align}
where $\Phi_{\a(j)\ad(j)}^{\perp}$ is given by \eqref{divextract}. For each vector-transverse superfield in \eqref{vecTsuper} we may repeat the irreducible decomposition described in section \ref{secFPSUSY}. The result of this procedure produces the reducible representation \eqref{sreducible} presented at the beginning of this section, and is depicted in figure \ref{fig1}. Note that when one gets to the scalar superfield, it decomposes only into three irreducible parts, $\Phi = \Phi^{\rT}+\Phi^{+}+\Phi^{-}$, corresponding to the trisection \eqref{trisection}, which carry superspins $\frac12, 0$ and $0$ respectively.

At this stage we can now easily deduce that the superprojectors which select each irreducible part of an unconstrained superfield $\Phi_{\a(s)\ad(s)}$ take the form
\begin{subequations}\label{superprojectors1}
	\begin{align}
	\Pi_{(s,j)}^{\rT}&= \frac{(\mb{Z}^2-\lambda_{j-\frac{1}{2}}\Box^2)}{(\lambda_{j+\frac{1}{2}}-\lambda_{j-\frac12})\Box^2}\mc{P}_{\rLAL} \Pi^{\perp}_{(s,j)}~,\qquad\qquad \Pi_{(s,j)}^{+}= \mc{P}_{+} \Pi^{\perp}_{(s,j)}~,\\[7pt]
	\Pi_{(s,j)}^{\rL}&= \frac{(\mb{Z}^2-\lambda_{j+\frac{1}{2}}\Box^2)}{(\lambda_{j-\frac{1}{2}}-\lambda_{j+\frac12})\Box^2}\mc{P}_{\rLAL}\Pi^{\perp}_{(s,j)}~,\qquad\qquad \Pi_{(s,j)}^{-}= \mc{P}_{-} \Pi^{\perp}_{(s,j)}~,
	\end{align}
\end{subequations}
for $0\leq j\leq s$. Note that from the above definition we have $\Pi_{(s,0)}^{\rT}=\mc{P}_{\rLAL}\Pi_{(s,0)}^{\perp}$ and $\Pi_{(s,0)}^{\rL}=0$ when acting on an unconstrained $\Phi_{\a(s)\ad(s)}$. Moreover, when acting on such a superfield, the superprojectors $\{\Pi_{(s,j)}^{\rT}, \Pi_{(s,j)}^{+}, \Pi_{(s,j)}^{-}, \Pi_{(s,j)}^{\rL} \}$ select the parts of $\Phi_{\a(s)\ad(s)}$ furnishing the UIRs $\{G(m,j+\frac12), G(m,j), G(m,j), G(m,j-\frac12)\}$ respectively, as is clear from figure \ref{fig1}.

Independent of the space on which they act, the collection of operators \eqref{superprojectors1} constitute a resolution of the identity 
\begin{subequations}
	\begin{align}
	\mathds{1}=\sum_{j=0}^{s}\Big(\Pi_{(s,j)}^{\rT}+\Pi_{(s,j)}^{+}+\Pi_{(s,j)}^{-}+\Pi_{(s,j)}^{\rL}\Big)~, \label{resol3}
	\end{align}
	which may be proven using \eqref{didres} and \eqref{resol2}.
	Furthermore,  when acting on the space $\textbf{V}_{(l)}$ of unconstrained superfields $\Phi_{\a(l)\ad(l)}$ of rank $l\leq s$, they are orthogonal to one another
	\begin{align}
	\Pi_{(s,j)}^{\mc{A}}\Pi_{(s,k)}^{\mc{B}} = \delta_{jk}\delta^{\mc{A}\mc{B}}\Pi_{(s,j)}^{\mc{A}}~,\qquad \text{on } \mathbf{V}_{(l)} \text{ with } 0\leq l\leq s~,
	\end{align}
\end{subequations}
where $\mc{A}=\{\rT,+,-,\rL\}$. This follows from the properties \eqref{orthog2}, \eqref{resol2}, and also
\begin{subequations}
	\begin{align}
	0&=\big(\mb{W}^2 - \lambda_j\Box \big)\Pi_{(s,j)}^{\mc{A}}\Phi_{\a(l)\ad(l)}~\qquad \forall ~\mc{A}~,\\
	0&=\big(\mb{Z}^2 - \lambda_{j+\frac12}\Box^2 \big)\Pi_{(s,j)}^{\rT}\Phi_{\a(l)\ad(l)}~,\\
	0&=\big(\mb{Z}^2 - \lambda_{j}\Box^2 \big)\Pi_{(s,j)}^{+}\Phi_{\a(l)\ad(l)}~,\\
	0&=\big(\mb{Z}^2 - \lambda_{j}\Box^2 \big)\Pi_{(s,j)}^{-}\Phi_{\a(l)\ad(l)}~,\\
	0&=\big(\mb{Z}^2 - \lambda_{j-\frac12}\Box^2 \big)\Pi_{(s,j)}^{\rL}\Phi_{\a(l)\ad(l)}~,
	\end{align} 
\end{subequations}
which holds for all $0\leq l \leq s$ (sometimes trivially, due to \eqref{annihilatesuper}).

Another important property of the superprojectors is that for $j\leq s$ and $j\leq l\leq s$
\begin{subequations}
	\begin{align}
	\Pi_{(s,j)}^{\mc{A}}\Phi_{\a(l)\ad(l)}  &=(\pa_{\a\ad})^{l-j}\O_{\a(j)\ad(j)}^{\mc{A}}~, \label{projirr}
	\end{align}
	for an irreducible superfield $\O_{\a(j)\ad(j)}^{\mc{A}}$ of type $\mc{A}$.
	Each superprojector $\Pi_{(s,j)}^{\mc{A}}$ acts as the identity operator on the space of superfields to which it projects 
	\begin{align}
	\Pi_{(s,j)}^{\mc{A}}\O_{\a(j)\ad(j)}^{\mc{A}} = \O_{\a(j)\ad(j)}^{\mc{A}}~.
	\end{align}
	On the other hand, for $j\leq s$ and $0\leq l \leq j-1$, one finds  
	\begin{align}
	\Pi_{(s,j)}^{\mc{A}}\Phi_{\a(l)\ad(l)} =0~, \label{annihilatesuper}
	\end{align}
\end{subequations}
i.e. they annihilate all superfields of low enough rank. As a consequence of \eqref{projirr} we have
\begin{subequations}
	\begin{align}
	0&=(\pa^{\b\bd})^{l-j+1}\Pi_{(s,j)}^{\mc{A}}\Phi_{\b(l-j+1)\a(j-1)\bd(l-j+1)\ad(j-1)} \qquad \forall~\mc{A}~,\\
	0&=(\pa^{\b\bd})^{l-j}D^{\b}\Pi_{(s,j)}^{\rT}\Phi_{\b(l-j+1)\a(j-1)\bd(l-j)\ad(j)}~,\label{Dbarproj}
	\end{align}
\end{subequations}
for $j\leq l\leq s$. A relation analogous to \eqref{Dbarproj} but involving $\bar{D}$  holds.  In addition it is obvious that $\Pi_{(s,j)}^{\rT}\Phi_{\a(l)\ad(l)} $ and $\Pi_{(s,j)}^{\rL}\Phi_{\a(l)\ad(l)}$ are LAL, whilst $\Pi_{(s,j)}^{+}\Phi_{\a(l)\ad(l)}$ is chiral and $\Pi_{(s,j)}^{-}\Phi_{\a(l)\ad(l)} $ is anti-chiral. It also follows that a complex unconstrained superfield $\Phi_{(s)}$ decomposes as
\begin{align}
\Phi_{\a(s)\ad(s)}=\sum_{t=1}^{s} &(\pa_{\a \ad} )^{s-t}\big(
\Phi^{\rT}_{\a(t)\ad(t)} 
+\Phi^{+}_{\a(t)\ad(t)}+\Phi^{-}_{\a(t)\ad(t)}
+\Phi^{\rL}_{\a(t)\ad(t)}\big) \non\\
+ \, &(\pa_{\a\ad}\big)^s \big(\Phi^{\rT} + \Phi^{+} + \Phi^{-})~,\label{decomp3}
\end{align}
where each superfield is complex (and unrelated by complex conjugation).  The first line of \eqref{decomp3} can be expressed entirely in terms of ultra-transverse superfields:\footnote{The AdS version of this irreducible decomposition was used, albeit without derivation,  in \cite{BKS} for the covariant quantisation of the massless supersymmetric higher-spin models in AdS${}_4$ \cite{KS94}. }
\begin{align}
\Phi_{\a(s)\ad(s)}=\sum_{t=1}^{s} &(\pa_{\a \ad} )^{s-t}\big(
\Phi^{\rT}_{\a(t)\ad(t)} 
+\bar{D}_{\ad}\z^{\rT}_{\a(t)\ad(t-1)}+D_{\a}\xi^{\rT}_{\a(t-1)\ad(t)}
+\big[D_{\a},\bar{D}_{\ad}\big]\O^{\rT}_{\a(t-1)\ad(t-1)}\big) \non\\
+ \, &(\pa_{\a\ad}\big)^s \big(\Phi^{\rT} + \Phi^{+} + \Phi^{-})~.\label{decomp4}
\end{align}
If the superfield $\Phi_{(s)}$ is real, then $\Phi_{(t)}^{\rT}$ and $\Phi_{(t)}^{\rL}$ would also be real (and hence so too would $\O_{(t-1)}^{\rT}$), whilst $\Phi^{-}_{(t)}=\bar{\Phi}^{+}_{(t)}$ and $\Phi^{-}=\bar{\Phi}^{+}$ (and hence $\xi^{\rT}_{\a(t-1)\ad(t)} = -\bar{\z}^{~\rT}_{\a(t-1)\ad(t)}$).

\subsection{Alternative form of the superprojectors}
In contrast to the non-supersymmetric projectors $\Pi_{(s,j)}^{\perp}$, which were constructed purely in terms of Casimir operators $\mb{W}^2$ and $\Box$ of $\mf{iso}(3,1)$, the above superprojectors are not constructed purely in terms of Casimir operators $\mb{Z}^2$ and $\Box$ of $\mf{iso}(3,1|4)$. It would be beneficial to derive equivalent forms of the superprojectors which have this property. Unfortunately, this is only possible for the superprojector selecting the UIR $G(m,s+\frac12)$ of maximal superspin, i.e. the superfield $\Phi^{\rT}_{\a(s)\ad(s)}$. This is because in the decomposition \eqref{sreducible} (see also figure \ref{fig1}), all other UIRs appear with multiplicity 2. In other words, for all superspins $\hat{s}\leq s$, the superspin operator $\mb{Z}^2$ is doubly degenerate, and we cannot use the recipe \eqref{linalg} to construct the corresponding two superspin-$\hat{s}$ projectors purely in terms of $\mb{Z}^2$ and $\Box$. 

Nevertheless, we can achieve the next best thing, by getting the superprojectors \textit{almost} purely in terms of the Casimir operators $\mb{Z}^2$ and $\Box$. To do this, we first make use of the algorithm \eqref{linalg} to select the eigenspace corresponding to a fixed superspin $\hat{s}$. For the superprojector $\Pi_{(s,s)}^{\rT}$ with maximal superspin $\hat{s}=s+\frac12$ this is all that needs to be done, and we have\footnote{Note that for $s=0$ only the first product in \eqref{SuperprojectorMaximal} is present, $\Pi_{(0,0)}^{\rT}=\frac{\mb{Z}^2-\lambda_0\Box^2}{(\lambda_{\frac12}-\lambda_0)\Box^2}=\frac{4}{3\Box}\mb{Z}^2$.}
\begin{align}
\Pi_{(s,s)}^{\rT} = 
\prod_{k=0}^{s}\frac{\big(\mb{Z}^2-\lambda_k\Box^2\big)}{(\lambda_{s+\frac12} - \lambda_{k})\Box^2}~
\prod_{l=0}^{s-1}\frac{\big(\mb{Z}^2-\lambda_{l+\frac12}\Box^2\big)}{(\lambda_{s+\frac12} - \lambda_{l+\frac12})\Box^2}~.\label{SuperprojectorMaximal}
\end{align}
For all other superspins $\hat{s}\leq s$ this will pick out an eigenspace which is doubly degenerate.
To resolve this, we break the degeneracy by composing with an operator which provides a bisection of the eigenspace into two disjoint spaces. Before elaborating, let us introduce some notation. We denote by $\mathbf{V}_{(s,j)}^{\mc{A}}$ the linear space of superfields of the form $(\pa_{\a\ad})^{s-j}\Phi_{\a(j)\ad(j)}^{\mc{A}}$, or simply $\pa_{(s-j)}\Phi_{(j)}^{\mc{A}}$, where $\Phi_{(j)}^{\mc{A}}$ is irreducible type $\mc{A}$. In the previous subsection we showed that $\mathbf{V}_{(s)}=\oplus_{j\in\{0,\dots,s\}}^{\mc{A}}\mathbf{V}_{(s,j)}^{\mc{A}}$.

The eigenspace corresponding to superspin $\hat{s} = j$, with $j\leq s$ an integer, may be selected using the operator
\begin{align}
\prod_{0 \leq k \leq s}^{k\neq j}\frac{\big(\mb{Z}^2-\lambda_k\Box^2\big)}{(\lambda_{j} - \lambda_{k})\Box^2}~
\prod_{l=0}^{s}\frac{\big(\mb{Z}^2-\lambda_{l+\frac12}\Box^2\big)}{(\lambda_{j} - \lambda_{l+\frac12})\Box^2}~.
\end{align} 
From figure \ref{fig1} we see that the corresponding eigenspace $\mathbf{V}_{(s,j)}^{+}\oplus \mathbf{V}_{(s,j)}^{-}$, consists of rank-$j$ transverse-chiral and transverse-anti-chiral superfields of the form $\pa_{(s-j)}\Phi_{(j)}^{+}+\pa_{(s-j)}\Phi_{(j)}^{-}$. 
We exploit this by introducing the bisection
\begin{subequations}
	\begin{align}
	\mathds{1} &= \hat{\mc{F}}^{+}_{(j)} + \hat{\mc{F}}^{-}_{(j)}~,\label{chiralbisection}\\[7pt]
	\hat{\mc{F}}^{+}_{(j)} = \frac{1}{16\Box} \big(\bar{D}^2D^2-D^{\g}\bar{D}^2&D_{\g}\big)~,\qquad \hat{\mc{F}}^{-}_{(j)} = \frac{1}{16\Box} \big(D^2\bar{D}^2-D^{\g}\bar{D}^2D_{\g}\big)~.
	\end{align}
\end{subequations}
The operator $\hat{\mc{F}}^{+}_{(j)}$ annihilates anti-chiral superfields and acts as the identity on chiral superfields, whilst the operator $\hat{\mc{F}}^{-}_{(j)}$ does the opposite.\footnote{The terms $D^{\g}\bar{D}^2D_{\g}$ may seem redundant, but without them \eqref{chiralbisection} does not hold, and nor will \eqref{resol3}.} Therefore, we arrive at the following alternative form of the $\Pi_{(s,j)}^{+}$ and $\Pi_{(s,j)}^{-}$ superprojectors
\begin{subequations}\label{superprojectorAlta}
	\begin{align}
	\hat{\Pi}_{(s,j)}^{+} & = \hat{\mc{F}}^{+}_{(j)}~\prod_{0 \leq k \leq s}^{k\neq j}\frac{\big(\mb{Z}^2-\lambda_k\Box^2\big)}{(\lambda_{j} - \lambda_{k})\Box^2}~
	\prod_{l=0}^{s}\frac{\big(\mb{Z}^2-\lambda_{l+\frac12}\Box^2\big)}{(\lambda_{j} - \lambda_{l+\frac12})\Box^2}~,\label{superprojector+}\\[5pt]
	\hat{\Pi}_{(s,j)}^{-} & = \hat{\mc{F}}^{-}_{(j)}~\prod_{0 \leq k \leq s}^{k\neq j}\frac{\big(\mb{Z}^2-\lambda_k\Box^2\big)}{(\lambda_{j} - \lambda_{k})\Box^2}~
	\prod_{l=0}^{s}\frac{\big(\mb{Z}^2-\lambda_{l+\frac12}\Box^2\big)}{(\lambda_{j} - \lambda_{l+\frac12})\Box^2}~,\label{superprojector-}
	\end{align}
\end{subequations}
which are almost purely in terms of $\mb{Z}^2$ and $\Box$.

Similarly, the eigenspace corresponding to superspin $\hat{s} = j+\frac12 $, with $j\leq s-1$ an integer, may be selected using the operator
\begin{align}
\prod_{k=0}^{s} \frac{\big(\mb{Z}^2-\lambda_k\Box^2\big)}{(\lambda_{j+\frac12} - \lambda_{k})\Box^2}
\prod_{0 \leq l \leq s}^{l\neq j}\frac{\big(\mb{Z}^2-\lambda_{l+\frac12}\Box^2\big)}{(\lambda_{j+\frac12} - \lambda_{l+\frac12})\Box^2}~.
\end{align} 
From figure \ref{fig1} we see that the corresponding eigenspace $\mathbf{V}_{(s,j+1)}^{\rL}\oplus \mathbf{V}_{(s,j)}^{\rT}$, consists of rank-$(j+1)$ ultra-longitudinal and rank-$j$ ultra-transverse superfields of the form $\pa_{(s-j-1)}\Phi^{\rL}_{(j+1)} + \pa_{(s-j)}\Phi_{(j)}^{\rT}$. To bisect we exploit the fact that, with respect to the operator $\mb{W}^2$, members of the space $\mathbf{V}_{(s,j+1)}^{\rL}$ have eigenvalue $\lambda_{j+1}\Box$ whilst those of $\mathbf{V}_{(s,j)}^{\rT}$ have eigenvalue $\lambda_{j}\Box$,
\begin{subequations}
	\begin{align}
	\mathds{1} = \hat{\mc{F}}^{\rL}_{(j+1)} &+ \hat{\mc{F}}^{\rT}_{(j)}~,\\[7pt]
	\hat{\mc{F}}^{\rL}_{(j+1)} = \frac{\mb{W}^2 - \lambda_{j}\Box}{(\lambda_{j+1}-\lambda_{j})\Box}~,&\qquad \hat{\mc{F}}^{\rT}_{(j)} = \frac{\mb{W}^2 - \lambda_{j+1}\Box}{(\lambda_{j}-\lambda_{j+1})\Box}~.
	\end{align}
\end{subequations}
The operator $\hat{\mc{F}}^{\rL}_{(j+1)}$ will annihilate elements of $\mathbf{V}_{(s,j)}^{\rT}$ and act as the identity on $\mathbf{V}_{(s,j+1)}^{\rL}$, whilst the operator $\hat{\mc{F}}^{\rT}_{(j)}$ will do the opposite.
Therefore, we arrive at the following alternative form of the $\Pi_{(s,j)}^{\rT}$ and $\Pi_{(s,j)}^{\rL}$ superprojectors (after shifting $j\rightarrow j-1$ for the latter)
\begin{subequations}\label{superprojectorAltb}
	\begin{align}
	\hat{\Pi}_{(s,j)}^{\rT} & = \hat{\mc{F}}^{\rT}_{(j)}~\prod_{k=0}^{s} \frac{\big(\mb{Z}^2-\lambda_k\Box^2\big)}{(\lambda_{j+\frac12} - \lambda_{k})\Box^2}
	\prod_{0 \leq l \leq s}^{l\neq j}\frac{\big(\mb{Z}^2-\lambda_{l+\frac12}\Box^2\big)}{(\lambda_{j+\frac12} - \lambda_{l+\frac12})\Box^2}~,\label{superprojectorT}\\[5pt]
	\hat{\Pi}_{(s,j)}^{\rL} & = \hat{\mc{F}}^{\rL}_{(j)}~\prod_{k=0}^{s} \frac{\big(\mb{Z}^2-\lambda_k\Box^2\big)}{(\lambda_{j-\frac12} - \lambda_{k})\Box^2}
	\prod_{0 \leq l \leq s}^{l\neq j-1}\frac{\big(\mb{Z}^2-\lambda_{l+\frac12}\Box^2\big)}{(\lambda_{j-\frac12} - \lambda_{l+\frac12})\Box^2}~.\label{superprojectorL}
	\end{align}
\end{subequations}
The superprojector $\hat{\Pi}_{(s,j)}^{\rL} $ is not defined for $j=0$.
We note that in the case $j=s$, the projector \eqref{superprojectorT} reduces to \eqref{SuperprojectorMaximal} when acting on $\mathbf{V}_{(s)}$ (this is not true when acting on $\mathbf{V}_{(l)}$ with $l>s$).

One may wonder if the two forms of the superprojectors \eqref{superprojectors1} and \eqref{superprojectorAlta} \& \eqref{superprojectorAltb} are really equivalent to each other. The answer is yes, as long as they are restricted to act on the spaces $\mathbf{V}_{(l)}$ with $0\leq l\leq s$.\footnote{In many cases this is trivially true, since   $\Pi_{(s,j)}^{\mc{A}}$ and $\hat{\Pi}_{(s,j)}^{\mc{A}}$ annihilate $\mathbf{V}_{(l)}$ with $0\leq l\leq j-1$, see \eqref{annihilatesuper}.} To prove this, one acts on some $\Phi_{(l)} \in \mathbf{V}_{(l)}$ with $\Pi_{(s,j)}^{\mc{A}}$ and then $\hat{\Pi}_{(s,j)}^{\mc{A}}$,
\begin{align}
\hat{\Pi}_{(s,j)}^{\mc{A}}\Pi_{(s,j)}^{\mc{A}} \Phi_{(l)} = \Pi_{(s,j)}^{\mc{A}} \Phi_{(l)}~,
\end{align}
where we have used that $\hat{\Pi}_{(s,j)}^{\mc{A}}$ acts as the identity on the space to which $\Pi_{(s,j)}^{\mc{A}}$ projects. Then we reverse the order of this operation,
\begin{align}
\Pi_{(s,j)}^{\mc{A}} \hat{\Pi}_{(s,j)}^{\mc{A}} \Phi_{(l)} = \hat{\Pi}_{(s,j)}^{\mc{A}} \Phi_{(l)}~.
\end{align}
Subtracting the second expression from the first, and using the fact that $\hat{\Pi}_{(s,j)}^{\mc{A}} $ and $\Pi_{(s,j)}^{\mc{A}}$ commute,\footnote{Since the $\hat{\Pi}_{(s,j)}^{\mc{A}}$ are almost purely constructed out of Casimir operators, the only non-trivial step in checking this is whether the $\hat{\mc{F}}_{(j)}^{\mc{A}}$ operator in $\hat{\Pi}_{(s,j)}^{\mc{A}}$ commutes with the $\mc{P}_{i}$ and $\Pi_{(s,j)}^{\perp}$ operators in \eqref{superprojectors1}. But this is true since $\mb{W}^2$ commutes with any scalar operator, even if it is constructed out of the spinor derivatives. } we find that they are equal on $\mathbf{V}_{(l)}$. It follows that on $\mathbf{V}_{(l)}$, superprojectors $\hat{\Pi}_{(s,j)}^{\mc{A}} $ possess all of the properties of $\Pi_{(s,j)}^{\mc{A}} $ that were listed in section \ref{secSupProj1}.

\section{Conclusion}\label{secCon}

In this paper we have derived the spin-$(s,j)$ projection operators $\Pi_{(s,j)}^{\perp}$ on any maximally symmetric $d$-dimensional background with $d\geq 3$. Denoting the isometry algebra of the latter by $\mf{g}$, the main purpose of $\Pi_{(s,j)}^{\perp}$ is to isolate the part of a traceless symmetric field $\phi_{a(s)}$ which furnishes the UIR of $\mf{g}$ with definite spin-$j$, where $0\leq j \leq s$. However, since the $\Pi_{(s,j)}^{\perp}$ are formulated in terms of the generators of $\mf{g}$ in an arbitrary field representation, they may act on a field of any rank. This allowed us to discover many useful properties belonging to $\Pi^{\perp}_{(s,j)}$, which in turn allowed us to systematically derive numerous gauge invariant actions. 

Some of these actions -- such as the (i) massless Fronsdal actions \eqref{dFMFronsdalAction} and \eqref{dAdSFMFronsdalAction}; (ii) the partially-massless actions \eqref{dAdSPMAction}; and (iii) the conformal actions in section \ref{secGenCHSMd} --  are well known, but our method nevertheless provided a different perspective on their origin and construction.  Other actions -- such as the (i) massless higher-depth actions \eqref{dFMDepthSJAction} and \eqref{dAdSFMDepthSJAction}; (ii) actions without compensators \eqref{dFMActionNoComp1} and \eqref{randact}; and (iii) higher-depth topologically massive actions \eqref{tma1}, \eqref{tma2} and \eqref{tma3} -- are novel, and were natural consequences of our analysis, or applications thereof. We have also derived closed-form expressions for the superprojectors in four-dimensional $\mc{N}=1$ Minkowski superspace. We expect the methods and results obtained in this paper to have the following applications and generalisations:
\begin{itemize}
\item \textbf{Complete the classification of two-derivative actions.} In this paper we have investigated all two-derivative actions for $\phi_{a(s)}$ in M$^d$ which have a propagating massless spin-$s$ mode, and which only require a single auxiliary field for gauge invariance. It would be interesting to investigate the possibility of allowing more compensating fields. Through this analysis, one should reproduce, e.g., the triplet \cite{Francia} and quartet \cite{Buchbinder07} formulations for massless higher-spin fields, and possibly derive higher-tuplet extensions of these. In (A)dS$_d$ there is more freedom to construct actions due to the non-vanishing curvature, and our analysis was not exhaustive. In section \ref{secPM} we investigated the specific case when $c_{s-1}=0$ and $d_{s-1}\neq 0$ in \eqref{dAdSGenActc}, and this lead to the partially-massless actions (which required more than one compensator in general). An intriguing question is; what other types of gauge-invariant actions lurk behind different choices for the constants $c_j$ and $d_j$ in (A)dS$_d$?

\item \textbf{Classification of all higher-derivative actions.} In this paper we analysed gauge invariant actions for $\phi_{a(s)}$ of order two in derivatives.\footnote{We also investigated some specific higher-derivative ones,  e.g. the conformal actions in section \ref{secGenCHSMd} which do not require compensators. This analysis was non-exhaustive.  } The method we have employed can accommodate for actions with more than two derivatives, and it would be interesting to pursue such a classification. For this task the results in \cite{JM12} will be useful.

\item \textbf{Mixed symmetry and fermionic fields.} From the onset we have restricted our analysis to totally symmetric rank-$s$ fields. It would be interesting to lift this restriction and allow for tensor(-spinors) transforming under diverse representations of $\mf{so}(d-1)$. For fermionic fields of the type $\psi_{a(s),\gamma}$ with $\gamma$ a spinor index, the spin projectors should have the same structural form, but the eigenvalues projected out via the recipe \ref{linalg} will change. For general mixed symmetry representations in $d>4$, one may have to incorporate higher-order Casimir operators (see e.g. \cite{BB06, SKAP}) into the projectors.

\item \textbf{Massive higher-spin actions.} Before the work of Singh and Hagen \cite{SH} (see also \cite{Sharapov, KF23} for $d>4$), one of the leading methods for constructing massive higher-spin actions was through use of the spin-$(s,s)$ projectors, see e.g. \cite{Fproj, Chang}. This analysis was only completed for spin $s\leq 4$, presumably due to computational complexity. The prescription utilised to construct actions in section \ref{Sec4Actions} was based on gauge-invariance, and thus cannot be used in its current form to construct Singh-Hagen action for a massive spin-$s$ field. 
However, it may be possible to adapt our approach and reproduce Klishevic and Zinoviev's gauge-invariant St\"{u}ckelberg reformulation of the massive spin-$s$ field \cite{Klishevich1997pd}.

The construction of massive spin-$s$ action in M$^d$ could provide valuable insight for the analogous problem with off-shell $\mc{N}=1$ supersymmetry in $d=4$. More specifically, the off-shell $\mc{N}=1$ supersymmetric actions describing massive representations of $\mf{iso}(3,1|4)$ with arbitrary half-integer superspin were recently derived in \cite{Koutrolikos}. However, the corresponding actions for integer superspin representations remain elusive, and the superspin projectors developed here could provide a way forward.

\item \textbf{Massive higher-spin propagators.} In \cite{Singh1} the propagators for free massive higher-spin fields were derived using the spin-$(s,s)$ projector in $d=4$ (see \cite{IP,Lindwasser2023,KF23} for an alternative method in generic $d$). It should be possible to extend this analysis to M$^d$ and (A)dS$_d$ using the results in this paper.

\item \textbf{Classification of two-derivative $\mc{N}=1$ actions.} In this paper we have only constructed the superspin projection operators, and have not used them to build gauge invariant actions. One could attempt to use the superprojectors to classify all two-derivative (and ultimately higher-derivative) actions for gauge supermultiplets with integer and half-integer superspin. In particular, it would be interesting to investigate the existence of off-shell formulations for $\mc{N}=1$ massless supermultiplets with arbitrary (half-)integer superspin, other than the original transverse and longitudinal formulations \cite{KSP, KS93}.\footnote{An alternative method to construct such massless actions was proposed in \cite{Gates2013rka}. Here, the superfield content of these theories arose from the requirement that there must be a smooth transition between the descriptions of massive and massless irreps of the super Poincar\'e group in the massless limit.} 
Such a classification has been performed \cite{SG2, GKP} for the low superspin multiplets ${\Phi}_{\a\ad}$ and ${\Phi}_{\a}$. 

We note that novel formulations of the massless half-integer superspin multiplet were computed in \cite{Buchbinder2020yip} at the level of the equations of motion. These descriptions are  higher-spin generalisations of new-minimal \cite{Akulov1976ck, Sohnius1981tp, Howe1981et, Gates1981tu} and new-new-minimal (also known as virial) \cite{Buchbinder2002gh, GKP, Nakayama2014kua} linearised supergravity. The action leading to these equations of motion should emerge from the complete classification of all two-derivative actions for gauge supermultiplets within the superprojector framework.

\item \textbf{Diverse superprojector extensions.} 
In 1981, Rittenberg and Sokatchev  \cite{RS} derived the superprojectors in four-dimensional $\cN$-extended Minkowski superspace $\text{M}^{4|4\cN}$, although their attention was restricted to superfields constrained to be (vector-)transverse. Around the same time, Gates and Siegel \cite{SG} created a powerful algorithm to obtain the $\mc{N}$-extended superprojectors for an unconstrained superfield, and gave some explicit examples for lower-rank superfields. It would be interesting to use our method to obtain closed-form expressions for the $\mc{N}$-extended superprojectors for an unconstrained superfield belonging to a generic irrep of $\sSU(\mc{N})\otimes \sSL(2,\mb{C})$.  Extensions of the superprojectors to anti-de Sitter superspace and to diverse dimensions are also desirable.\footnote{Some progress has already been made in this direction in \cite{BHKP, HKP, Hutchings2023,BuchbinderHutchingsHutomoKuzenko2019,HutchingsKuzenkoRaptakis2023}.} 

\item \textbf{$\mc{N}=2$ massless higher-spin actions.} An off-shell formulation for actions describing massless $\mc{N}=2$ supersymmetric representations  with arbitrary integer superspin in $d=4$ were found recently \cite{BIZ}. They were obtained within the framework of $\mc{N}=2$ harmonic superspace. It would be interesting to investigate the existence of analogous models within the conventional $\mc{N}=2$ superspace, and their half-integer superspin cousins. The $\mc{N}=2$ superspin projection operators could facilitate such an analysis. 

\end{itemize}

\noindent
{\bf Acknowledgements:} \\
The authors are grateful to Sergei Kuzenko for collaboration at an early stage of this project and for comments on the manuscript, and to the referee for useful suggestions.  DH would like to thank Mirian Tsulaia for useful discussions and Ruben Manveylan for encouraging the formulation of the (A)dS$_d$ projectors for $d > 4$. A preliminary version of this work was presented by DH at the CQUeST-APCTP workshop \textit{Gravity beyond Riemannian Paradigm} (Jeju island, South Korea).  DH is grateful to the organisers of this workshop for the stimulating scientific atmosphere and for the generous support. DH would also like to acknowledge the warm hospitality extended to him by the Quantum Gravity Unit at Okinawa Institute of Science and Technology where final revisions were completed.  This work was supported in part by the Australian Research Council, project No. DP200101944 and DP230101629. 

\appendix

\section{Helicity projection operators in $\text{M}^3$}\label{AppHel}

In the introduction we noted that the massive spin-$s$ UIR $D(m,s)$ of $\mf{iso}(d-1,1)$ can be realised on a rank-$s$ field satisfying the Klein-Gordon equation 
\begin{align}
0 &= (\Box - m^2) \phi^{\perp}_{a(s)}~, \label{aKG1}
\end{align}
and the transverse  and traceless  constraints 
\begin{subequations} \label{AppCons}
\begin{align}
0 &= \pa^{b}\phi^{\perp}_{ba(s-1)}~,\label{aC1}\\ 
0 &= \phi^{\perp}_{a(s-2)b}{}^{b}~. \label{aC2}
\end{align}
\end{subequations}
In $d=3$ dimensions, when $\phi_{a(s)}^{\perp}$ satisfies constraints \eqref{aC1} and \eqref{aC2}, the Klein-Gordon equation \eqref{aKG1} factorises according to
\begin{align}
0=(\mb{W}+sm)(\mb{W}-sm)\phi_{a(s)}^{\perp}~, \label{Factor}
\end{align}
where $\mb{W}$ is the Pauli-Lubanski pseudo-scalar\footnote{We adopt the convention whereby $\ve_{012}=-1$.}
\begin{align}
\mb{W}:=-\frac{1}{2}\ve^{abc}\pa_a M_{bc}~,\qquad \big[\mb{W},\pa_a\big]=0~.
\end{align}
The latter is a quadratic Casimir operator of $\mf{iso}(2,1)$ and is related to the quartic Casimir operator $\mb{W}^2$ defined in \eqref{spinOp} through $\mb{W}^2=\mb{W}\mb{W}$.

The factorisation \eqref{Factor} suggests that the representation $D(m,s)$ is reducible and the KG equation should be replaced with \cite{Kuzenko96,GitmanTyutin1996}
\begin{align}
0 = (\mb{W} - \sigma s m) \phi^{\perp}_{a(s)} \qquad \Longleftrightarrow \qquad \ve^{bc}{}_{a}\pa_b\phi^{\perp}_{ca(s-1)}=\sigma m \phi_{a(s)}^{\perp}~, \label{heleq}
\end{align} 
where $\sigma=\pm 1$. Indeed, in $d=3$ the (universal cover of the) massive little group is $\mb{R}$. This means that the spin, or, more precisely, the helicity of a massive UIR of $\mf{iso}(2,1)$ can take any real number.  
In accordance with this observation, a field satisfying \eqref{AppCons} and \eqref{heleq}   is said to carry mass $m>0$, helicity $\sigma s$ and spin $s=0,1,\cdots$ and furnishes the UIR $D(m,s,\s)$. For $s=0$ there is only the single helicity value $0$.

It follows from \eqref{Factor}  that a field satisfying  \eqref{aKG1} and \eqref{AppCons} furnishes the reducible representation $D(m,s)=D(m,s,+)\oplus D(m,s,-)$.\footnote{The representation $D(m,s)$ is  reducible with respect to the \textit{connected} Poincar\'{e} group, but irreducible with respect to the \textit{full} Poincar\'{e} group. This is because parity transformations flip the sign of helicity. }
A traceless symmetric field $\phi_{a(s)}$ satisfying the KG equation \eqref{aKG1} furnishes the reducible  representation 
\begin{align}
\bigoplus_{j=0}^s  D(m , j ,+)\oplus D(m ,  j ,-) ~.
\end{align}
In section \ref{secProjMd} we constructed the spin-$(s,j)$ projection operators $\Pi_{(s,j)}^{\perp}$ which select the representation $D(m , j,+)\oplus D(m , j,-)$ from this decomposition. Here we construct the helicity-$(s,j,\s)$ projection operators $\Pi_{(s,j,\s)}^{\perp}$ which extract the part of $\phi_{a(s)}$ on the mass-shell \eqref{aKG1} carrying definite helicity $\sigma j$, hence furnishing the UIR $D(m, j,\sigma)$. Such projectors could be useful, for example, in building parity violating actions. 

The procedure for extracting $D(m,j)$ from $\phi_{a(s)}$, with $0\leq j \leq s$, employed in the body is as follows: (i) assume $\phi_{a(s)}$ satisfies the massive KG equation; (ii) construct the projection operator $\Pi_{(s,j)}^{\perp}$; and (iii) impose the constraint $\phi_{a(s)}=\Pi_{(s,j)}^{\perp}\phi_{a(s)}$. The last step is equivalent to imposing $\phi_{a(s)}=\pa_{\{a_1}\cdots\pa_{a_{s-j}}\phi^{\perp}_{a_{s-j+1}\cdots a_s\}}$ which implies $0=(\pa^b)^{s-j+1}\phi_{b(s-j+1)a(j-1)}$. To select the UIR $D(m,j,\sigma)$ we must amend the third step in this process. In particular, we first bisect the operator $\Pi_{(s,j)}^{\perp}$ according to 
\begin{align}
\Pi_{(s,j)}^{\perp} = \Pi_{(s,j,+)}^{\perp}+\Pi_{(s,j,-)}^{\perp}~,
\end{align}
where we have defined 
\begin{align}
\Pi_{(s,j,\sigma)}^{\perp} := \frac{1}{2}\Big(\mathds{1}+\s\frac{\mb{W}}{j\sqrt{\Box}}\Big)\Pi_{(s,j)}^{\perp}~.
\end{align}
Step three is then replaced with; (iii) impose the constraint $\phi_{a(s)}=\Pi_{(s,j,\s)}^{\perp}\phi_{a(s)}$.

The helicity projector $\Pi_{(s,j,\s)}^{\perp}$ inherits most of the properties that $\Pi_{(s,j)}^{\perp}$ possesses, in addition to some others. In particular, they resolve the identity on $\mathscr{V}_{(l)}$ with $l$ arbitrary
\begin{align}
\mathds{1} = \sum_{j=0}^s\sum_{\s=\pm}\Pi_{(s,j,\s)}^{\perp}~,
\end{align}
and are orthogonal on $\mathscr{V}_{(l)}$ with $0\leq l \leq s$
\begin{align}
\Pi_{(s,j,\s)}^{\perp}\Pi_{(s,k,\s')}^{\perp}=\delta_{jk}\delta_{\s\s'}\Pi_{(s,j,\s)}^{\perp}~.
\end{align}
However, unlike $\Pi_{(s,j)}^{\perp}$ (see equation \eqref{actid2}), $\Pi_{(s,j,\s)}^{\perp}$ does not act as the identity operator on the space of transverse rank-$j$ fields. Properties \eqref{equivalence} - \eqref{did53} remain valid for $\Pi_{(s,j,\s)}^{\perp}$, and they have the additional property 
\begin{align}
0=\big(\mb{W}-\s j \sqrt{\Box}\big) \Pi_{(s,j,\s)}^{\perp} \phi_{a(s)}~.
\end{align}
This ensures $\Pi_{(s,j,\s)}^{\perp}\phi_{a(s)}$ has definite helicity $\s j$ and furnishes $D(m, j,\s)$ when restricted to the mass-shell. For $j=s$ the helicity projectors were first derived in \cite{BKLFP} (see also \cite{HKP} for a derivation using Casimir operators).

Finally, we point out the following equivalent expression for the helicity projectors
\begin{align}
\P^{\perp}_{(s,j,\s)} = \frac{\big (\mb{W} + \s j\sqrt{\Box}\big )}{\big ( \s j + \s j \big ) \sqrt{\Box}} \prod_{0 \leq k \leq s}^{k \neq j}\frac{\big ( \mb{W} - k\sqrt{\Box}  \big ) \big (\mb{W} + k\sqrt{\Box}   \big )}{\big ( \s j - k \big ) \big ( \s j + k \big )\Box} ~. 
\end{align}
This is in agreement with the general algorithm \eqref{linalg}, where all eigenvalues of $\mb{W}$ are projected out except for the desired one.

\section{Towards an explicit representation of $\Pi_{(s,j)}^{\perp}$ on $\mathscr{V}_{(s)}$}\label{AppExp}

In this section we convert the spin-$(s,j)$ projectors in M$^d$ from the representation 
\begin{align}
\Pi_{(s,j)}^{\perp} = \prod_{0\leq k \leq s}^{k \neq j} \frac{\big(\mb{W}^2-\lambda_k\Box\big)}{(\lambda_j-\lambda_k)\Box}~, \label{Appsjproj}
\end{align}
which is in terms of Casimir operators, to one using only $\pa_a$ and $\eta_{ab}$. To do this we must have $\Pi_{(s,j)}^{\perp}$ act on a traceless symmetric field $\phi_{a(s)}$, i.e. restrict $\Pi_{(s,j)}^{\perp}$ to the space $\mathscr{V}_{(s)}$, and evaluate each occurrence of $\mb{W}^2 = -\frac{1}{2}\Box \mb{M}^2+\pa^a\pa^bM_{a}{}^{c}M_{bc}$ using \eqref{W2red}.

Before evaluating \eqref{Appsjproj} by brute force, we first recall from \eqref{Longi2a} that 
\begin{align}
\Pi_{(s,j)}^{\perp} \phi_{a(s)} &= \pa_{\{a_1}\cdots\pa_{a_{s-j}}\phi^{\perp}_{a_{s-j+1}\dots a_s\}}~,
\end{align}
where we have defined 
\begin{align}
\phi_{a(j)}^{\perp} &= \Box^{j-s}\Big[\prod_{k=1}^{s-j}\frac{(j+k)(2j+2k+d-4)}{k(2j+k+d-3)}\Big]\Pi_{(j,j)}^{\perp}(\pa^b)^{s-j}\phi_{a(j)b(s-j)}~. \label{Applow}
\end{align}
To obtain \eqref{Applow} from \eqref{divextract} we have used the equivalence relations \eqref{equivalence} to replace $\Pi_{(s,j)}^{\perp}$ with $\Pi_{(j,j)}^{\perp}$. We see that the problem reduces to finding the explicit expression for $\Pi^{\perp}_{(j,j)}\psi_{a(j)}$ (here $\psi_{a(j)}$ is arbitrary). This is a simpler problem, and it is not difficult to show that 
\begin{align}
\Pi^{\perp}_{(j,j)}\psi_{a(j)} = \sum_{l=0}^j(-1)^l\binom{j}{l}\frac{(2j-l+d-4)!(2j+d-4)!!}{(2j+d-4)!(2j-2l+d-4)!!}\Box^{-l}\pa_{\{a_1}\cdots\pa_{a_l} (\pa^b)^l\psi_{a_{l+1}\dots a_j\}b(l)}~,
\end{align}
where $k!! = k(k-2)(k-4)\cdots 1$ if $k$ is odd, and $k!! = k(k-2)(k-4)\cdots 2$ if $k$ is even. 
Combining these results, we arrive at the following expression for $\phi_{a(j)}^{\perp}$
\begin{align}
\phi_{a(j)}^{\perp}&=\sum_{l=0}^j(-1)^l\binom{s}{j}\binom{j}{l}\frac{(2j-l+d-4)!(2s+d-4)!!(2j+d-3)}{(s+j+d-3)!(2j-2l+d-4)!!}\non\\
&\phantom{=} \times\Box^{j-s-l}\pa_{\{a_1}\cdots\pa_{a_{l}}(\pa^{b})^{s-j+l}\phi_{a_{l+1}\dots a_j\}b(s-j+l)}~,\label{divstructure}
\end{align}
and thus $\Pi_{(s,j)}^{\perp} \phi_{a(s)}$  is given by
\begin{align}
\Pi_{(s,j)}^{\perp} \phi_{a(s)} &=  \sum_{l=0}^j(-1)^l\binom{s}{j}\binom{j}{l}\frac{(2j-l+d-4)!(2s+d-4)!!(2j+d-3)}{(s+j+d-3)!(2j-2l+d-4)!!}\non\\
&\phantom{=} \times\Box^{j-s-l}\pa_{\{a_1}\cdots\pa_{a_{s-j+l}}(\pa^{b})^{s-j+l}\phi_{a_{s-j+l+1}\dots a_s\}b(s-j+l)}~.\label{explicit}
\end{align}

In the form \eqref{explicit} the symmetric and traceless nature of the right hand side is manifest. However, one can forego this property and eliminate all occurrences of $\{\cdots\}$, in which case trace terms involving $\eta_{aa}$ will arise. To achieve this we need to derive an expression for $\pa_{\{a_1}\cdots\pa_{a_j}\psi_{a_{j+1}\cdots a_{j+k}\}}$ with $j$ and $k$ arbitrary. We recall that $\{\cdots\}$ stretching over multiple derivatives and tensors, such as in this expression, is defined as a nested operation (see eq. \eqref{nested}) that is to be evaluated from outside to inside, allowing repetitive use of 
\begin{align}
\pa_{\{a}\psi_{a(k)\}} = \pa_a \psi_{a(k)} -\frac{k}{2k+d-2}\eta_{a(2)}\pa^{b}\psi_{ba(k-1)}~. \label{trsym1}
\end{align}
For example, using \eqref{trsym1} and $\pa_{\{a}\pa_a\psi_{a(k)\}} = \pa_{\{a}\hat{\psi}_{a(k+1)\}}$ with $\hat{\psi}_{a(k+1)} = \pa_{\{a}\psi_{a(k)\}}$ we have 
\begin{align}
\pa_{\{a}\pa_a\psi_{a(k)\}}&= \pa_a\pa_a\psi_{a(k)} -\frac{2k}{(2k+d)}\eta_{aa}\pa_a\pa^b\psi_{a(k-1)b}-\frac{1}{2k+d}\eta_{aa}\Box\psi_{a(k)}\non\\
&\phantom{=}+\frac{k(k-1)}{(2k+d)(2k+d-2)}\eta_{aa}\eta_{aa}\pa^b\pa^c\psi_{a(k-2)bc}~.
\end{align}
These are the relevant expressions for $j=1$ and $j=2$.

For $\pa_{\{a_1}\cdots\pa_{a_j}\psi_{a_{j+1}\cdots a_{j+k}\}}\equiv (\pa_{\{a})^j\psi_{a(k)\}} $ with $j$ arbitrary we propose the ansatz\footnote{Strictly speaking, the lower bound for the summation over $m$ should be max$(\{0,\lceil \frac12 (j-l-k)\rceil\})$. This is because the rank $k$ of $\psi_{a(k)}$ places an upper bound on the number of divergences that can be taken.}
\begin{align}
(\pa_{\{a})^j\psi_{a(k)\}} = \sum_{l=0}^j\sum_{m=0}^{\lfloor \frac12 (j-l) \rfloor}\Gamma_{(l,m)}(\eta_{aa})^{j-l-m}\Box^m(\pa_{a})^l(\pa^b)^{j-l-2m}\psi_{a(k-j+l+2m)b(j-l-2m)}~. \label{transatz}
\end{align} 
The constants $\Gamma_{(l,m)}$ may be determined via the following steps: (i) take the trace of \eqref{transatz} and require this to vanish; and (ii) use the inductive property 
\begin{align}
\pa_{\{a_1}\cdots\pa_{a_{j+1}}\psi_{a_{j+2}\cdots a_{j+k+1}\}} = \pa_{\{a_1}\cdots\pa_{a_j}\hat{\psi}_{a_{j+1}\cdots a_{j+k+1}\}}~.
\end{align}
This will yield two separate recurrence relations on $\Gamma_{(l,m)}$ that should enable one to solve for $\Gamma_{(l,m)}$. For arbitrary $j$ this is a tedious task which we will not complete here.


\begin{footnotesize}

\end{footnotesize}

\end{document}